%% file: 3D.tex
\documentclass[12pt]{article}
\usepackage{algorithm}
\usepackage{algorithmic}
\usepackage{amsmath}
\usepackage{verbatim}
\usepackage{amssymb}
\usepackage{latexsym}
\usepackage{fullpage}
\usepackage{color}
\usepackage{boxit}
\usepackage{epsfig}

\newtheorem{definition}{Definition}[section]
\newtheorem{theorem}{Theorem}[section]
\newtheorem{corollary}{Corollary}[section]
\newtheorem{lemma}{Lemma}[section]
\newtheorem{claim}{Claim}[section]
\newtheorem{proposition}{Proposition}[section]
\newtheorem{conjecture}{Conjecture}[section]
\newtheorem{remark}{Remark}[section]

\newcommand{\de}{\mathrm{\,d}}
\input{psfig}

\begin{document}

\def\eps{\varepsilon}
\def\D{\cal D}
\def\b{\mathbf{b}}
\def\bb{\mathbf{b_1}}
\def\f{\mathbf{f}}
\def\C{{\cal C}}
\def\D{{\cal D}}
\def\L{{\cal L}}
\def\T{{\cal T}}
\def\I{{\cal I}}
\def\V{{D_\eps}}
\def\A{{\cal A}}
\def\l{{\ell}}
\def\w{\dot{w}}
\def\dx{\dot{x}}
\def\dy{\dot{y}}
\def\tp{\tilde{\pi}}
\def\x{\mathbf{x}}
\def\y{\mathbf{y}}
\def\a{\mathbf{a}}
\def\k2{\kappa^2}
\def\s2{\sin^2\varphi}
\def\c2{\cos^2\alpha}
\def\f{\mathbf{f}}
\def\boldrho{\boldsymbol{\rho}}
\def\sign{{\rm sign}}

\title{An Approximation Algorithm for Computing\\ Shortest Paths in Weighted 3-d Domains\thanks{Research
supported by NSERC. Preliminary results have appeared in \cite{AMS00}.}}

\author{
Lyudmil Aleksandrov\thanks{ Bulgarian Academy of Sciences, IPP,
Acad. G. Bonchev Str. Bl. 25-A, 1113 Sofia, Bulgaria. {\sf lyualeks@bas.bg}}
\and Hristo Djidjev \thanks{ Los Alamos National Laboratory, Los Alamos, NM 87544, U.S.A}
\and Anil Maheshwari \thanks{ School of Comp. Sci., Carleton U., Ottawa, Ontario K1S5B6, Canada.
{\sf anil@scs.carleton.ca}}
\and  J\"org-R\"udiger Sack \thanks {School of Comp. Sci., Carleton U., Ottawa, Ontario K1S5B6, Canada.
{\sf sack@scs.carleton.ca}} } \maketitle
\sloppy
\begin{abstract}
\label{CONTRIBUTIONS}
We present the first polynomial time approximation algorithm for computing shortest paths
in weighted three-dimensional domains. Given a polyhedral domain $\D$, consisting of $n$
tetrahedra with positive weights, and a real number $\eps\in(0,1)$, our algorithm constructs
paths in $\D$ from a fixed source vertex  to all vertices of $\D$, whose costs are at most $1+\eps$
times the costs of (weighted) shortest paths, in
$O(\C(\D)\frac{n}{\eps^{2.5}}\log\frac{n}{\eps}\log^3\frac{1}{\eps})$ time, where
$\C(\D)$ is a geometric parameter related to the aspect ratios of tetrahedra.

The efficiency of the proposed algorithm is based on an in-depth study of the local
behavior of geodesic paths and additive
Voronoi diagrams in weighted three-dimensional domains, which are of independent interest.
The paper extends the results of Aleksandrov et al. \cite{AMS05} to three dimensions.
\end{abstract}

\section{Introduction}
\subsection{Motivation}
The computation of shortest paths 
is a key problem arising in a number of diverse application areas
including geographic information systems, robotics, computer graphics,
computer-aided design, medical computing and others. This has motivated 
the study and subsequent design of efficient algorithms for solving shortest
path problems in different settings based on the geometric nature of the problem domain
(e.g., two-dimensional (2-d), three-dimensional (3-d), surfaces, presence/absence of obstacles) and the cost function/metric
(e.g., Euclidean, $L_p$, link distance, weighted/unweighted, multi-criteria). In addition to
its driver - the applications - the field has provided, and continues to do so, exciting
challenges from a theoretical perspective. As a result, shortest path problems have become
fundamental problems in areas of  Computer Science such as Computational Geometry and
Algorithmic Graph Theory.

The standard 3-d Euclidean shortest path problem of computing a shortest
path between pair of points avoiding a set of polyhedral obstacles, denoted as the ESP3D problem,
is known to be $NP$-hard even when the obstacles are parallel triangles in the space.  It is not
difficult to see that the number of combinatorially distinct shortest paths from
a source point to a destination point may be exponential in the input size.
Canny and Reif \cite{CR87} used this to establish the $NP$-hardness of the ESP3D problem,
for any $L_p$ metric, $p\geq 1$. In addition to this combinatorial hardness result,
Bajaj \cite{Baj88} has provided an algebraic hardness argument that an exponential number
of bits may be required. More recently, Mitchell and Sharir \cite{MS04} gave
$NP$-completeness proofs for  the problem of computing Euclidean shortest paths among sets of stacked axis-aligned
rectangles, and computing $L_1$-shortest paths among disjoint balls.
Given the $NP$-hardness of the ESP3D problem, work has focused on exploiting the geometric structure
of the obstacles and/or on providing approximation algorithms. We will mention some of
these approaches in Section \ref{Prev}.

In many applications, the Euclidean metric does not capture adequately
the nature of the problem, for instance when the problem domain is not homogeneous.
This motivates the weighted versions of the shortest path problem. For example in the 2-d case,
consider triangulated regions where each triangle represents a particular terrain type such
as water, rock, or forest. Here different weights capture the cost of traveling a Euclidean
unit-length through each face.
Incorporating weights makes the solution more difficult to obtain even in 2-d, but it does
provide more realistic answers.  It is known that  light and other types of waves  (e.g., seismic
and sonic) travel along the shortest paths in heterogeneous media. Hence, algorithms solving
the weighted shortest path problem (WSP3D) can be used for modeling wavefront propagation in such media.
In the 3-d, a number of applications are non-homogeneous in nature and can be expressed using the weighted model.
Next, we list some  of such potential applications.

\begin{itemize}
\item
In \emph{geology},  seismic refraction and reflection methods are used based on measurements of the travel time of seismic waves refracted at the interfaces between subsurface layers of different
densities. As such waves propagate along shortest paths and weighted shortest path algorithms may be used to produce more accurate and more efficient estimation of subsurface layer characteristics, e.g., the amount of oil contained in the subsurface \cite{IL94}. Another related application is the assessment of garbage dumps' health. When a new
garbage dump is built, sensors are placed at the bottom, and when the garbage dump starts to fill, waves from the top passing through the garbage to these sensors are used in order to determine the decomposition rate of the garbage  \cite{IL94}.
\item
Computation of 3-d  shortest path have also been used to compute \emph{fastest routes} for aircrafts
between designated origin and destination points while avoiding hazardous, time-varying
weather systems. Krozel et al.\ \cite{Krozel2004446} investigate synthesizing weather
avoidance routes in the transition airspace.
Our weighted 3-d region model can be used to generalize that approach: instead of
totally avoiding undesirable regions, one can assign penalty weights to them and then
search for routes that minimize travel through such regions, while also avoiding
unnecessarily long detours.
\item
In \emph{medical applications} simulation of sonic wavefront propagation is used
when performing imaging methods as  photoacoustic tomography or ultrasound imaging through heterogeneous tissue \cite{CT10,VT05}. In radiation therapy, domain features include densities of tissue, bone, organs, cavities, or risk to radiation exposure, and optimal radiation treatment planning takes this non-homogeneity into consideration.
%
%
\item
The problem of time-optimum \emph{movement planning} in 2-d and 3-d for a point
robot that has bounded control velocity through a set of $n$ polygonal regions of given
translational flow velocities has been studied by Reif and Sun \cite{Reif2004127}. They
state that  this intriguing geometric problem has immediate applications to macro-scale
motion planning for ships, submarines, and airplanes in the presence of significant
flows of water or air.
Also, it is a central motion planning problem for many of the meso-scale
and micro-scale robots that have environments with significant flows that can affect their
movement. They establish the computational hardness for the 3-d version of this problem
by showing the problem to be PSPACE hard. They give a decision algorithm for the 2-d flow
path problem, which has very high computational complexity, and they also design an efficient
approximation algorithm with bounded error. The determination of the exact computational complexity
of the 3-d flow path problem is posed as an open problem.
Although, our weighted 3-d model does not apply directly to this setting, it can be used to construct
initial approximations by assigning appropriate weights depending on the velocity and direction of the
flows in different regions. In addition,  the discretization scheme and the algorithmic
techniques developed here can be used for solving the 3-d flow path problem.
\end{itemize}

\subsection{Problem formulation}
In this paper, we consider the following problem. Let $\D$ be a connected 3-d domain consisting of $n$
tetrahedra with a positive real weight associated to each of them. The 3-d weighted shortest path
problem (WSP3D) is to compute minimum-cost paths in $\D$ from a fixed source vertex to all vertices of $\D$.
The cost of  a path in $\D$ is defined as the weighted sum of the Euclidean lengths of the sub-paths within each crossed tetrahedron. We will describe and analyze an approximation algorithm for this problem that, for any real number 
$\eps\in (0,1)$, computes paths whose costs are at most $1+\eps$ times greater
than the costs of the minimum cost paths. In Section \ref{model}, we describe our model in detail.

Note that the WSP3D problem can be viewed as a generalization of the ESP3D problem. Namely, given an instance
of the ESP3D problem, one can find a large enough cube containing all the obstacles, tetrahedralize the free-space (i.e., exterior of the obstacles, but in the interior of the cube) and set equal weights to the resulting tetrahedra obtaining an instance of the WSP3D problem.

\subsection{Challenges}
A key difference between Euclidean shortest path computation in 2-d and 3-d weighted domain is the $NP$-hardness
already mentioned. Underlying this is the fact that, unlike in 2-d, Euclidean 3-d shortest paths are not discrete.
Specifically, in 2-d, the edges of a shortest path (e.g., Euclidean shortest paths among obstacles in the plane)
are edges of a graph, namely, the visibility graph of the obstacles including the source and the destination points.
In contrast, in polyhedral 3-d domains, the bending points of shortest paths  on obstacles may lie in the interior
of the obstacles' edges. Moreover, in weighted 3-d settings, bending points may even belong to the interior of the faces.

Furthermore, even in the case of weighted triangulated planar domains, the (weighted) shortest path
may cross each of the $n$ cells $\Theta(n)$ times and may be composed of $\Theta(n^2)$ segments in total.
Not only is the path complexity higher, but the computation of weighted shortest paths in 2-d turns out to be substantially more involved than in the Euclidean setting. In fact, there is not even an exact algorithm known,
and the first $(1+\eps)$ approximation algorithm due to \cite{Mit91} had an $O(n^8\log(\frac{n}{\eps}))$
time bound, where $n$ is the number of triangles in the subdivision. This problem has been actively researched since then, and currently the best known algorithm for the weighted region problem on planar subdivisions (as well as on polyhedral surfaces) runs in $O(\frac{n}{\sqrt\varepsilon}\log\frac{n}{\varepsilon}\log\frac{1}{\varepsilon})$ time \cite{AMS05}. (Also, see \cite{AMS05} for a detailed literature review for the planar case.)

One of the classical tools of Computational Geometry is the Voronoi Diagram. This structure finds numerous applications (see e.g., \cite{AK00}). It is also a key ingredient in several efficient shortest path algorithms.
Researchers have studied these diagrams under several metrics (including Euclidean, Manhattan, weighted, additive, convex, abstract) and for different types of objects (including points, lines, curves, polygons), but somehow the computation of these diagrams in media with different densities (i.e., the refractive media) remained elusive.
One of the main ingredients in solving the problem studied here is to compute (partial) additive Voronoi
diagrams of points in refractive media. The generic techniques of Klein \cite{Kl89,KLN09} and L\^e \cite{Le04}
do not apply in this case, as the bisecting surfaces do not satisfy the required conditions.
In this paper, we make an important step towards the  understanding and computation of  these diagrams.

\subsection{Previous related work} \label{Prev}
By now, shortest path problems in 2-d are fairly well understood. Efficient algorithms have been
developed for many problem instances and surveys are readily available describing the state of the art in
the field.

In 3-d, virtually all the work has been devoted to the ESP3D problem.
Papadimitriou \cite{Pap85}  suggested the first polynomial time approximation scheme
for that problem.  It runs in  $O(\frac{n^4}{\eps^2}(L+\log(n/\eps))$ time, where $L$ is the number of
bits of precision in the model of computation.
Clarkson \cite{Clk87a} provided an algorithm running in
$O(n^2\lambda(n)\log(n/{\varepsilon})/{(\varepsilon^4)}+n^2\log n\rho\log(n\log \rho))$ time,
where $\rho$ is the ratio of the longest obstacle edge to the distance between the source and the target
vertex, $\lambda(n)={\alpha(n)}^{O(\alpha(n))^{O(1)}}$, and $\alpha(n)$ is
the inverse Ackermann's function.

Papadimitriou's algorithm was revised and its analysis was refined by  Choi et al.\ \cite{Cho97} under the bit complexity framework. Their algorithm runs roughly in $O(\frac{n^4L^2}{\eps^2}\mu(X))$  time, where $\mu(X)$ represents the time (or bit) complexity of multiplying $X$-bit integers and $X=O(\log(\frac{n}{\varepsilon})+L)$.
In \cite{CSY}, the same authors further developed their ideas and proposed a precision-sensitive algorithm for the ESP3D problem. In \cite{AKY04}, Asano et al.\ proposed and studied a technique for computing approximate solutions to optimization problems and obtained another precision-sensitive approximation algorithm for the ESP3D problem with
improved running time in terms of $L$.

Har-Peled \cite{HP99} proposed an algorithm that invokes Clarkson's algorithm
as a subroutine $O(\frac{n^2}{\eps^2}\log\frac{1}{\eps})$ times to build a data structure  for
answering approximate shortest path queries from a fixed source in $O(\log\frac{n}{\eps})$ time.
The data structure is constructed in roughly $O(\frac{n^6}{\eps^4})$ time.
Agarwal et al. \cite{ASY09} considered the ESP3D problem for the case of convex obstacles and proposed an approximation algorithm running in $O(n+\frac{k^4}{\varepsilon^7}\log^2\frac{k}{\varepsilon}\log\log k)$ time, where $k$ is the number of obstacles. In contrast to all other algorithms discussed here, the complexity of  this algorithm does not depend on the geometric features of the obstacles. In the same paper, the authors describe a data structure for answering approximate shortest path queries from a fixed source in logarithmic time.

In the weighted (non-Euclidean) 3-d case no previous algorithms have been reported by other authors.
In \cite{AMS00}, we have announced and sketched a polynomial time approximation scheme for WSP3D problem
that runs in $O(\frac{n}{\varepsilon^{3.5}}\log\frac{1}{\varepsilon}(\frac{1}{\sqrt\varepsilon}+\log n))$ time.
The run-time improves to $O(\frac{n}{\varepsilon^3}\log \frac{1}{\varepsilon}\log n)$  when all weights are equal. This algorithm can be used to efficiently solve the ESP3D problem. In this paper, we apply that approach, but develop the required details,  apply new techniques, improve the complexity bounds, and provide a rigorous  mathematical analysis.

\subsection{Contributions of this paper}
In this paper, we make several contributions to the fields of shortest path computations and the
analysis of weighted 3-d regions model, as listed below.
\begin{itemize}
\item
We provide an approximation algorithm for solving the WSP3D problem in a
polyhedral domain $\D$ consisting of $n$ weighted tetrahedra. The algorithm computes approximate
weighted shortest paths from a source vertex to all other vertices of $D$  in
$O(\C(\D)\frac{n}{\eps^{2.5}}\log\frac{n}{\eps}\log^3\frac{1}{\eps})$ time,  where $\eps\in(0,1)$ is the user-specified approximation parameter and
$\C(\D)$ is a geometric parameter related to the aspect ratios of tetrahedra\footnote{See Lemma \ref{number-of-points} for details on the value of $\C(\D)$.}.
The cost of the computed paths are within a factor of $1+\eps$ of the cost of the corresponding shortest paths.

As we stated above, the ESP3D problem, i.e., the unweighted version of this problem, is already $NP$-hard even when the obstacles are parallel triangles in the space \cite{CR87}.
The time complexity of our algorithm, which is  designed for the more general weighted setting, compares favorably even when applied in the Euclidean setting to the existing approximation algorithms.

\item
Our detailed analysis, especially the results on additive Voronoi diagrams derived in Section \ref{model}, provides valuable insights into the understanding of Voronoi diagrams in heterogeneous media. This may open new avenues, for example,  for designing an algorithm to compute  discretized Voronoi diagrams in such settings.

\item
Our approximation algorithms in 2-d have proven to be easily implementable and of practical value \cite{LMS01}. Our algorithm for WSP3D presented here, in spite of being hard to analyze, essentially uses  similar primitives, and thus has the potential to be implementable, practical, and applicable in different areas.

\item
Our work provides further evidence that discretization is a powerful tool when solving shortest
path-related problems in both Euclidean and weighted settings. We conjecture that the discretization
methodology  used here generalizes to any fixed dimension.

Furthermore, our discretization scheme is  independent of the source vertex and can be used with no changes to approximate paths from other source vertices. This feature makes it applicable for solving  all pairs shortest paths problem and for designing  data structures for answering shortest path queries in weighted 3-d domains.

\item
The complexity of our algorithm does not depend on the weights assigned to the tetrahedra composing $\D$, but it depends on their geometry. We analyze and evaluate that dependence in detail. Geometric dependencies arise also in Euclidean settings and in most of the previous papers. For example, in Clarkson \cite{Clk87a}, the running time of the algorithm depends on the ratio of the longest edge to the distance between the source and the target vertex.
Applying known techniques (see e.g., \cite{ASY09}), such dependency can  often be removed. Here, this  would be possible provided that an upper bound on the number of segments on weighted shortest paths in 3-d is known. However, the increase in the dependency on $n$ in the time complexity that these techniques suffer from, which is of order $\Omega(n^2)$, appears not to justify such an approach here.
In our approach, the dependency on the geometry is proportional to the average of the reciprocal squared sinuses of the dihedral angles of the tetrahedra composing $\D$. Thus, when $n$ is large, many tetrahedra would have to be fairly degenerate so that this average to play a major role. We therefore conjecture that in typical applications, the approach presented here would work well.
\end{itemize}

\subsection{Organization of the paper}
In Section \ref{model}, we describe the  model used throughout this paper, formulate the problem,
present some properties of shortest paths in 3-d, and derive a key result on additive Voronoi diagrams in refractive media. In Section \ref{3D-Discretization}, we describe our discretization scheme which is a generalization of the 2-d scheme introduced in \cite{AMS05}. In Section \ref{discretepath-section}, we construct a weighted graph, estimate the number of its nodes and edges and prove that shortest paths in $\D$ can be approximated by paths in the graph so constructed. In Section \ref{Algorithms}, we present our algorithm for the WSP3D problem. In Section \ref{conclusions}, we conclude this paper.

\section{Problem formulation and preliminaries}
\label{model}

\subsection{Model} \label{model-section}
Let ${\cal D}$ be a connected polyhedral domain in  3-d
Euclidean space. We assume that ${\cal D}$ is partitioned into $n$ tetrahedra
$T_1, \dots, T_n$, such that ${\cal D}=\cup_{i=1}^n T_i$ and the intersection
of any pair of different tetrahedra is either empty or a common element (a face, an edge,
or a vertex) on their boundaries. We call these tetrahedra {\em cells}.
A positive weight $w_i$ is associated with each cell $T_i$ representing the cost of traveling in it.
The cost of traveling along a boundary element of a cell is the minimum of the weights of the cells
incident to that boundary element. We consider \emph{paths} (connected rectifiable curves) that
stay in ${\cal D}$.  The cost of a path $\pi$ in ${\cal D}$ is defined by $ \|\pi\|=\sum_{i=1}^n w_i|\pi_i|$,
where $|\pi_i|$ denotes the Euclidean length of the intersection $\pi_i=\pi\cap T_i$. Boundary edges and
faces are assumed to be part of the cell from which they inherit their weight.

Given two distinct points $u$ and $v$ in ${\cal D}$, the shortest
path problem in ${\cal D}$ is to find a minimum cost path
$\pi(u,v)$ between $u$ and $v$ that stays in ${\cal D}$. We refer
to the minimum cost paths as {\em shortest paths}. For a given
approximation parameter $\eps>0$, a path $\pi_\eps=\pi_\eps(u,v)$ is an
$\eps$-approximation of the shortest path $\pi=\pi(u,v)$, if
$\|\pi_\eps\|\leq(1+\eps)\|\pi\|$. Without loss of generality, we
may assume that the points $u$ and $v$ are vertices of $\D$, since,
if they are not, we can make them such by partitioning the cells
where they belong. In this paper, we present an algorithm that, for
a given source vertex $u$ and an approximation parameter
$\eps\in(0,1)$, computes $\eps$-approximate shortest paths from $u$
to all vertices of $\D$.

In this setting, it is well known \cite{Mit91}\footnote{The
2-d case was treated there, but the arguments readily
apply to the 3-d model considered in this paper.} that shortest paths are
simple (non self-intersecting) and consist of a sequence of
segments, whose endpoints are on the cell boundaries. The
intersection of a shortest path with the interior of a cell, a face,
or an edge is a set of disjoint segments. More precisely, each
segment on a shortest path  is of one of the following three
types:

(i) {\it cell-crossing} -- a segment that crosses a cell joining two points on its boundary;

(ii) {\it face-using} -- a segment lying along a face of a cell;

(iii) {\it edge-using} -- a segment along an edge of a cell.\\[1ex]

We define {\it linear paths} to be paths consisting of cell-crossing, face-using, and
edge-using segments exclusively. A linear path $\pi(u,v)$ can be represented as the sequence
of its segments $\{s_1, \dots, s_{l+1}\}$ or, equivalently, as the sequence of points
$\{a_0,\dots ,a_{l+1}\}$, lying on the cell boundaries that are endpoints of these segments, i.e., $s_i=(a_{i-1},a_i)$, $u=a_0$, and $v=a_{l+1}$. The points $a_i$ that are not vertices of
cells are called {\it bending points} of the path.

\begin{figure}[tbh]
\begin{center}
\resizebox{8cm}{!}{
\input{snells.pstex_t}}
\caption{An illustration of the Snell's law of refraction.} \label{snells}
\end{center}
\end{figure}
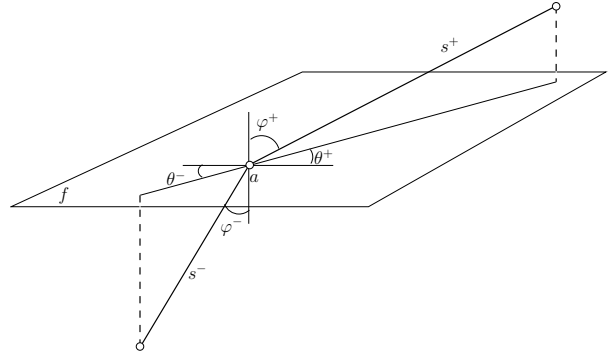

The local behavior of a shortest path around a bending point $a$, lying in the interior of a face $f$,
is fully described by the directions of the two segments of the shortest path, $s^-$ and $s^+$, that
are incident to $a$. The direction of each of these two segments is described by a pair of angles,
which we denote by $(\varphi^-, \theta^-)$ and $(\varphi^+, \theta^+)$, respectively.
The {\em in-angle} $\varphi^-$ is defined to be the acute angle between the direction normal to $f$
and the segment $s^-$. Similarly, the {\em out-angle} $\varphi^+$ is the acute angle between the normal
and the segment $s^+$. The angles $\theta^-$ and $\theta^+$ are the acute angles between the
orthogonal projections of $s^-$ and $s^+$ with a reference direction in the plane containing the face $f$, 
respectively (see Figure \ref{snells}).

It is well known that when $\pi$ is a shortest path it is a linear path such that the angles
$(\varphi^-, \theta^-)$ and $(\varphi^+, \theta^+)$ are related by Snell's law as follows:\\

\noindent{\bf Snell's Law of Refraction}:
{\em Let $a$ be a bending point on a geodesic path $\pi$ lying in the interior of a face $f$ of {\cal D}.
Let the segments preceding and succeeding $a$ in $\pi$ be $s^-$ and $s^+$, respectively. Let the weights
of the cells containing $s^-$ and $s^+$ be $w^-$ and $w^+$, respectively. Then $s^+$ belongs to the plane
containing $s^-$ and perpendicular to $f$ and the angles $\varphi^-$ and $\varphi^+$ are
related by $w^-\sin\varphi^- =w^+\sin\varphi^+$.}

We refer to linear paths that are locally optimal (i.e., satisfy the Snell's law) as {\em geodesic paths}.
Hence, the shortest path between pair of vertices $u$ and $v$ is the geodesic path of smallest cost joining
them. In the following we discuss some of the implications of  Snell's law on the local behavior of geodesic paths.

Hereafter, we denote by $\kappa$ the ratio $w^+/w^-$. Without loss of generality, we assume that
$w^-\geq w^+$, i.e., $\kappa\leq 1$. Let $\varphi^*$ be the acute or right angle for which
$\sin\varphi^*=\kappa$. We refer to this angle as the {\em critical angle} for the face $f$. From 
Snell's law, it follows that $\varphi^-\leq\varphi^*$. The case where $\varphi^-=\varphi^*$ deserves a special attention. In this case, $\varphi^+$ must be a right angle and therefore the segment $s^+$ is a face-using
segment. Furthermore, if the second endpoint $a_1$ of $s^+$ is in the interior of $f$, then the segment following $s^+$ is inside the tetrahedron containing $s^-$, and the out-angle at $a_1$ is equal to $\varphi^*$ (see Figure
\ref{snells1} (b)). In summary, if $s$ is a face-using segment, then it is preceded and followed by segments
lying in the cell with bigger weight and their corresponding in-angle and out-angles are equal to the critical
angle $\varphi^*$.

In the next subsection, we study the properties of simple geodesic paths joining points in neighboring cells or
in the same cell through a face-using segment. We define a function related to the cost of these geodesic paths
and prove a number of properties that it possesses. These properties are essential to the design and the analysis
of our algorithm.

\subsection{Weighted distance function} \label{wdf-section}
Let $F$ be a plane in the three-dimensional Euclidean space. We denote the two half-spaces defined by $F$ by
$F^-$ and $F^+$ and assume that positive weights $w^-$ and $w^+$ have been associated with them, respectively.
We extend our model by assigning a weight $w$ to $F$, so that $w=\min(w^-,w^+)$ if $w^-\not=w^+$, and
$0<w = w^+(=w^-)$ if $w^-=w^+$. The latter case models the situation where the geodesic path joins two points in the same cell through a face-using segment on the boundary of that cell.

\begin{figure}[tbh]
\begin{center}
\resizebox{14cm}{!}{
\input{f_wdf.pstex_t}}
\caption{The geodesic path $\bar{\pi}(v,\x)$
joining $v$ and $x$ is illustrated. The weighted distance function $c(v,x)$ equals to the cost
of $\bar{\pi}(v,\x)$, i.e.\ $c(v,x)=\|\bar{\pi}(v,\x)\|=w^-|va|+w^+|a\x|$.}
\label{fig:wdf}
\end{center}
\end{figure}

We refer to the half-spaces $F^-$ and $F^+$ as the {\em lower} and the {\em upper} half-space, respectively.
Let $v$ be a point in the lower half-space $F^-$ at distance $z^-$ from $F$ and $\ell$ be a line parallel to
$F$ in the upper half-space $F^+$ at distance $z^+$ from $F$. Let $O_{xyz}$ be a Cartesian coordinate system
such that the plane $O_{xy}$ coincides with $F$, $v$ has coordinates $(0,y,-z^-)$, and the line $\ell$ is
described by $\{\ell :\ y=0,\ z=z^+\}$.

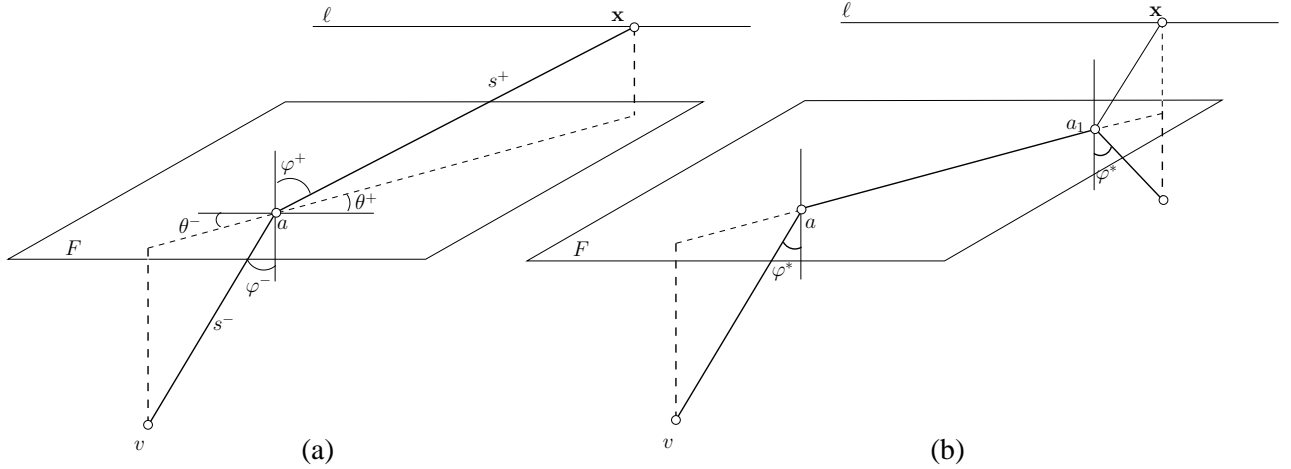
\begin{figure}[tbh]
\begin{center}
\resizebox{17cm}{!}{
\input{snells1.pstex_t}}
\caption{The diagrams illustrate the local structure of geodesic paths in different cases.}
\label{snells1}
\end{center}
\end{figure}

We consider a point $\x=(x,0,z^+)$ on $\ell$ and denote by $\bar{\pi}(v,\x)$ the geodesic
path between $v$ and $\x$. In this setting, the geodesic path is unique and thus coincides with the
shortest path. We denote the cost of this path by $c(v,x)$, where $x$ is the $x$-coordinate of $\x$.
 So, for fixed $l$, $c(v,x)$ can be viewed as a function defined for any real $x$. We call $c$ the
 {\em weighted distance function} from $v$ to $l$ (Figure \ref{fig:wdf}).

The local structure of the geodesic path $\bar{\pi}(v,\x)$  is governed by Snell's law. In the case where $w^-\not=w^+$, the shortest path  between $v$ and $\x$ consists of two segments $(v,a)$ and $(a,\x)$, where
the bending point $a$ is uniquely determined by Snell's law (Figure \ref{snells1} (a)).
In the case where $w^-=w^+$, the structure of the path $\bar{\pi}(v,\x)$ is as follows. It is a single segment $(v,\x)$, provided that the angle $\varphi$ between $(v,\x)$ and the direction normal to the plane $F$ is smaller
than or equal to the critical angle $\varphi^*$ defined by $\sin\varphi^*=w/w^-$. Or, if $\varphi>\varphi^*$,
it is in the plane perpendicular to $F$ containing $v$ and $\x$ and consists of three segments $(v,a)$, $(a,a_1)$
and $(a_1,\x)$, where the acute angles between the segments  $(v,a)$ and $(a_1,\x)$, and the direction normal to
the plane $F$ are equal to the critical angle $\varphi^*$, and the segment $(a,a_1)$ is in $F$
(Figure \ref{snells1} (b)).

From these observations, it follows that, in all cases, weighted distance function can equivalently be defined by
\begin{equation} \label{c(v,x)}
c(v,x)=\|\bar{\pi}(v,\x)\|=\min_{a,a_1\in F}(w^-|va| + w|aa_1| + w^+|a_1\x|).
\end{equation}
In the case where $w^-\not=w^+$, the minimum is achieved when $a=a_1=(\tau x,(1-\tau)y,0)$, where $\tau$ is the
unique solution in $(0,1)$ of the equation
\begin{equation} \label{tau}
\frac{w^-\tau}{\sqrt{\tau^2(x^2+y^2)+(z^-)^2}} =
\frac{w^+(1-\tau)}{\sqrt{(1-\tau)^2(x^2+y^2)+(z^+)^2}},
\end{equation}
where $v=(0,y,z^-)$ and $\x=(x,0,z^+)$. The latter leads to an algebraic equation of degree four and it is
infeasible\footnote{Although the roots of a quartic can be expressed as a rational function of radicals over its coefficients, they are too complex to be analytically manipulated and used here.}
to evaluate $c(v,x)$ explicitly.

The case where $w^-=w^+$ is easier, as in that case the geodesic path is either a straight line,
or a three segment path as described above and illustrated in Figure \ref{snells1} (b) and the function
$c(v,x)$ has an explicit representation, which is
\begin{equation} \label{eqn-explicit}
c(v,x)=\left\{
\begin{array}{lcl}
w^-\sqrt{x^2+y^2+\bar{z}^2} & \quad {\rm if} \quad & \sqrt{x^2+y^2}\leq \bar{z}/\bar{w}\\
w(\sqrt{x^2+y^2} - \bar{z}\bar{w} ) & \quad {\rm if} \quad &
\sqrt{x^2+y^2} > \bar{z}/\bar{w} ,
\end{array}
\right.
\end{equation}
where $\bar{z}=z^-+z^+$ and $\bar{w}=\sqrt{(w^-/w)^2-1}$. We refer to this case as the {\em explicit case}.
In the next lemma we state and prove some useful properties of the function $c(v,x)$.

\begin{lemma} \label{wdf-convex}
For a fixed $v$, the weighted distance function $c(v,x)$ has the following properties:\\[0.5ex]
\indent{\bf (a)} It is continuous and differentiable.\\[0.5ex]
\indent{\bf (b)} It is symmetric, i.e.\ $c(v,x)=c(v,-x)$.\\[0.5ex]
\indent{\bf (c)} It is strictly increasing for $x>0$.\\[0.5ex]
\indent{\bf (d)} It is convex.\\[0.5ex]
\indent{\bf (e)} It has asymptotes for $x\rightarrow\pm\infty$ as follows:\\
\hspace*{1cm}(e1) if $w^+ < w^-$ then the asymptotes are $w^+(z^-\cot\varphi^* \pm x)$,\\
\hspace*{1cm}(e2) if $w^- < w^+$ then the asymptotes are $w^-(z^+\cot\varphi^* \pm x)$,\\
\hspace*{1cm}(e3) in the explicit case $w^+=w^-\geq w$ the asymptotes are $\pm wx$,\\
where $\varphi^*$ is the critical angle.
\end{lemma}
\noindent{\bf Proof:} In the explicit case ($w^-=w^+$), all these properties follow straightforwardly from the
explicit representation (\ref{eqn-explicit}). So, we consider the case $w^-\not=w^+$

From (\ref{c(v,x)}) and $a_1=a=(\tau x, (1-\tau) y, 0)$ it follows that
$c(v,x)=w^-\sqrt{\tau^2(x^2+y^2)+(z^-)^2} + w^+\sqrt{(1-\tau)^2(x^2+y^2)+(z^+)^2}$,
where $\tau$ is the root of the equation (\ref{tau}). The root $\tau$ can be viewed as a function of $x$,
which by the implicit function theorem is continuous and differentiable. Hence property $\bf (a)$ holds.

The property {\bf (b)} follows from the observation that the value of the function $c(v,x)$ is
determined by the distance between the projections of the points $v$ and $\x$ on $F$, which is
$\sqrt{y^2+x^2}$ where $y$ is fixed.

To prove {\bf (c)} we consider a point $\x'=(x',0,z^+)$ such that $x'>x\geq 0$ and
denote by $\tau'$ the corresponding root of (\ref{tau}). We have
$c(v,x')=w^-\sqrt{\tau'^2(x'^2+y^2)+(z^-)^2} + w^+\sqrt{(1-\tau')^2(x'^2+y^2)+(z^+)^2}$.
Using the fact that the function $c(v,x)$  is defined as the cost of the shortest path
joining $v$ and $\x$ we have
\begin{eqnarray*}
c(v,x)\leq w^-\sqrt{\tau'^2(x^2+y^2)+(z^-)^2} + w^+\sqrt{(1-\tau')^2(x^2+y^2)+(z^+)^2}\\
< w^-\sqrt{\tau'^2(x'^2+y^2)+(z^-)^2} + w^+\sqrt{(1-\tau')^2(x'^2+y^2)+(z^+)^2}=c(v,x').
\end{eqnarray*}

In order to prove {\bf (d)}, we show that for any three equidistant points $\x_1<\x_0<\x_2$ on $\l$,
i.e., such that $2x_0=x_1+x_2$, the second finite difference
$\triangle_2(c;x_1,x_0,x_2) = c(v,x_1) -2 c(v,x_0) +c(v,x_2)$ of the function $c(v,x)$ is positive.
We denote by $a_1$ and $a_2$ the bending points of the shortest paths from $v$ to $\x_1$ and $\x_2$, respectively.
Let $a'_0$ be the middle point of the segment $(a_1,a_2)$. Then, using the definition of  $c(v,x_0)$ and the
convexity of the Euclidean distance function we obtain
$2c(v,x_0)\leq 2(w^-|va'_0|+w^+|a'\x_0|)<w^-(|va_1|+|va_2|)+w^+(|a_1\x_1|+|a_2\x_2|)=c(v,x_1)+c(v,x_2)$,
which implies $\triangle_2(c;x_1,x_0,x_2)>0$ and {\bf (d)}.

Finally, we prove {\bf (e)}. Let us assume that $w^+<w^-$. In this case, using Snell's law we observe that
when $x\rightarrow +\infty$ the bending point $a(x)$ of the shortest path $\bar{\pi}(v,\x)$
converges to the point $(z^-\tan\varphi^*,y,0)$ (see Figure \ref{fig:wdf}). Hence, we have
$\lim_{x\rightarrow +\infty}(w^+\sqrt{(x-z^-\tan\varphi^*)^2 + y^2 + (z^+)^2}+w^-z^-/cos\varphi^*-c(v,x))= 0$.
On the other hand
\begin{eqnarray*}
\lim_{x\rightarrow +\infty}(w^+\sqrt{(x-z^-\tan\varphi^*)^2 + y^2 + (z^+)^2}+w^-z^-/\cos\varphi^* -w^+(z^-\cot\varphi^* + x))\\
=w^+\lim_{x\rightarrow +\infty}(\sqrt{(x-z^-\tan\varphi^*)^2 + y^2 + (z^+)^2}-(x-z^-\tan\varphi^*))\\
=w^+\lim_{x\rightarrow +\infty}\frac{y^2+(z^+)^2}{(\sqrt{(x-z^-\tan\varphi^*)^2 + y^2 + (z^+)^2}+(x-z^-\tan\varphi^*)}=0.
\end{eqnarray*}
Combining these two limits we obtain $\lim_{x\rightarrow +\infty}(c(v,x)-w^+(z^-\cot\varphi^* + x))= 0$ and
thus (e1) is valid for $x\rightarrow +\infty$. The case where $x\rightarrow -\infty$ is symmetric.

In the case where $w^-<w^+$ we use Snell's law and observe that the bending point $a(x)$ of the shortest path
$\bar{\pi}(v,\x)$ converges to $x-z^+\tan\varphi^*$, that is  $\lim_{x\rightarrow +\infty}(a(x)-x-z^+\tan\varphi^*)=0$. Then (e2) is established analogously to (e1). \hfill $\Box$

\subsection{Refractive Additive Voronoi diagram} \label{rvd-subsec}
Next we study Voronoi diagrams under the weighted distance metric defined above. Given a set $S$ of $k$ points $v_1,\dots,v_k$ in $F^-$, called \emph{sites}, and nonnegative real numbers $C_1,\dots ,C_k$, called
{\em additive weights}, the \emph{additive Voronoi diagram} for $S$ is a subdivision of $F^+$ space into regions
${\cal V}(v_i, F^+)=\{x\in F^+~|~\mathrm{dist}(x,v_i)+C_i\leq \mathrm{dist}(x,v_j) + C_j \mbox{ for } j\neq i\}$,
where ties are resolved in favor of the site with smaller additive weight. The regions ${\cal V}(v_i,F^+)$ are
called \emph{Voronoi cells}.
In the classic case, $\mathrm{dist}(\cdot,\cdot)$ has been defined as the Euclidean distance. Here, for $\mathrm{dist}(\cdot,\cdot)$, we use the weighted distance function $c(v,x)$. 

Let $v$ and $v'$ be two points in $F^-$. 
We wish to study the intersection of
the additive Voronoi diagram of $v$ and $v'$ with $\ell$ with respect to the weighted distance function. Without
loss of generality, we assume that $C'=0$ and $C\geq 0$, where $C$ and $C'$  are the additive weights assigned to
$v$ and $v'$, respectively. We denote the intersection of the Voronoi cell of $v$ with $\l$ by ${\cal V}(v,v',\l;C)$, or simply by ${\cal V}(v)$ when no ambiguity arises. We have
$${\cal V}(v,v',\l;C)={\cal V}(v)=\{ \x\in\l\ :\ c(v,x) + C < c(v',x) \}.$$
In Theorem \ref{the_thm}, we will show that if $v$ and $v'$ are at the same distance to $F$, then the Voronoi cell ${\cal V}(v)$ restricted to the line $\l$ has a very nice structure (i.e., it is an interval). Furthermore, in Remark \ref{Rem:1}, we will show that if $v$ and $v'$ are not at the same distance to $F$ then Theorem \ref{the_thm} does not hold.
In our algorithm, presented in Section 5, we use the information about the shape of ${\cal V}(v)$ in
order to propagate approximate shortest paths in $\D$ and it turns out that we need to only consider the sites that are restricted to be within a half-space of $F$ and at the  same distance to $F$.
But, as we will see, this case in itself 
is mathematically challenging  and provides valuable insights into the combinatorial structure of these diagrams.
%
%
%
%

\begin{theorem}\label{the_thm}
The Voronoi cell ${\cal V}(v,v',\l;C)$ is an interval on $\l$ -- possibly empty, finite or infinite.
\end{theorem}
{\bf Proof:} First, consider  the case when $C=0$. We denote the set of points $\x$ in $F^+$ such that $c(v,\x)=c(v',\x)$ by $B(v,v')$ and observe that it is a half-plane perpendicular to $F$. Therefore, the set of
points $\x$ on $\l$  for which $c(v,x)=c(v',x)$ is either a single point, the whole line $\l$, or empty.
Correspondingly, the Voronoi cell ${\cal V}(v,v',\l;0)$ is either a half-line, empty, or the whole line $\l$ and
the theorem holds for $C=0$.

So, we assume that $C>0$. We consider the equation $c(v',x)-c(v,x)=C$ and claim that it cannot have more than two
solutions. Before we prove that claim (Claim \ref{two-solutions} below), we argue that it implies the theorem.

Assume that the equation $c(v',x)-c(v,x)=C$ has at most two solutions. If it does not have any or has just one solution, then the theorem follows straightforwardly. In the case where it has exactly two solutions, the cell
${\cal V}(v,v',\l;C)$ has to be either a finite interval on $\l$, or a complement of a finite interval on $\l$. From the
definition of the Voronoi cell ${\cal V}(v,v',\l;C)$ and $C>0$ it follows that ${\cal V}(v,v',\l;C)\subset {\cal V}(v,v',\l;0)$.
We know that  ${\cal V}(v,v',\l;0)$ is either empty, a half-line, or the whole line. If it is either empty or a
half-line then ${\cal V}(v,v',\l;C)$ must be either empty or a finite interval and the theorem holds.

It remains to consider the case where ${\cal V}(v,v',\l;0)$ is the whole line $\l$. We argue that ${\cal V}(v,v',\l;C)$ can
not be a complement to a finite interval. We have ${\cal V}(v,v',\l;0)=\l$ and therefore the line $\l$ must be parallel
to the half-plane $B(v,v')$. Furthermore, the plane containing $B(v,v')$ is a perpendicular bisector of the segment $(v,v')$ and thus the point $v'$ must have coordinates $(0,y',z^-)$ (Figure \ref{fig:wdf}).
In this setting, using Lemma \ref{wdf-convex}(e), we observe that $c(v,x)$ and $c(v',x)$ have same asymptotes
at infinity and thus $\lim_{x\rightarrow\infty}(c(v',x)-c(v,x))=0$. Therefore, the cell  ${\cal V}(v,v',\l;C)$ must be finite. The theorem follows.\hfill $\Box$

Next we establish the validity of the claim used in the proof of Theorem~\ref{the_thm}.
\begin{claim}\label{two-solutions}
The equation $c(v',x)-c(v,x)=C$ has at most two solutions.
\end{claim}
We will prove the claim by showing that the function $g(x)=c(v',x)-c(v,x)$ is unimodal, i.e., it has at most one
local extremum. We establish this property in two steps. First, we prove a characterization property of a local extremum of $g$ (Proposition~\ref{prop:p1}). Then, we show that there may be no more than one point possessing
that property.

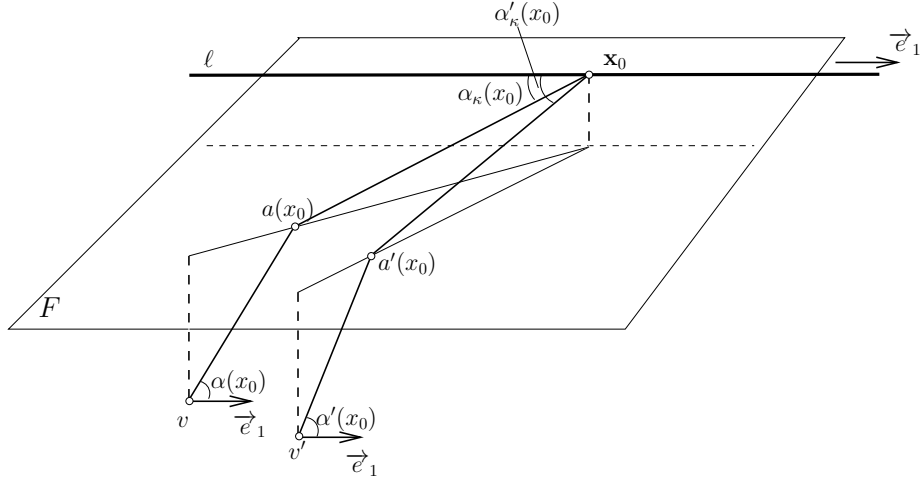
\begin{figure}
\begin{center}
\resizebox{12cm}{!}{
\input{alpha.pstex_t}}
\caption{If the function $g(x)$ has a local extremum at the point $x_0$ then the angles  $\alpha(x_0)$ and $\alpha'(x_0)$ must be equal.} \label{alpha}
\end{center}
\end{figure}

We focus our discussion on the case $w^-\not=w^+$, since the other case is either simpler or can be treated analogously. We denote by $a(x)$ and $a'(x)$ the bending points defining the shortest paths from $v$ and $v'$
to $\x$, respectively. We assume that $\l$ is oriented and denote by $\overrightarrow{e_1}$ the positive
direction unit vector on $\l$. Furthermore, let $\alpha(x)$ and $\alpha'(x)$ be the angles between vectors
$\overrightarrow{va(x)}$, $\overrightarrow{v'a'(x)}$ and $\overrightarrow{e_1}$, respectively (see Figure \ref{alpha}).

These angles are completely defined by the angles $\varphi$ and $\theta$ defining the corresponding shortest
paths at the bending points $a(x)$ and $a'(x)$. Precisely, we have $\cos\alpha = \sin\varphi \cos\theta$. Next,
we prove that the angles $\alpha(x_0)$ and $\alpha'(x_0)$  must be equal at any local extremum $x_0$.

\begin{figure}
\begin{center}
\resizebox{14cm}{!}{
\input{diagonals.pstex_t}}
\caption{Illustration of the proof of inequality (\ref{eq:diag_inequality}).} \label{fig:diagonals}
\end{center}
\end{figure}
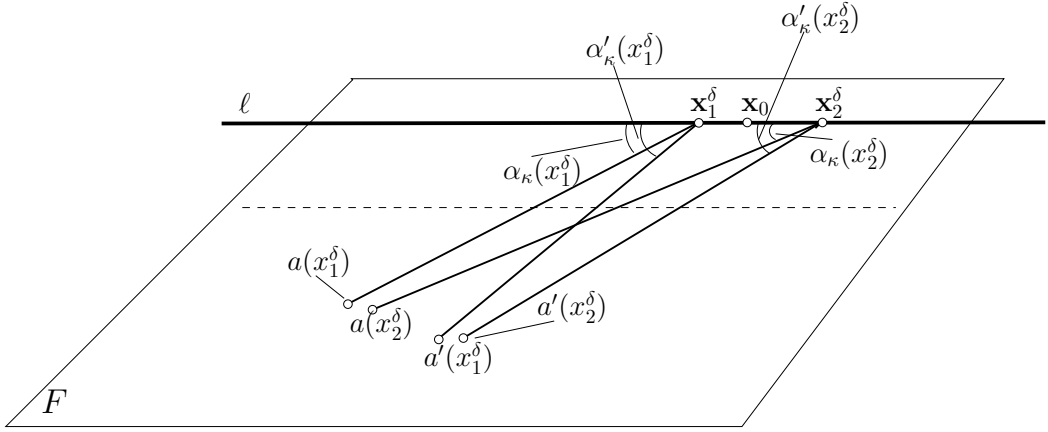

\begin{proposition}\label{prop:p1}
If  $x_0$ is a local extremum of the function $g$, then $\alpha(x_0)=\alpha'(x_0)$.
\end{proposition}
{\bf Proof:}
The proof is by contradiction. Let us assume that $\alpha(x_0) \not= \alpha'(x_0)$. We denote by $\alpha_{\kappa}(x_0)$ the angle between vectors $\overrightarrow{a(x_0)\x_0}$ and $\overrightarrow{e_1}$. Similarly, $\alpha'_{\kappa}(x_0)$ denotes the angle between vectors $\overrightarrow{a'(x_0)\x_0}$ and $\overrightarrow{e_1}$
(Figure \ref{alpha}). The relation $\cos\alpha=\cos\theta\sin\varphi$ and Snell's law readily imply that $\kappa\cos\alpha_\kappa(x_0)=\cos\alpha(x_0)$ and $\kappa\cos\alpha'_\kappa(x_0)=\cos\alpha'(x_0)$, where $\kappa=w^+/w^-$. Thus, we have that $\alpha_\kappa(x_0)\not=\alpha'_\kappa(x_0)$. Without loss of generality, we assume that $\alpha_\kappa(x_0)<\alpha'_\kappa(x_0)$. Under these assumptions, we show the existence of two points on $\x_1$ and $\x_2$ on $\ell$, such that:
\begin{equation} \label{a2-20}
x_1<x_0<x_2,\quad g(x_1)=g(x_2),\quad {\rm and} \quad |a(x_2)\x_2|+|a'(x_1)\x_1| > |a(x_2)\x_1|+|a'(x_1)\x_2|.
\end{equation}
By the assumption that $x_0$ is a local extremum of $g$, it follows that, for any positive real number $\delta>0$ inside the interval $(x_0-\delta,x_0+\delta)$, there are reals $x_1^\delta$ and $x_2^\delta$, such that
$$x_0-\delta<x_1^\delta<x_0<x_2^\delta<x_0+\delta \mbox{ and } g(x_1^\delta)=g(x_2^\delta).$$
On the other hand, if $\delta$ converges to zero, then $a(x_2^\delta)$ converges to $a(x_0)$, and $a'(x_1^\delta)$ converges to $a'(x_0)$. Therefore, the inequality $\alpha_\kappa(x_0)<\alpha'_\kappa(x_0)$ implies that for a small
enough $\delta_0$, the inequalities
$$\alpha_\kappa(x_1^{\delta_0})<\alpha'_\kappa(x_1^{\delta_0}) \quad \mbox{ and} \quad \alpha_\kappa(x_2^{\delta_0})<\alpha'_\kappa(x_2^{\delta_0})$$
hold. From these inequalities, it follows that if we make the quadrilateral $a(x_2^\delta)a'(x_1^\delta)x_2^\delta x_1^\delta$ planar by rotation of the point $a(x_2^\delta)$ around $\ell$, then the obtained planar quadrilateral
will be convex (Figure~\ref{fig:diagonals}).
Therefore we have
\begin{equation}\label{eq:diag_inequality}
|a(x_2^{\delta_0})\x_2^{\delta_0}|+|a'(x_1^{\delta_0})\x_1^{\delta_0}|>
|a(x_2^{\delta_0})\x_1^{\delta_0}|+|a'(x_1^{\delta_0})\x_2^{\delta_0}|,
\end{equation}
which proves (\ref{a2-20}) for $\x_1=\x_1^{\delta_0}$ and $\x_2=\x_2^{\delta_0}$.

Next, we estimate the sum $c(v,x_1)+c(v',x_2)$ we show that there may be no more than one point possessing
that property. We use (\ref{a2-20}) and obtain
\begin{align*}
c(v,x_1)+c(v',x_2)&=c(v,x_2)+c(v',x_1)\\
&= w^-|va(x_2)| +w^+ |a(x_2)\x_2| + w^-|v'a'(x_1)| + w^+|a'(x_1)\x_1|\\
&= w^-(|va(x_2)|+|v'a'(x_1)|) + w^+(|a(x_2)\x_2| + |a'(x_1)\x_1|)\\
&> w^-(|va(x_2)|+|v'a'(x_1)|) + w^+(|a(x_2)\x_1| + |a'(x_1)\x_2|)\\
&= w^-|va(x_2)|+w^+|a(x_2)\x_1| + w^-|v'a'(x_1)| + w^+|a'(x_1)\x_2|.
\end{align*}

On the other hand, by the definition of the weighted distance function (\ref{c(v,x)}), we have the inequalities
\begin{align*}
c(v,x_1)\leq w^-|va(x_2)|+w^+|a(x_2)\x_1|\quad{\rm and}\quad
c(v',x_2)\leq w^-|v'a'(x_1)| + w^+|a'(x_1)\x_2|,
\end{align*}
that contradict the previous strict inequality. Therefore, the angles $\alpha_\kappa(x_0)$ and $\alpha'_\kappa(x_0)$ and consequently $\alpha(x_0)$ and $\alpha'(x_0)$ must be equal.
\hfill $\Box$

Our next step is to show that there cannot be two points on $\l$ satisfying Proposition \ref{prop:p1}. To do that, we study in more detail the relationship between the position of points $v$ and $\x$, and the angle
$\alpha(x)$. We observe that, for any fixed $y$ (the $y$-coordinate of $v$), there is a one-to-one correspondence between the real numbers $x$ and the angles $\alpha$. That is, for a fixed $v$, there is a one-to-one correspondence between the points $\x$ on $\ell$ and the angle between the shortest path $\bar{\pi}(v,\x)$ and the
positive direction on $\ell$. Hence, we may consider and study the well defined function $x=x(y,\alpha)$. We prove the following:

\begin{proposition} \label{prop:p2}
The second mixed derivative of the function $x=x(y,\alpha)$ exists and is negative, i.e., $x_{y\alpha}<0$.
\end{proposition}

Let us first show that Proposition \ref{prop:p2} implies Claim \ref{two-solutions}.

\medskip\noindent
\textbf{Proof of Claim \ref{two-solutions}:}
We assume that Proposition \ref{prop:p2} holds and will show that the function $g(x)$ has at most one
local extremum. Recall that the point $v$ has coordinates $(0,y,z^-)$ and let us denote the coordinates of
the point $v'$ by $(x',y',z^-)$.

We first consider the case where $y=y'$. In this case, we observe that $\alpha'(x)=\alpha(x+x')$. In addition, the function $\alpha(x)$ is strictly monotone and, therefore, for any $x$, the angels $\alpha(x)$ and $\alpha'(x)$ are different. From Proposition \ref{prop:p1} it follows that in this case the function $g(x)$ has no local extremum.

Next, we consider the case $y\not=y'$. We assume, for the sake of contradiction, that $g(x)$
has two local extrema, say  $x_1$ and $x_2$. By Proposition \ref{prop:p1}, $\alpha(x_1)=\alpha'(x_1)$ and $\alpha(x_2)=\alpha'(x_2)$. We denote $\alpha_1=\alpha(x_1)=\alpha'(x_1)$ and $\alpha_2=\alpha(x_2)=\alpha'(x_2)$.
Then, the difference $x_2-x_1$ can be represented, using the function $x(y,\alpha)$, in two ways
\begin{equation}
x_2-x_1=\int_{\alpha_1}^{\alpha_2}x_\alpha(y,\alpha)\de\alpha \qquad {\rm and}
\qquad  x_2-x_1=\int_{\alpha_1}^{\alpha_2}x_\alpha(y',\alpha)\de\alpha. \label{eq:x2-x1}
\end{equation}

Subtracting the last two equalities, we get
\begin{equation}
0=\int_{\alpha_1}^{\alpha_2}x_\alpha(y,\alpha)\de\alpha -
\int_{\alpha_1}^{\alpha_2}x_\alpha(y',\alpha)\de\alpha =
\int_{y'}^y\int_{\alpha_1}^{\alpha_2}x_{y\alpha}(y,\alpha)\de\alpha \de y.\label{eq:integral}
\end{equation}
The integral on the right side is negative since by Proposition \ref{prop:p2},
the derivative $x_{y\alpha}$ is negative,  $\alpha_1\neq \alpha_2$, and $y\neq y'$.
Hence, we have a contradiction and Claim \ref{two-solutions} follows.\hfill $\Box$\\

The proof of Proposition \ref{prop:p2} is rather long and uses elaborate mathematical techniques and manipulations. On the other hand, the rest of the paper is independent of the details in that proof. So, we present the full proof for the interested readers in  Appendix \ref{appendix1}.

\begin{corollary}
 Consider a plane $H$ in $F^+$ parallel to $F$ and  two points $v$ and $v'$ in $F^-$ lying at the same distance from $F$. For any non-negative constant $C$, the Voronoi cell  ${\cal V}(v,v',{\cal H};C)=\{x\in {\cal H}: c(v,x)+C < c(v',x)\}$  is convex.
\end{corollary}
What can we say about the Voronoi cell of $v$ in $F^+$? The above corollary implies that the intersection between the Voronoi cell and  any plane, parallel to $F$,  is convex. This,  as such, is not sufficient to conclude that the cell is convex.  We close this section with the following conjecture.
\begin{conjecture}
In the above setting, the Voronoi cell of $v$ in $F^+$, ${\cal V}(v,v',F^+;C)=\{x\in F^+: c(v,x)+C < c(v',x)\}$,  is convex.
\end{conjecture}

\begin{remark}  \label{Rem:1}
Examples showing that equal distance of the points $v$ and $v'$ from the bending plane $F$ is
a necessary condition for $c(v',x)-c(v,x)$ to be unimodal in Theorem \ref{the_thm} is
not difficult to construct. 
In fact, if we take arbitrary
points $v$ and $v'$ in $F^-$ at different distances from $F$, it is very likely that $c(v',x)-c(v,x)$  will have more 
than one local extrema and hence, for a 
proper choice of $C$,  the equation $c(v', x) - c(v,x)= C$ will have more than two solutions.  
Such examples can be constructed by choosing arbitrary points $v$ in $F^-$ and $x_1$, $x_2$ on $\ell$ and
then computing a point $v' \in F^-$ so that $\alpha(v, x_i)=\alpha(v', x_i)$,  for $i= 1, 2$, 
where $\alpha$'s are the angles between the line $\ell$ and the
shortest paths coming from $v$ and $v'$, respectively.
As a result $c(v',x)-c(v,x)$ will have local extrema at $x_1$ and $x_2$. 
\end{remark}
%
%

\section{Discretization of $\D$} \label{3D-Discretization}
In this section, we describe the definition of a carefully chosen set of additional points placed in ${\cal D}$, called \emph{Steiner points}. These Steiner points collectively  form a discretization of  ${\cal D}$, which is later used to approximate geodesic paths in  ${\cal D}$. Steiner points are placed on the edges of ${\cal D}$ and on the bisectors of the dihedral angles of the tetrahedra in ${\cal D}$. While it may seem more natural to place the Steiner points on the faces of the tetrahedra, placing them on the bisectors proves to be more efficient, leading to a speed up of approximately $\eps^{-1}$ compared to the alternate placement. Recall that $\eps$ is an approximation parameter in $(0,1)$.  We provide a precise estimate on the number of Steiner points which depends on $\eps$ and aspect ratios of tetrahedra of
${\cal D}$.

\subsection{Placement of Steiner points}
We  use the following definitions:
\begin{definition}\label{D(x)}
\mbox{}\\
(a) For a point $x\in {\cal D}$, we define $D(x)$ to be the union of the tetrahedra incident to $x$. We denote by
$\partial D(x)$ the set of faces  on the boundary of $D(x)$ that are not incident to $x$.\\[1ex]
(b) We define $d(x)$ to be the minimum Euclidean distance from $x$ to any point on  $\partial D(x)$.\\[1ex]
(c) For each vertex $v\in{\cal D}$, we define a radius $r(v)=d(v)/14$.\\[1ex]
(d) For any internal point $x$ on an edge in ${\cal D}$, we define a radius $r(x)=d(x)/24$. The radius of an edge $e\in {\cal D}$ is $r(e)=\max_{x\in e} r(x)$.
\end{definition}
Using radii $r(v)$ and $r(x)$ and our approximation parameter $\eps$, we define ``small'' regions around vertices and edges of ${\cal D}$, called vertex and edge vicinities, respectively.
\begin{definition} \label{vertex-vicinity}
\mbox{}\\
(a) The convex hull of the intersection points of the ball $B(v,\varepsilon r(v))$ having center $v$ and radius $\varepsilon r(v)$ with the edges incident to $v$ is called the {\em vertex-vicinity} of $v$ and is denoted
by $D_\eps (v)$.\\[1ex]
(b) The convex hull of the intersections  between the ``spindle'' $\cup_{x\in e}B(x,\varepsilon r(x))$ and the faces incident to $e$ is called  the {\em edge-vicinity} of $e$ and is denoted by $D_\eps (e)$.
\end{definition}
On each edge $e=AB$ of $\cal D$, we define a set of Steiner points as follows. Denote by $AA'$ and $BB'$ the intersections of $e$ with vertex vicinities ${D_\eps}(A)$ and ${D_\eps}(B)$,
respectively. Points $A'$ and $B'$ are Steiner points. All other Steiner points on $e$ are placed between $A'$ and $B'$. Let $M_e$ be the point on $e$, such that $d(M_e)=\max_{x\in e} d(x)$. The point $M_e$ is defined to be a
Steiner point. Next, we define a sequence of points $M_i$, for $i=0,1,\dots$ on $M_eA'$, by
\begin{equation}\label{SP-on-an-edge}
M_0 = M_e \quad {\rm and}\quad |M_{i-1}M_i| = \varepsilon r(M_i) \quad {\rm for}\quad i=1,2,...
\end{equation}
%
%
All such points $M_i$ between $M_e$ and $A'$ are defined as Steiner points. Analogously, we define the set of Steiner points on $M_eB'$. The number of Steiner points defined in this way on $e$ is bounded by
$$C_e\frac{1}{\eps}\log\frac{2}{\eps},\quad {\rm where}\quad  C_e<\frac{33}{\sin\alpha(e)}\log\frac{|AB|}{\sqrt{r(A)r(B)}}$$ and $\alpha(e)$ is the minimum angle between $e$ and  the faces on  $\partial D(e)$. (All logarithms with unspecified base are assumed to be base $2$.) The total number of Steiner points placed on the edges of $\D$ is bounded by $6\Gamma_1\frac{n}{\eps}\log\frac{2}{\eps}$, where $\Gamma_1$ is the average of the constants $C_e$ over all edges of $\D$.

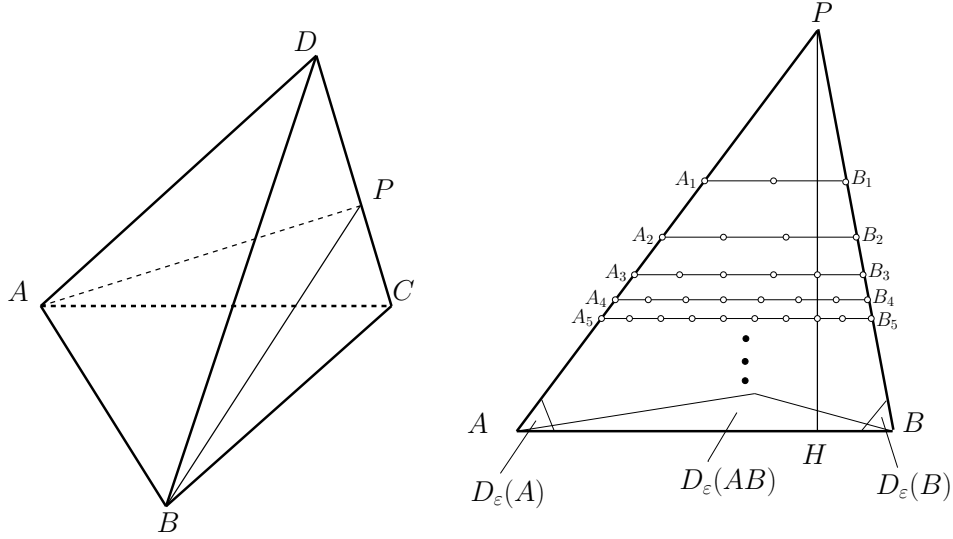
\begin{figure}
\begin{center}
\resizebox{12cm}{!}{
\input{f20.pstex_t}}
\caption{(a) A tetrahedron $ABCD$ and one of its bisectors $ABP$.  (b) Placement of Steiner points on $ABP$.\label{3D-Steiner_points-figure}
}
\end{center}
\end{figure}

The remaining Steiner points lie on the bisectors of the dihedral angles of tetrahedra in ${\cal D}$. Steiner points in any tetrahedron $T$ of $\cal D$ are  defined to lie on the six bisectors of the dihedral angles of $T$. Let the vertices of $T$ be $A$, $B$, $C$, and $D$ and let us consider one of the bisectors of the dihedral angles of
$T$, say $ABP$ (see Figure \ref{3D-Steiner_points-figure}(a)). Next, we describe the placement of Steiner points in the triangle $ABP$.

Let the dihedral angle at $AB$ of $T$ be $\gamma$ and let $PH$ be the height (altitude) of $ABP$  (see Figure \ref{3D-Steiner_points-figure}(b)).
First, we define an infinite sequence of points $P_0,P_1, \dots $ on $PH$ by
\begin{eqnarray}\label{lines}
P_0=P, \quad |P_{i-1}P_i| = \sqrt{\varepsilon/8} |HP_i|\sin (\gamma/2),\quad {\rm for}\ i=1,2,\dots
\end{eqnarray}
Then, we consider the  sequence of lines $L_i$ in the plane $ABP$, parallel to $AB$, and containing $P_i$, for
$i=1,2,\dots $. Let the intersection points of these lines with $AP$ and $BP$ be $A_i$ and $B_i$, respectively.  Points $A_i$ and $B_i$ lying outside of the vertex vicinities ${D_\eps}(A)$ and ${D_\eps}(B)$ are defined to be Steiner points, respectively. The intersection points of these lines with the boundary of the union of the edge vicinity ${D_\eps}(AB)$ and vertex vicinities ${D_\eps}(A)$ and ${D_\eps}(B)$, are defined to be Steiner points.
On each of the segments $A_iB_i$, we define a set of $k_i$ equidistantly placed Steiner points $P_{i,j}$,
$j=1,\dots ,k_i$, where
\begin{equation}\label{e15}
k_i=\left\lfloor\frac{|A_iB_i|}{|P_iP_{i+1}|}\right\rfloor\quad {\rm and} \quad
|P_{i,j}P_{i,j+1}|= \frac{|A_iB_i|}{k_i+1}, \quad {\rm  for}\quad
j=0,1,\dots , k_i.
\end{equation}
In the above expression, we have assumed that $P_{i,0}=A_i$ and $P_{i,k_i+1}=B_i$.
\begin{definition}
The set of Steiner points in the triangle $ABP$ consists of\\
(a) all points $P_{i,j}$  outside the union ${D_\eps}(AB)\cup {D_\eps}(A)\cup {D_\eps}(B)$,\\
(b) the intersection points of the lines $L_i$ with the boundary of that union.
\end{definition}
Next, we estimate the number of Steiner points placed in the triangle $ABP$. We denote  $h=p_0=|PH|$  and  $p_i=|P_iH|$  for $i=1,2,\dots$ In this notation, we have $p_{i-1}-p_i= |P_iP_{i-1}|$ and $|P_iP_{i-1}| =  p_i\sqrt{\varepsilon/8} \sin(\gamma/2)$, which implies
\begin{eqnarray}\label{e20}
p_i=h\lambda^i,\quad {\rm where}\quad \lambda =(1+\sqrt{\varepsilon/8} \sin(\gamma/2))^{-1}.
\end{eqnarray}

Let $i_1$ be the smallest index such that the line $L_{i_1}$ is at distance smaller than $\varepsilon r(e)$ from $AB$. We denote by $K_1$ the number of Steiner points lying on lines $L_i$, with $i<i_1$, and by $K_2$ the number of the remaining Steiner points in $ABP$. Let us estimate the number $K_1$ first. The number of Steiner points  on  a line $L_i$, with $i<i_1$, is $k_i+2$. Using (\ref{e15}) and  (\ref{e20}), we have
\begin{eqnarray*}
k_i = \left\lfloor\frac{|A_iB_i|}{|P_iP_{i+1}|}\right\rfloor
=\left\lfloor\frac{(h-p_i)|AB|}{h(p_i-p_{i+1})}\right\rfloor
=\left\lfloor\frac{(1-\lambda^i)|AB|}{h\lambda^i(1-\lambda)}\right\rfloor.
\end{eqnarray*}
Thus, for the number $K_1$, we obtain
\begin{eqnarray}\label{e30}
K_1&=&\sum_{i=1}^{i_1-1}(2+k_i) \leq 2(i_1-1) + \frac{|AB|}{h(1-\lambda)}
\sum_{i=1}^{i_1-1}\frac{1-\lambda^i}{\lambda^i}\\
&=&2(i_1-1) + \frac{|AB|}{h(1-\lambda)}
\left(\frac{1-\lambda^{i_1}}{(1-\lambda)\lambda^{i_1-1}}-i_1\right)\leq
2(i_1-1)+\frac{|AB|}{h(1-\lambda)^2}
\frac{1-\lambda^{i_1-1}}{\lambda^{i_1-1}}\ .\nonumber
\end{eqnarray}
From the definition of  $i_1$ and (\ref{e20}), we have $h\lambda^{i_1}<\varepsilon r(e)\leq h\lambda^{i_1-1}$. Therefore,
\begin{equation}\label{e40}
i_1 - 1 = \left\lfloor\log_{\lambda} \frac{\varepsilon r(e)}{h}\right\rfloor.
\end{equation}
From  (\ref{e30}) and (\ref{e40}),  we obtain
\begin{equation}\label{e50}
K_1<\frac{|AB|}{\varepsilon r(e)(1-\lambda)^2} + 2\log_{\lambda^{-1}}\frac{h}{\varepsilon r(e)}.
\end{equation}
Next, we estimate $K_2$, that is the number of Steiner points lying on segments $A_iB_i$ with $i\geq i_1$. By our definition, on the segment $A_{i_1}B_{i_1}$ there is a point $M_1$, such that the triangle $ABM_1$ lies entirely inside the edge vicinity. Let $M_2'$ be the intersection point of the boundaries of the edge vicinity ${D_\eps}(AB)$ and the vertex vicinity ${D_\eps}(A)$ that lies in the triangle $ABP$. Similarly, let $M_2''$ be the intersection point of the boundaries of the edge vicinity ${D_\eps}(AB)$, the vertex vicinity ${D_\eps}(B)$, and the triangle $ABP$.
Furthermore, let $i'_2$ be the smallest index such that the segment $A_{i_2'}B_{i_2'}$ is closer to $AB$ than $M'_2$ and similarly, let $i''_2$ be the smallest index so that the segment $A_{i_2''}B_{i_2''}$ is closer to $AB$ than $M''_2$. All  Steiner points on segments $A_iB_i$, with $i\geq i_1$, lie in the quadrilaterals $A_{i_1}A_{i_2'}M_2'M_1$ and $B_{i_1}B_{i_2''}M_2''M_1$. We denote the number of Steiner points in  these two quadrilaterals by $K'_2$ and $K_2''$, respectively.

To  estimate $K_2'$, we  show an upper bound on the number of Steiner points on $A_iB_i$, $i_1\leq i < i_2'$ that lie inside the quadrilateral  $A_{i_1}A_{i_2'}M_2'M_1$. Namely, if we denote this number by $k'_i$ and by $M_i$ the intersection point between $A_iB_i$ and $AM_1$, we have,
\begin{eqnarray*}
k'_i\leq 2 + \frac{|A_iM_i|}{p_i-p_{i+1}}= 2 + \frac{|A_{i_1}M_1|p_i}{p_{i_1}(p_i-p_{i+1})}= 2
+\frac{|A_{i_1}M_1|}{h\lambda^{i_1}(1-\lambda)}.
\end{eqnarray*}
Thus,  the number of Steiner points inside  the quadrilateral $A_{i_1}A_{i_2'}M_2'M_1$ is bounded by
$K_2'\leq (i_2'-i_1)( 2 + \frac{|A_{i_1}M_1|}{h\lambda^{i_1}(1-\lambda)})$.
Analogously, for the number of Steiner points inside the quadrilateral $B_{i_1}B_{i_2''}M_2''M_1$, we obtain
$K_2''\leq (i_2''-i_1)( 2 + \frac{|B_{i_1}M_1|}{h\lambda^{i_1}(1-\lambda)})$.
We sum the estimates on $K_2'$ and $K_2''$, use (\ref{e20}), (\ref{e40}) and obtain
\begin{eqnarray}\label{e60}
K_2=K_2'+K_2''\leq (i_2'+i_2''-2 i_1)\left(2+\frac{|A_{i_1}B_{i_1}|}
{h\lambda^{i_1}(1-\lambda)}\right)\leq\nonumber\\
(i_2'+i_2''-2 i_1)\left(2+\frac{|AB|(1-\lambda^{i_1})}{h\lambda^{i_1}(1-\lambda)}\right)<
(i_2'+i_2''-2 i_1)
\left(2+\frac{|AB|}{\varepsilon r(e)\lambda(1-\lambda)}\right).
\end{eqnarray}
From the definitions of the indices $i_2'$, $i_2''$, we easily derive that
$$p_{i_2'-1}>\frac{\sqrt{2}\eps^2 r(e) r(A)}{|AM_e|}\cos\frac{\angle BAP}{2}, \qquad
 p_{i_2''-1}>\frac{\sqrt{2}\eps^2 r(e) r(B)}{|BM_e|}\cos\frac{\angle PBA}{2},$$
 where $M_e$ is the point on $AB$ where the radius $r(e)$ is achieved. These inequalities and (\ref{e20}) imply
\begin{eqnarray*}
i_2'\leq 1+\log_{\lambda^{-1}}\frac{h|AM_e|}{\sqrt{2}\varepsilon^2 r(e) r(A)\cos\frac{\angle BAP}{2}}\quad
{\rm and} \quad i_2''\leq 1+\log_{\lambda^{-1}}\frac{h|M_eB|}{\sqrt{2}\varepsilon^2 r(e) r(B)
\cos\frac{\angle PBA}{2}}.
\end{eqnarray*}
Then, we use (\ref{e40}) and obtain
\begin{eqnarray} \label{e70}
i_2'+i_2''-2 i_1 & \leq & 2 + 2\log_{\lambda^{-1}}\frac{h|AB|}{\varepsilon^2 r(e)\sqrt{r(A)r(B)}}
- 2 \log_{\lambda^{-1}}\frac{h}{\varepsilon r(e)}=\nonumber\\
& & 2 + 2\log_{\lambda^{-1}}\frac{|AB|}{\varepsilon \sqrt{r(A)r(B)}}=
2\log_{\lambda^{-1}}\frac{|AB|}{\varepsilon \lambda \sqrt{r(A)r(B)}}\ .
\end{eqnarray}
Combining  (\ref{e50}), (\ref{e60}) and (\ref{e70}), we obtain
\begin{eqnarray}\label{e80}
K_1+K_2 \leq 2\frac{|AB|}{\varepsilon r(e)\lambda(1-\lambda)}
\log_{\lambda^{-1}}\frac{|AB|}{\varepsilon \lambda \sqrt{r(A)r(B)}}
&+&
\frac{|AB|}{\varepsilon r(e)(1-\lambda)^2}\\
&+&
2\log_{\lambda^{-1}}\frac{h}{\eps r(e)} + 4\log_{\lambda^{-1}}\frac{|AB|}{\varepsilon
\lambda \sqrt{r(A)r(B)}}.\nonumber
\end{eqnarray}
From the last equation, we easily derive that
\begin{eqnarray*}
K_1+K_2 = C_{ABP}(T)\frac{1}{\varepsilon^2}\log\frac{2}{\varepsilon},
\end{eqnarray*}
where the constant $C_{ABP}(T)$ depends on the geometry of the tetrahedron $T$ and is bounded by (see Appendix \ref{appendix2} for details)
\begin{eqnarray}\label{e90}
C_{ABP}(T)\leq 23 \frac{|AB|}{r(e)\sin^2(\gamma/2)}\log\frac{4|AB|^2 h}{r(e)r(A)r(B)}.
\end{eqnarray}
Our discussion is summarized in the following lemma.
\begin{lemma} \label{number-of-points}
(a) The number of Steiner points placed on a bisector $ABP$ of a dihedral angle $\gamma$ in a tetrahedron $T$, is bounded by $C_{ABP}(T)\frac{1}{\varepsilon^2}\log\frac{2}{\varepsilon}$, where the constant $C_{ABP}(T)$ depends on the geometric features of ${\cal D}$ around the edge $AB$ and is bounded by
$23\frac{|AB|}{r(e)\sin^2(\gamma/2)}\log \frac{4|AB|^2 h}{r(e)r(A)r(B)}$.

(b) The number of segments that are parallel to $AB$ on a bisector  $ABP$, containing Steiner points, is bounded by
$C^1_{ABP}(T)\frac{1}{\sqrt{\eps}}\log\frac{2}{\eps}$, where
$C^1_{ABP}(T)<\frac{4}{\sin(\gamma/2)}\log_2 \frac{4|AB|^2 h}{r(e)r(A)r(B)}$.

(c) The total number of Steiner points is bounded by $C({\cal D})\frac{n}{\varepsilon^2}\log\frac{2}{\varepsilon}$,
where $n$ is the number of tetrahedra in ${\cal D}$ and $C({\cal D})$ is the average of $C_{ABP}(T)$ over
all $6n$ bisectors in ${\cal D}$.
\end{lemma}
By placing Steiner points in this way, in the next lemma, we show that it is possible to approximate the cell crossing segments that have their endpoints outside the vertex and the edge vicinities.
\begin{lemma}\label{angle-lem}
Let $ABP$ be the bisector of a dihedral angle $\gamma$ formed by the faces $ABC$ and $ABD$ of a tetrahedron $ABCD$. Let $x_1$ and $x_2$ be points on the faces $ABC$ and $ABD$, respectively, that lie outside  of the union ${D_\eps}(AB)\cup {D_\eps}(A)\cup {D_\eps}(B)$. Then, there exists a Steiner point $q$ on $ABP$, such
that $\max (\angle x_2x_1q, \angle x_1x_2q) \leq \sqrt{\frac{\eps}{2}}$ and $|x_1q|+|qx_2|\leq(1+\eps/2)|x_1x_2|$.
\end{lemma}
\noindent {\bf Proof: } Clearly, the segment $x_1x_2$ intersects the bisector triangle $ABP$ in a point $x_0$ lying outside the vertex vicinities ${D_\eps}(A)$, $ {D_\eps}(B)$, and the edge vicinity ${D_\eps}(AB)$. Recall that Steiner points in $ABP$ are placed on a set of lines $L_i$ parallel to $AB$ and passing through the sequence of points $P_i$ on the altitude $PH$ of $ABP$. Let $i_0$ be the maximum index such that the line $L_{i_0}$ is farther away from $AB$ than from $x_0$. We define $q$ to be the closest Steiner point to $x_0$ on the line $L_{i_0}$.

\begin{figure}[htb]
\begin{center}
\resizebox{5.cm}{!}{
\input{f30.pstex_t}}
\end{center}
\caption{Illustrates Lemma \ref{angle-lem}.} \label{fig:f30}
\end{figure}
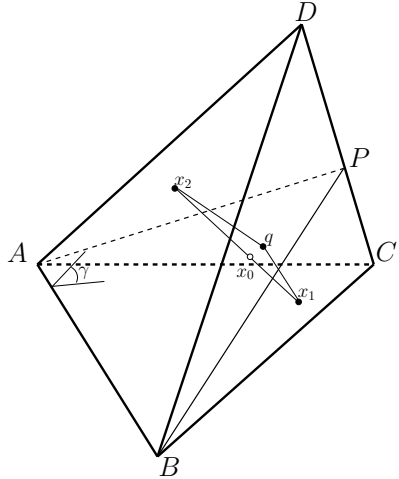

First, we estimate the angles $\angle x_2x_1q = \angle x_0x_1q$ and $\angle x_1x_2q = \angle x_0x_2q$. By our definition of the Steiner points and Pythagorean theorem, it follows that
\begin{equation} \label{e250}
|x_0q|\leq\frac{\sqrt{5}}{4}h\lambda^{i_0}\sqrt{\frac{\eps}{2}}\sin\frac{\gamma}{2},
\end{equation}
where $h$ and $\lambda$ are as defined above (see (\ref{e20})).
Let $\rho$ be the radius of  the smallest sphere containing $x_0$ and $q$ and touching the face $ABC$. It is easily observed that
\begin{equation} \label{e260}
2\rho>(h\lambda^{i_0}-\frac{|x_0q|}{2})\sin\frac{\gamma}{2}>\frac{8\sqrt{2}-\sqrt{5}}{8\sqrt{2}}
h\lambda^{i_0}\sin\frac{\gamma}{2}\ .
\end{equation}
If we denote the angle $\angle x_0x_1q$ by $\theta_1$, then $\sin\theta_1 \leq \frac{|x_0q|}{2\rho}$, and using (\ref{e250}) and (\ref{e260}), we obtain
\begin{equation}\label{e270}
\sin \theta_1 \leq \frac{|x_0q|}{2\rho}<\sqrt{\frac{\eps}{2}} .
\end{equation}
The same estimate applies to  angle  $\angle x_0x_2q$. Hence, the first inequality of the lemma holds.
Next, we prove the second inequality. We denote by $\theta$, $\theta_1$, and $\theta_2$ the angles of the triangle
$qx_1x_2$ at $q$, $x_1$ and $x_2$, respectively (Figure \ref{fig:f30}). By a trigonometric equality valid in any triangle, we have
$$|x_1q|+|qx_2|=\left(1+\frac{2\sin(\theta_1/2)\sin(\theta_2/2)}{\sin(\theta/2)}\right)|x_1x_2|.$$
Thus, it suffices to prove that
$\frac{2\sin(\theta_1/2)\sin(\theta_2/2)}{\sin(\theta/2)}\leq \eps/2$. By (\ref{e270}), it follows that $\sin\theta_1$ and $\sin\theta_2$ are smaller than $\sqrt{\eps/2}$ and from $\eps\leq 1$ we have $\theta\geq\pi/2$. Therefore, we obtain
\begin{eqnarray*}
\frac{2\sin(\theta_1/2)\sin(\theta_2/2)}{\sin(\theta/2)} &=&
\frac{\sin\theta_1\sin\theta_2}{2\sin(\theta/2)\cos(\theta_1/2)\cos(\theta_2/2)}
\leq \frac{\eps}{4\sin(\theta/2)\cos(\theta_1/2)\cos(\theta_2/2)}\\
&=&\frac{\eps}{4\sin(\theta/2)(\sin(\theta/2)+\sin(\theta_1/2)\sin(\theta_2/2))}
\leq\frac{\eps}{4\sin^2(\theta/2)} \leq\frac{\eps}{2}.\hfill \Box
\end{eqnarray*}

\section{Discrete paths} \label{discretepath-section}
In this section, we use the Steiner points introduced above for the construction of a weighted graph
$G_{\eps}=(V(G_{\eps}),E(G_{\eps}))$. We estimate the number of its nodes and edges and then establish that shortest paths in $\D$ can be approximated by paths in $G_\eps$. We follow the approach laid out in \cite{AMS05}, but the details are substantially different, as we have to handle both the vertex and edge vicinities, as well as the bisectors in 3-d space.

The set of nodes $V(G_\eps)$ consists of the vertices of $\D$, the Steiner points placed on the edges of $\D$ and the Steiner points placed on the bisectors. The edges of the graph $G_\eps$ join nodes lying on {\em neighboring} bisectors as defined below. A bisector is a neighbor to itself. Two different bisectors are neighbors if the dihedral angles they split share a common face. We say that a pair of bisectors sharing a face $f$ are neighbors with respect to $f$. (So, a single bisector $\b$ is a neighbor to itself with respect to both faces forming the dihedral angle it splits.)

First, we define edges joining pairs of Steiner points on neighboring bisectors. Let $p$ and $q$ be nodes corresponding to Steiner points lying on neighboring bisectors $\b$ and $\bb$, respectively, that share a common face $f$. We consider the shortest weighted path between $p$ and $q$ of the type $\{p,x,y,q\}$, where $x$ and
$y$ belong to $f$ (points $x$ and $y$ are not necessarily different). We refer to this shortest path as a {\em local
shortest path} between $p$ and $q$ crossing $f$ and denote it by $\hat{\pi}(p,q;f)$. Nodes $p$ and $q$ are joined by an edge in $G_\eps$ if none of the points $x$ or $y$ are on an edge of $f$. Such an edge is said to {\em cross}
the face $f$. In the case where $p$ and $q$ lie on the same bisector, say $\b$, splitting an angle between faces $f_1$ and $f_2$, we define two parallel edges in $G_\eps$ joining $p$ and $q$ -- one crossing $f_1$ and another crossing $f_2$.

The cost of an edge $(p,q)$ in $G_\eps$ that crosses a face $f$ is defined as the cost of the local shortest path $\hat{\pi}(p,q;f)$ and is denoted by $c(p,q;f)$, or simply by $c(p,q)$ when no ambiguity arises. Formally, we have
\begin{equation} \label{local-path}
c(p,q)=c(p,q;f)= \|\hat{\pi}(p,q;f)\|= \min_{x,y\in f}(\|px\|+\|xy\|+\|yq\|).
\end{equation}

Next, we consider a node $p$ of $G_\eps$ lying on an edge $e$ of $\D$. The node $p$ can be either a Steiner point on $e$ or a vertex of $\D$ incident to $e$. It is adjacent to nodes lying in tetrahedra in $D(e)$. The edges of $G_\eps$
incident to $p$ are associated with pairs of neighboring bisectors as follows. We consider a tetrahedron $t$ in $D(e)$, and describe edges incident to $p$ in $t$. Let $f_1$ and $f_2$ be the two faces of $t$ incident to $e$, and let $\b$ be the bisector of the dihedral angle formed by $f_1$ and $f_2$. We define edges between $p$ and nodes lying on bisectors in $t$ that are neighbors of $\b$. There are four such bisectors -- two with respect to $f_1$ and two with respect to $f_2$. For a node $q$ on a neighboring bisector $\bb$ sharing, say, the face $f_1$ with $\b$, we consider the local shortest path $\hat{\pi}(p,q;f_1)$. By definition,  $\hat{\pi}(p,q;f_1)=\{p,x,q\}$, where $x\in f_1$. We define an edge between $p$ and $q$ if and only if the point $x$ defining the local shortest path is in the interior of $f_1$. The cost of the edge $(p,q)$ equals the cost of the local shortest path $\hat{\pi}(p,q;f_1)$, i.e.,
$$
c(p,q)=c(p,q;f_1)=\|\hat{\pi}(p,q;f_1)\|=\min_{x\in f_1}(\|px\|+\|xq\|).
$$
We associate the edge $(p,q)$ to $\b$, $\bb$ and $f_1$ and say that it crosses $f_1$. Furthermore, $p$ is joined to nodes on $\b$ by pair of parallel edges, provided that the corresponding local shortest paths do not touch the edges of $\D$ -- one crossing $f_1$ and the other crossing $f_2$.
\begin{lemma} \label{size_of_G}
We have $|V(G_\eps)|=O(\frac{n}{\eps^2}\log\frac{1}{\eps})$ and $|E(G_\eps)|=O(\frac{n}{\eps^4}\log^2\frac{1}{\eps})$.
\end{lemma}
\noindent{\bf Proof:}
The estimate on the number of nodes follows directly from Lemma \ref{number-of-points} and the fact that $\D$ has $O(n)$ vertices. The number of edges in $G_\eps$ can be estimated as follows.  There are $O(n)$ faces in $\D$ and  at most 21 pairs of neighbor bisectors with respect to a fixed face in $\D$. By Lemma
\ref{number-of-points}(a), there are $O(\frac{1}{\eps^4}\log^2\frac{1}{\eps})$ pairs of nodes lying on two fixed neighboring bisectors. When combined, these three facts  prove the estimate on the number of edges of $G_\eps$.
\hfill $\Box$

Paths in $G_\eps$ are called {\em discrete paths}. The cost, $c(\pi)$, of a discrete path $\pi$ is the sum of the costs of its edges. Note that if we replace each of the edges in a discrete path $\pi$ by the corresponding
(at most three) segments forming the shortest path used to compute its cost we obtain a path in $\D$ with cost $c(\pi)$. Next, we state the main theorem of this section.

\begin{theorem} \label{approx}
Let $\tp(v_0,v)$ be a shortest path between two different vertices $v_0$ and $v$ in $\D$. There exists a discrete path $\pi(v_0,v)$, such that $ c(\pi(v_0,v)) \leq (1+\varepsilon)\|\tp(v_0,v)\|$.
\end{theorem}
\noindent{\bf Proof:} We prove the theorem by constructing a discrete path $\pi(v_0,v)$ whose cost is as required.
Recall that the shortest path $\tp(v_0,v)$ is a linear path consisting of cell-crossing, face-using, and edge-using segments that satisfy Snell's law at each bending point. We construct the discrete path $\pi$ by successive modifications of $\tp$ described below as steps.

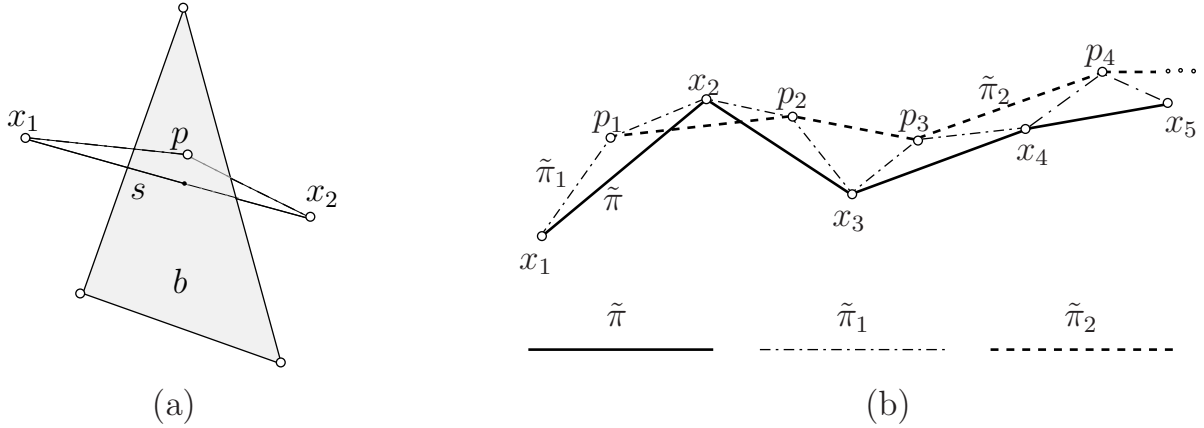
\begin{figure}
\begin{center}
\resizebox{16cm}{!}{
\input{Step1.pstex_t}}
\caption{(a) Replacement of a cell-crossing segment $s=(x_1,x_2)$ by a two-segment path $\{x_1,p,x_2\}$.
(b) Replacement of $\tp_1$ by a path $\tp_2$ joining Steiner points $p_i$. Note that edges $(p_ip_{i+1})$
denote local shortest paths, rather than straight-line segments.\label{fig:Step1}}
\end{center}
\end{figure}

\medskip\noindent {\bf Step 1:} In this step, we replace each of the cell-crossing segments of $\tp$, which satisfy the conditions of Lemma \ref{angle-lem}, by a two-segment path through a Steiner point. Precisely, let
$s=(x_1,x_2)$ be a cell-crossing segment in  $\tp$ (Figure~\ref{fig:Step1} (a)). Let $f_1$ and $f_2$ be the faces containing $x_1$ and $x_2$, respectively. Let $e=(A,B)$ be the common edge between $f_1$ and $f_2$. Assume that $s$ is outside of the union of the edge and vertex vicinities $D_\eps(e)\cup D_\eps(A)\cup D_\eps(B)$. We refer to such segment as {\em vicinity-free}\footnote{Note that such a segment still can have an end-point in a vertex or edge-vicinity related to other vertices or edges incident to $f_1$ and $f_2$.}. Then, according to
Lemma \ref{angle-lem}, there is a Steiner point $p$ on the bisector $b$ splitting the dihedral angle formed by $f_1$ and $f_2$ such that $|x_1p|+|px_2|\leq (1+\eps/2)|x_1x_2|$. So, in this step, each cell-crossing and vicinity-free segment $s=(x_1,x_2)$ is replaced by two-segment path $\{x_1,p,x_2\}$, where $p$ is the approximating Steiner point as described above. Clearly, after this step, we obtain a path joining $v_0$ and $v$, whose cost does not exceed $(1+\eps/2)\|\tp\|$. We denote this path by $\tp_1$ (see Figure \ref{fig:Step1} (b)).

\medskip\noindent {\bf Step 2:} In this step, we consider the sequence of Steiner points added as new bending points along $\tp_1$ in Step 1. In the case where two consecutive Steiner points are split by a single bending point or a face-using segment on  $\tp_1$, we replace the sub-path between them by the corresponding local shortest path. Precisely, assume that $p_1$ and $p_2$ are consecutive Steiner points along $\tp_1$ and the sub-path between them is
either $\{p_1, \tilde{x}, p_2\}$ or $\{p_1,\tilde{x},\tilde{y},p_2\}$, $\tilde{x}$ and $\tilde{y}$ are bending points on the face $f$, shared by the two neighboring tetrahedra containing $p_1$ and $p_2$, respectively. So, in Step 2, we replace all such sub-paths by the local shortest paths $\hat{\pi}(p_1,p_2;f)=\{p_1,x,y,p_2\}$,  using (\ref{local-path}). We denote the obtained path by $\tp_2$ (Figure~\ref{fig:Step1} (b)).
Clearly, $\tp_2$ is a path joining $v_0$ and $v$, whose cost does not exceed that of $\tp_1$. Hence,
\begin{equation}\label{e110}
\|\tp_2\|\leq \|\tp_1\|\leq (1+\eps/2)\|\tp\|\ .
\end{equation}

In the following two steps, we identify the portions of $\tp_2$ that lie inside  the vertex and edge vicinities and
replace them with discrete paths using the corresponding vertices and edges.

\medskip\noindent {\bf Step 3:} Follow $\tp_2$ from $v_0$ to $v$ and let $a_0$ be the last bending point on $\tp_2$ that lies inside the vertex vicinity $\V (v_0)$. Next, let $b_1$ be the first bending point after $a_0$ that is in the vertex vicinity, say $\V (v_1)$. Likewise, let $a_1$ be the last bending point in $\V (v_1)$. Continuing in this
way, we define a sequence of, say $k+1$ for some $k\geq 1$, different vertices $v_0,v_1,\dots, v_k=v$ and a sequence of bending points $a_0,b_1,a_1,\dots, a_{k-1},b_k$ on $\tp_2$, such that for $i=0,\dots , k$, points $b_i, a_i$ are in $\V (v_i)$ (we assume $b_0=v_0$, $a_k=v$). Furthermore, by our definition, portions of $\tp_2$ between $a_i$ and $b_{i+1}$ do not intersect any vertex vicinities. We partition  $\tp_2$ into portions
\begin{equation}\label{partition}
\tp_2(v_0,a_0),\tp_2(a_0,b_1),\tp_2(b_1,a_1),\dots ,\tp_2(b_k,v).
\end{equation}
The portions $\tp_2(a_i,b_{i+1})$, for $i=0,\dots, k-1$, are called the {\em between-vertex-vicinities} portions, while the portions $\tp_2(b_i,a_i)$, for $i=0,\dots, k$, are called the {\em vertex-vicinity} portions.

We define path $\tp_3$ by replacing each of the vertex-vicinities portions by a two segment path trough the corresponding vertex and show that the cost of $\tp_3$ is bounded by $(1+\eps/6)\|\tp_2\|$.
Consider a between-vertex-vicinities portion $\tp_2(a_i,b_{i+1})$ for some $0 \leq i < k-1$. If this portion consists of a single segment $(a_i,b_{i+1})$, then the  vertices $v_i$ and $v_{i+1}$ must be adjacent in $\D$ and we define $\tp_3(v_i,v_{i+1})$ to be the segment $(v_i,v_{i+1})$. The length of $(v_i,v_{i+1})$ is estimated by using the triangle inequality and the definition of the vertex-vicinities as follows:
\begin{eqnarray} \label{e115}
|v_iv_{i+1}|\leq |v_ia_i|+|a_ib_{i+1}|+|b_{i+1}v_{i+1}|\leq |a_ib_{i+1}| + \eps(r(v_i)+r(v_{i+1}))\leq\nonumber\\
|a_ib_{i+1}| + \frac{\eps}{14}(d(v_i)+d(v_{i+1}))\leq |a_ib_{i+1}| + \frac{\eps}{7}|v_iv_{i+1}|.
\end{eqnarray}
To estimate the cost of the segment $(v_i,v_{i+1})$, we observe that  $(a_i,b_{i+1})$ lies inside a tetrahedron
incident to  $(v_i,v_{i+1})$. Thus, the weight of $(v_i,v_{i+1})$ is at most the weight of $(a_i,b_{i+1})$.
This observation and (\ref{e115}) readily imply
\begin{equation} \label{e120}
\|\tp_3(v_i,v_{i+1})\| = \|v_iv_{i+1}\| \leq (1+\frac{\eps}{6}) \|a_ib_{i+1}\| = (1+\frac{\eps}{6}) \|\tp_2(a_i,b_{i+1})\|.
\end{equation}
In the general case, where $\tp_2(a_i,b_{i+1})$ contains at least two segments, we follow the bending points along
$\tp_2(a_i,b_{i+1})$ and define $X$ to be the last bending point on the boundary $\partial D(v_i)$ (see Definition \ref{D(x)}). If the path $\tp_2(a_i,b_{i+1})$ lies entirely in $D(v_i)$, then we set $X=b_{i+1}$. Thus, the bending points on the path $\tp_2$ between $a_i$ and $X$ lie in the tetrahedra incident to $v_i$. Let $\w_i$ be the minimum weight among the segments of the path $\tp_2(a_i,X)$ and let $x$ be the first bending point after $a_i$
incident to a segment, whose weight is $\w_i$.
Analogously, define the bending points $Y$ and $y$, by following the bending points of the backward path
$\tp_2(b_{i+1},a_i)$ from $b_{i+1}$.  Note that $x$ precedes $y$ on the path $\tp_2(a_i,b_{i+1})$.
We define the path $\tp_3(v_i,v_{i+1})$ as the concatenation of the segments $(v_i,x)$, $(y,v_{i+1})$ and the portion
$\tp_2(x,y)$, i.e.,
$$\tp_3(v_i,v_{i+1})=\{(v_i,x),\tp_2(x,y),(y,v_{i+1})\}.$$
Next, we estimate the cost of  $\tp_3(v_i,v_{i+1})$. First, we observe that the weight of the segment $(v_i,x)$ cannot exceed $\w_i$. Then, we use the triangle inequality and the fact that $a_i$ is inside the vertex vicinity $\V(v_i)$, obtaining
\begin{eqnarray*}
\|v_ix\|&\leq&\w_i|v_ia_i| + \w_i|\tp_2(a_i,x)|\leq \w_i|v_ia_i| + \|\tp_2(a_i,x)\|\leq \\
& & \w_i\eps r(v_i) + \|\tp_2(a_i,x)\|\leq \w_i\frac{\eps}{14} d(v_i) + \|\tp_2(a_i,x)\|.
\end{eqnarray*}
Analogously, for the cost of the segment $(y,v_{i+1})$, we have
$$
\|yv_{i+1}\| \leq \w_{i+1}\frac{\eps}{14} d(v_{i+1}) + \|\tp_2(y,b_{i+1})\|.
$$
Using these estimates, and the way we defined the path $\tp_3(v_i,v_{i+1})$, the weights $\w_i$, $\w_{i+1}$,
the distances $d(v_i)$, $d(v_{i+1})$, and the points $X$, $Y$, we obtain
\begin{eqnarray}\label{e125}
\|\tp_3(v_i,v_{i+1}\| \leq \|\tp_2(a_i,b_{i+1})\| + \frac{\eps}{14}(\w_i d(v_i) + \w_{i+1} d(v_{i+1})) \leq \\
\|\tp_2(a_i,b_{i+1})\| + \frac{\eps}{14}(\|\tp_3(v_i,X)\| + \|\tp_3(Y,v_{i+1})\|)\leq \|\tp_2(a_i,b_{i+1})\| +
\frac{\eps}{7}(\|\tp_3(v_i,v_{i+1})\| \nonumber ,
\end{eqnarray}
which implies the estimate (\ref{e120}) in the general case.
Applying the above construction to each pair of consecutive vertices in the sequence $v_0,v_1,\dots ,v_k=v$, we obtain a linear path
\begin{equation*}
\tp_3(v_0,v)=\{ \tp_3(v_0,v_1), \tp_3(v_1,v_2), \dots , \tp_3(v_{k-1},v)\},
\end{equation*}
that has no bending points inside vertex vicinities except for the vertices $v_0,v_1,\dots ,v_k=v$. We estimate the cost of this path by summing up (\ref{e120}), for $i=0,\dots, k-1$, and obtain
\begin{equation} \label{e140}
\|\tp_3(v_0,v)\|\leq(1+\frac{\eps}{6})\sum_{i=0}^{k-1}\|\tp_2(a_i,b_{i+1})\| \leq(1+\frac{\eps}{6})\|\tp_2(v_0,v)\|.
\end{equation}
Observe that the path $\tp_3$ constructed above may contain self intersections (e.g.,\ if one and the same vertex
vicinity is visited twice by  $\tp_2$). It is also possible that  $\tp_3$ may contain consecutive
face-using segments. Hence, at the end of Step 3, we traverse the obtained path and {\em compress} it. That is, we remove the loops in case of self intersections. We replace the consecutive face-using segments  (which obviously lie in the same face) by the single face-using segment joining their free end-points. We denote the compressed path again by $\tp_3$. Clearly, compressing reduces the cost of the path and hence the estimate (\ref{e140}) remains true for  $\tp_3$.

Next, in Step 4, using a similar approach as above, we further  partition each vertex-vicinity-portion $\tp_3(v_i,v_{i+1})$ into {\em between-edge-vicinities} portions and  {\em edge-vicinity} portions. Then, we replace each  edge-vicinity portion by an edge-using segment plus 2 additional  segments and estimate the
cost of the resulting path $\tp_4$.\\

\noindent {\bf Step 4:}
First we define analogues of vertex and between-vertex vicinities for edges. Let $(v_i,a)$ be the first segment of the path $\tp_3(v_i,v_{i+1})$. If $a$ is not inside an edge-vicinity, we define $a_{i,0}=v_i$. Otherwise, if $a$ is inside an edge-vicinity, say $\V(e_0)$, and let $a'$ be the first bending point on the path $\tp_3(v_i,v_{i+1})$ after $v_i$ lying on $\partial D(e_0)$, then we define $a_{i,0}$ to be the last bending point on $\tp_3(v_i,a')$ that is inside $\V(e_0)$. Next, let $b_{i,1}$ be the first bending point on $\tp_3(a_{i,0},v_{i+1})$ that is inside an edge-vicinity, say $D_\eps (e_1)$ and let $b'$ be the first bending point on $\tp_3(b_{i,1},v_{i+1})$ that is on $\partial D(e_1)$. We define $a_{i,1}$ as the last bending  point on $\tp_3(b_{i,1},b')$ that is in the same edge vicinity as $b_{i,1}$. Assume that, following this approach, the sequence of bending points $a_{i,0}, b_{i,1}, a_{i,1},\dots, a_{i,k_i-1},
b_{i,k_i}$ has been defined. They  partition the portion $\tp_3(v_i,v_{i+1})$ into sub-portions
$\tp_3(v_i,v_{i+1})=\left\{ \tp_3(v_i,a_{i,0}),\dots ,\tp_3(a_{i,j-1},b_{i,j}), \tp_3(b_{i,j},a_{i,j}), \dots ,
\tp_3(b_{i,k_i},v_{i+1})\right\}$. Portions  between $a_{i,j}$ and $b_{i,j+1}$, for $j=0, \dots ,k_i-1$, are called the {\em between-edge-vicinity} portions. Portions between $b_{i,j}$ and $a_{i,j}$, for $j=0,\dots ,k_i$, are called the {\em edge-vicinity} portions ($b_{i,0}=v_i$ and $a_{i,k_i}=v_{i+1}$).

According to our construction, the bending points $a_{i,0}, b_{i,1}, a_{i,1},\dots, a_{i,k_i-1}, b_{i,k_i}$, defining the above partition lie inside edge vicinities. Moreover, consecutive points $b_{i,j}$ and $a_{i,j}$, for $j=0,\dots ,k_i$,  are in one and the same edge-vicinity $D_\eps (e_j)$.

\begin{figure}
\begin{center}
\resizebox{10.5cm}{!}{
\input{Step4.pstex_t}}
\caption{Replacement of subpath $\tp_3(b_{i,j},a_{i,j})$ by the edge-using segment $(q_{i,j}p_{i,j})$. \label{fig:Step4}}
\end{center}
\end{figure}
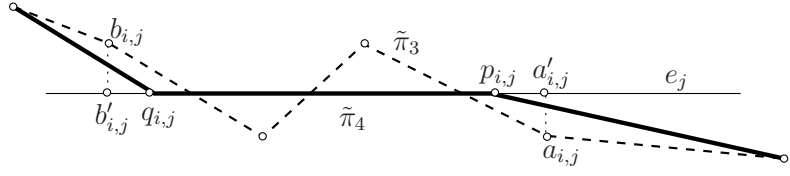

For $j=0,\dots ,k_i$, let $b'_{i,j}$  and $a'_{i,j}$ be the orthogonal projections of the points  $b_{i,j}$  and $a_{i,j}$ onto the edge $e_j$, respectively (Figure~\ref{fig:Step4}). Let $p_{i,j}$ and $q_{i,j}$  be the Steiner points on $e_j$ defining the largest sub-interval of the interval $(a'_{i,j},b'_{i,j})$ on $e_j$ and assume that $p_{i,j}$ is between $a'_{i,j}$ and $q_{i,j}$. (In the case where the interval $(a'_{i,j},b'_{i,j})$ contains no Steiner points, we define $p_{i,j}=q_{i,j}$ to be  the closest Steiner point to $a'_{i,j}$ on $e_j$.)
In $\tp_4$, the edge-using segment $(q_{i,j}p_{i,j})$ will replace in $\tp_3$ the subpath $\tp_3(b_{i,j},a_{i,j})$. Let us estimate the resulting error. From the definition of the edge vicinity $D_\eps (e_j)$, the Steiner points on $e_j$, and the radii $r(p_{i,j})$ and $r(q_{i,j})$, it is easy to derive that
\begin{equation}\label{e153}
|p_{i,j}a_{i,j}|\leq \frac{3}{2}\eps r(p_{i,j}) \quad {\rm and} \quad
|b_{i,j}q_{i,j}|\leq \frac{3}{2}\eps r(q_{i,j}).
\end{equation}
Furthermore, by our construction and the fact that $(q_{i,j},p_{i,j})$ is an edge-using segment, it follows that
\begin{equation}\label{e155}
\|q_{i,j}p_{i,j}\| \leq \|\tp_3(b_{i,j},a_{i,j})\|.
\end{equation}

Next, we modify between-edge-vicinities portions $\tp_3(a_{i,j},b_{i,j+1})$, into paths $\tp_4(p_{i,j},q_{i,j+1})$, joining Steiner points $p_{i,j}$ and $q_{i,j+1}$, and not intersecting any vertex or edge vicinities. We apply a construction  analogous to the one  used in Step 3 to define the paths $\tp_3(v_i,v_{i+1})$.

We fix $j$ and consider the between-edge-vicinities portion $\tp_3(a_{i,j},b_{i,j+1})$. We first consider the special case where $\tp_3(a_{i,j},b_{i,j+1})$ is the segment $(a_{i,j},b_{i,j+1})$. In this case, we observe that $e_j$ and
$e_{j+1}$ must be edges of the tetrahedron containing the segment $(a_{i,j},b_{i,j+1})$ and define
$\tp_4(p_{i,j},q_{i,j+1})=(p_{i,j},q_{i,j+1})$. We estimate the length of this segment using the triangle inequality, the estimates (\ref{e153}) and Definition \ref{D(x)} as follows
\begin{eqnarray*}
|p_{i,j}q_{i,j+1}| &\leq& |p_{i,j}a_{i,j}| + |a_{i,j}b_{i,j+1}|
+|b_{i,j+1}q_{i,j+1}|\leq \frac{3\eps}{2}
(r(p_{i,j})+r(q_{i,j+1})) + |a_{i,j}b_{i,j+1}| \leq \\
&& \frac{\eps}{16}(d(p_{i,j})+d(q_{i,j+1})) +
|a_{i,j}b_{i,j+1}|\leq \frac{\eps}{8}|p_{i,j}q_{i,j+1}| +
|a_{i,j}b_{i,j+1}|.
\end{eqnarray*}
Using this estimate and the observation that the weight of the segment $(p_{i,j},q_{i,j+1})$ cannot exceed the weight of $(a_{i,j},b_{i,j+1})$, we obtain
\begin{equation}\label{e160}
\|\tp_4(p_{i,j},q_{i,j+1})\|=\|p_{i,j}q_{i,j+1}\|\leq
(1+\frac{\eps}{7})\|a_{i,j}b_{i,j+1}\|=(1+\frac{\eps}{7})\|\tp_3(a_{i,j},b_{i,j+1})\|.
\end{equation}

Next, we consider the general case, where $\|\tp_3(a_{i,j},b_{i,j+1})\|$ consists of at least two segments.
Let $X$ be the first bending point after $a_{i,j}$ that is on the boundary $\partial D(e_j)$. If the path $\tp_3(a_{i,j},b_{i,j+1})$ is entirely inside $D(e_j)$, then we set $X=b_{i,j+1}$. Furthermore, let $\w'_j$ be the minimum weight among the segments in $\tp_3(a_{i,j},X)$, and let $x$ be the first bending point after $a_{i,j}$ that is an end-point of a segment whose weight is $\w'_j$.
We define the weight $\w'_{j+1}$, and the bending points $Y$, and $y$, analogously with respect to $b_{i,j+1}$ and the edge vicinity $\V(e_{j+1})$. It follows that the point $x$ precedes $y$ along $\tp_3(a_{i,j},b_{i,j+1})$.
We define the portion of the path $\tp_4$ joining $p_{i,j}$ and $q_{i,j+1}$ by $\tp_4(p_{i,j},q_{i,j+1})=\{(p_{i,j},x),\tp_3(x,y),(y,q_{i,j+1})\}$ and estimate its cost.
Let us first estimate the cost of the segment $(p_{i,j},x)$. We observe that $\|p_{i,j}x\|\leq \w'_j|p_{i,j}x|$,
that $|p_{i,j},x|\leq |p_{i,j}a_{i,j}| + |\tp_3(a_{i,j},x)|$ (by triangle inequality), and that the segments on the path $\tp_3(a_{i,j},x)$ have weight greater than or equal to $\w'_j$. Using these observations, (\ref{e153}), and Definition \ref{D(x)}, we obtain
\begin{eqnarray}\label{e162}
\|p_{i,j}x\| &\leq& \w'_j|p_{i,j}a_{i,j}| +
\w'_j|\tp_3(a_{i,j},x)|\leq \w'_j|p_{i,j}a_{i,j}| + \|\tp_3(a_{i,j},x)\| \leq \nonumber \\
&& \frac{3\eps}{2}\w'_j r(p_{i,j}) +  \|\tp_3(a_{i,j},x)\| =\frac{\eps}{16} \w'_j d(p_{i,j}) + \|\tp_3(a_{i,j},x)\|.
\end{eqnarray}
Analogously, we have
\begin{equation}\label{e163}
\|yq_{i,j+1}\| \leq \frac{\eps}{16} \w'_{j+1} d(q_{i,j+1}) + \|\tp_3(y,b_{i,j+1})\|.
\end{equation}
Using the definition of the path $\tp_4(p_{i,j},q_{i,j+1})$, the estimates (\ref{e162}) and (\ref{e163}), and the definition of the distances $d(p_{i,j})$ and $d(q_{i,j+1})$, the weights $\w'_j$ and $\w'_{j+1}$, and the points $X$ and $Y$, we obtain
\begin{eqnarray}\label{e165}
&&\|\tp_4(p_{i,j},q_{i,j+1})\|= \|\tp_3(a_{i,j},b_{i,j+1})\|+\frac{\eps}{16}(\w'_j d(p_{i,j}) +
\w'_{j+1} d(q_{i,j+1})) \\ \nonumber
&&\leq \|\tp_3(a_{i,j},b_{i,j+1})\|+\frac{\eps}{16}(\|\tp_3(p_{i,j},X)\|
+ \|\tp_3(Y,q_{i,j+1})\|)\\ \nonumber
&& \leq \|\tp_3(a_{i,j},b_{i,j+1})\|+\frac{\eps}{8}\|\tp_3(p_{i,j},q_{i,j+1})\|,
\end{eqnarray}
which implies  estimate (\ref{e160}), in the general case.
Finally, combining segments $(q_{i,j},p_{i,j})$ and  paths $\tp_4(p_{i,j},q_{i,j+1})$, for $j=0,\dots , k_i$, we construct a path
$$
\tp_4(v_i,v_{i+1})=\{ (v_i,p_{i,0}), \tp_4(p_{i,0},q_{i,1}),(q_{i,1},p_{i,1}),
\tp_4(p_{i,1},q_{i,2}),\dots , \tp_4(p_{i,k_i-1},q_{i,k_i}), (q_{i,k_i},v_{i+1})\}.
$$
This path has no bending points in any of the edge or vertex vicinities. Its cost can be bounded using (\ref{e155}) and (\ref{e160}) as follows
\begin{eqnarray} \label{e195}
 \|\tp_4(v_i,v_{i+1})\|&=&
\sum_{j=0}^{k_i}\|q_{i,j},p_{i,j}\| +
\sum_{j=0}^{k_i-1}\|\tp_4(p_{i,j},q_{i,j+1})\|\leq\\
&&\sum_{j=0}^{k_i} \|\tp_3(b_{i,j},a_{i,j})\| + \sum_{j=0}^{k_i-1}
(1+\frac{\eps}{7})\|\tp_3(a_{i,j},b_{i,j+1})\|\leq
 (1+\frac{\eps}{7})\|\tp_3(v_i,v_{i+1})\|,\nonumber
\end{eqnarray}
where we assume $v_i=b_{i,0}=q_{i,0}$ and $v_{i+1}=a_{i,k_i}=p_{i,k_i}$.

The paths $\tp_4(v_i,v_{i+1})$, for $i=0,\dots k-1$, form a linear path $\tp_4(v_0,v)$, whose cost is estimated using (\ref{e195}), (\ref{e120}), and (\ref{e110}) by 
\begin{equation} \label{e205}
\tp_4(v_0,v)\leq
\sum_{i=0}^{k-1}(1+\frac{\eps}{7})\|\tp_3(v_i,v_{i+1})\| = (1+\frac{\eps}{7})\|\tp_3(v_0,v)\| \leq
(1+\frac{\eps}{3})\|\tp_2(v_0,v)\|\leq (1+\eps)\|\tp(v_0,v)\|.
\end{equation}
As in Step 3, it is possible for $\tp_4$ to contain self-intersections and consecutive face-using segments. Hence, we
traverse $\tp_4$ and compress it by removing loops and by replacing consecutive face-using segments. The obtained
path is denoted again by $\tp_4$, and estimate (\ref{e205}) is  valid.

The bending points defining $\tp_4$ can be partitioned into two groups. The first group consists of bending points
corresponding to nodes of the graph $G_\eps$, i.e.,  Steiner points on bisectors, Steiner points on edges, and vertices of $\D$. The second group consists of the remaining bending points of $\tp_4$, which are bending points inside the faces of $\D$. We complete the proof of the theorem by showing that the sequence of the nodes in the first group defines a discrete path $\pi(v_0,v)$ whose cost $c(\pi(v_0,v))\leq \|\tp_4(v_0,v)\|$.
It suffices to show that any two consecutive nodes (bending points in the first group) along the path $\tp_4$ are adjacent in the approximation graph $G_\eps$.

To show this, we review closely the structure of the path $\tp_4$. In Step 3, portions of  $\tp_2$ related to vertex
vicinities have been replaced by two segment portions through-vertices of $\D$. Furthermore, we observe that the segments $(v_i,x)$ created in Step 3 are either a face-using segments or join $v_i$ to a Steiner point on a bisector. The same applies to segments $(y,v_{i+1})$. Similarly, in Step 4, portions related to edge-vicinities have been replaced by three segment portions visiting corresponding edges. Again segments $(p_{i,j},x)$ are either face-using segments or join $p_{i,j}$ to a node on a bisector, that is a neighbor of the bisector incident to the edge
containing $p_{i,j}$. The same applies to the segments $(y,q_{i,j+1})$. In summary,  the segments created in Steps 3
and 4 are of one of the following two types:
\begin{enumerate}
\item  A face-using segment with one of its endpoint being a (node) vertex of $\D$ or a Steiner point on an edge of $\D$.
\item A segment joining  two nodes, at least one of them being a Steiner point on an edge of $\D$ or a vertex of $\D$.
    \end{enumerate}
The remaining segments in  $\tp_4$ are cell-crossing and face-using segments, whose endpoints are outside any vertex or edge vicinity. All the cell-crossing segments in $\tp_4$ were created during Steps 1 and 2. Hence, one of their endpoints is a (node) Steiner point on a bisector of a tetrahedron. Finally, due to the compressing, there are no consecutive face-using segments in $\tp_4$.

Now, let $p$ and $q$ be two consecutive nodes along the path $\tp_4$. We show that $p$ and $q$ are adjacent in $G_\eps$.
We consider, first, the case where at least one of the nodes, say $p$, is a vertex of $\D$. Let $x$ be the bending point following $p$ along the path $\tp_4$. By the definition of bending points adjacent to the vertices
(in Step 3), we know that $(p,x)$ is a face-using segment followed by a cell-crossing segment $(x,x_1)$, joining $x$ to a (node) Steiner point on a bisector lying in one of the tetrahedra incident to the face that contains $(p,x)$. So, $q=x_1$ and $q$ is inside a tetrahedron incident to $p$.  Thus, $p$ and $q$ are adjacent in $G_\eps$. The
case where at least one of the nodes $p$ or $q$ is a Steiner point on an edge of $\D$ can be treated analogously.

Assume now that both $p$ and $q$ are Steiner points on bisectors. Let $x$ and $x_1$ be the bending points following $p$ along $\tp_4$. The point $x$ has to be a bending point on a face of the tetrahedron containing $p$. The segment $(x,x_1)$ is either a cell-crossing or a face-using segment. In the first case, $q$ must coincide with $x_1$ and is adjacent to $p$ in $G_\eps$, since it lies in a tetrahedron that is a neighbor to the one containing $p$. In the second case, where $(x,x_1)$ is a face-using segment, we consider the bending point $x_2$ that follows $x_1$ along the path $\tp_4$.
The segment $(x_1,x_2)$ must be a cell-crossing segment. Thus, in this case, $q=x_2$ is adjacent to $p$, because the tetrahedra containing $p$ and $q$ are neighbors.

We have shown that any pair $p$ and $q$ of consecutive nodes on the path $\tp_4$ are adjacent in  $G_\eps$. Hence, we define a discrete path $\pi(v_0,v)$ to be the path in $G_\eps$ following the sequence of nodes along  $\tp_4$.
Finally, we observe that the sub-paths of $\tp_4(p,q)$ joining pairs of consecutive nodes stay in the union of the tetrahedra containing these nodes and cross faces shared by the bisectors containing them. Hence, by the definition of the cost of the edges in $G_\eps$, we have $c(p,q)\leq \|\tp_4(p,q)\|$. Summing these estimates, for all edges of $\pi(v_0,v)$, and  using (\ref{e205}), we obtain,
$ c(\pi(v_0,v))\leq \|\tp_4(v_0,v)\|\leq (1+\eps)\|\tp(v_0,v)\|. $\hfill $\Box$

\section{An algorithm for computing SSSP in $G_\eps$}\label{Algorithms}
In this section we present our algorithm for solving the Single Source Shortest Paths (SSSP) problem in the approximation graph $G_\eps=(V(G_\eps),E(G_\eps))$. Straightforwardly, one can apply Dijkstra's algorithm,
which runs in $O(|E(G_\eps)| + |V(G_\eps)|\log |V(G_\eps)|)$ time. By Lemma \ref{size_of_G} we have $|V(G_\eps)|=O(\frac{n}{\eps^2}\log \frac{1}{\eps})$ and $|E(G_\eps)|=O(\frac{n}{\eps^4}\log^2 \frac{1}{\eps})$.
Thus, the SSSP problem in $G_\eps$ can be solved in $O(\frac{n}{\eps^4}\log \frac{n}{\eps}\log \frac{1}{\eps})$ time.

In the remainder of this section, we demonstrate how geometric properties of our model can be used to obtain a more efficient algorithm for solving the SSSP problem. More precisely, we present an algorithm that runs in $O(|V_\eps|(\log |V_\eps|+\frac{1}{\sqrt{\eps}}\log^3\frac{1}{\eps}))=
O(\frac{n}{\eps^{2.5}}\log\frac{n}{\eps}\log^3 \frac{1}{\eps})$
time.

Informally, the idea is to avoid consideration of large portions of the edges of the graph $G_\eps$ when
searching for shortest paths. We achieve that by applying the strategy proposed first in \cite{SR01,RS06} and developed further in \cite{AMS05} and by using the properties of the weighted distance function and
additive Voronoi diagrams studied in Section \ref{wdf-section}.
We maintain a priority queue containing candidate shortest paths. At each iteration of the algorithm,
a shortest path from the source $s$ to some node $u$ of $G_\eps$ is found. Then, the algorithm constructs edges adjacent to $u$ that can be continuations of the shortest path from  $s$ to $u$ and inserts them in the priority queue as new candidate shortest paths. In general, one needs to consider all edges adjacent to $u$ as possible continuations. In our case, we divide the edges adjacent to $u$ into $O(\frac{1}{\sqrt{\eps}}\log\frac{1}{\eps})$ groups related to the segments containing Steiner points in the neighboring bisectors and demonstrate that we can consider just a constant number of edges in each group.
The latter is possible due to the  structure of the Voronoi cell ${\cal V}(u)$ of the node $u$
in the additive Voronoi diagram related to a fixed group (see Theorem \ref{the_thm}).

This section is organized as follows: In the next subsection, we describe the general structure of the algorithm.
In Subsection \ref{implementation}, we show how this strategy can be applied in our case and present an outline  of the algorithm. We provide details of the implementation of the algorithm and analyze its complexity. Finally, at the end we establish the main result of the paper.

\subsection{General structure of the algorithm}\label{Algorithms_1}

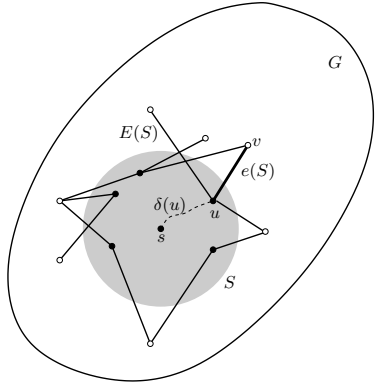
\begin{figure}
\begin{center}
\resizebox{5cm}{!}{
\input{F6.pstex_t}}
\end{center}
\caption{\small The sets $S$, $S^a$, and an optimal edge $e(S)=(u,v)$  in $E(S)$ are illustrated.
A shortest path from $s$ to $u$ is illustrated by a dashed curve. } \label{fig:F6}
\end{figure}

Let $G(V,E)$ be a directed graph with positive costs (lengths) assigned to its edges and $s$ be a fixed node of $G$, called the {\em source}.  A standard greedy approach for solving the SSSP problem works as follows: a subset, $S$, of nodes  to which the shortest path has already been found and a set, $E(S)$, of edges connecting $S$ with
$S^a\subset V\setminus S$ are maintained. The set $S^a$ consists of nodes not in $S$ but adjacent to $S$. In each iteration,  an {\em optimal} edge $e(S)=(u,v)$ in $E(S)$ is selected, with source $u$ in $S$ and target $v$ in $S^a$ (see Figure \ref{fig:F6}). The target vertex $v$ is added to $S$ and $E(S)$ is updated correspondingly. An edge $e=e(S)$ is optimal if it  minimizes the value $\delta(u) + c(e)$, where $\delta(u)$ is the distance from $s$ to $u$ and $c(e)$ is the cost of $e$.
%

Different strategies for maintaining information about $E(S)$ and finding an optimal edge $e(S)$ during each iteration result in different algorithms for computing SSSP. For example, Dijkstra's algorithm maintains only a subset $Q(S)$ of $E(S)$, which, however, always contains an optimal edge. Alternatively, as in \cite{AMS05}, one may maintain a subset  of $E(S)$ containing one edge per node $u$ in $S$. The target node of this edge is called the {\em representative} of $u$ and is denoted by $\rho(u)$. The node $u$ itself is called {\em predecessor} of its representative. The
representative $\rho(u)$ is defined to be the target of the minimum cost edge in the {\em propagation set} $I(u)$ of $u$, where $I(u)\subset E(S)$ consists of all edges $(u,v)$ such that $\delta(u)+c(u,v) < \delta(u') + c(u',v)$ for all  nodes $u'\in S$ that have entered $S$ before $u$. The union of propagation sets forms a subset $Q(S)$ of $E(S)$ that always contains an optimal edge. Propagation sets $I(u)$, for $u\in S$, form a partition of $Q(S)$. The propagation sets of the vertices in $S$ form a partition of $E(S)$, which is called {\em propagation diagram},
and is denoted by ${\cal I}(S)$.

The set of representatives $R\subset S^a$ can be organized in a priority queue, where the key of the node $\rho(u)$ in $R$ is defined to be $\delta(u) + c(u,\rho(u))$.  Observe that the edge corresponding to the minimum in $R$ is an optimal edge for $S$. In each iteration, the minimum key node $v$ in $R$ is selected and the following three steps are carried:\\[.5ex]
{\bf Step 1.} {\em The node  $v$ is  moved from $R$ into $S$. Then, the propagation set $I(v)$ is computed and  the propagation diagram ${\cal I}(S)$ is updated accordingly.}\\[.5ex]
{\bf Step 2.} {\em Representative $\rho(v)$  of $v$ and a new representative, $\rho(u)$, for the predecessor  $u$ of $v$  are computed.}\\[.5ex]
{\bf Step 3.} {\em The new representatives, $\rho(u)$ and $\rho(v)$, are either  inserted  into $R$ together with their corresponding keys, or (if they are already in $R$) their keys are updated.}\\[.5ex]

Clearly, this leads to a correct algorithm for solving the SSSP problem in $G$.
The total time for the priority queue operations
\footnote{Note that we do not need a priority queue based on elaborated data structures such as Fibonacci heaps. Any priority queue with
logarithmic time per operation suffices.}
is $O(|V|\log |V|)$. Therefore, the efficiency of this strategy depends on the maintenance of the propagation diagram,  the complexity of the propagation sets, and the efficient updates of the new representatives.
In the next subsection, we address these issues and provide necessary details.

\subsection{Implementation details and analysis}\label{implementation}
\subsubsection{Notation and algorithm outline}
Our algorithm   follows the general strategy as described in the previous subsection. First, we convert $G_\eps$ into a directed graph by replacing each of its edges by a pair of oppositely oriented edges with cost equal to the cost of the original edge.

Let, as above, $S$ be the set of the nodes to which the shortest path has already been found and $E(S)$ be the set of the edges joining $S$ with $S^a\subset V\setminus S$.
We partition the edges of $G_\eps$  (and respectively $E(S)$) into groups so that the propagation sets and the corresponding propagation diagrams, when restricted to a fixed group, have a simple structure and can be updated efficiently. Then, for each node $u$ in $S$, we will keep multiple representatives in $R$ -- a constant number on the average, for each group where edges incident to $u$ participate and where its propagation set is non-empty. A node in $S^a$ will have  multiple predecessors -- at most as many as the number of the groups where edges incident to it participate. We will show that the number of the groups, where edges incident to $u$ can participate, is bounded by
$O(\frac{1}{\sqrt{\eps}}\log\frac{1}{\eps})$ times the number of bisectors incident to $u$. In a fixed group, we will be able to compute new representatives in $O(\log\frac{1}{\eps})$ time and update propagation diagrams in $O(\log^2\frac{1}{\eps})$ time.

\begin{figure}
\begin{center}
\resizebox{7cm}{!}{
\input{F65.pstex_t}}
\end{center}
\caption{Two Steiner segments $\ell$ and $\ell_1$ lying on neighboring bisectors $\b=\triangle ABP$
and $\bb=\triangle ACP_1$  respectively, that share a face $\f=\triangle ABC$ are illustrated.
Steiner segments on $\b$ and $\bb$ are parallel to the shared face $\f$. The edges joining nodes on
$\ell$ and $\ell_1$ form the group of edges corresponding to the triple $(\ell,\ell_1,\f)$.} \label{fig:F65}
\end{figure}
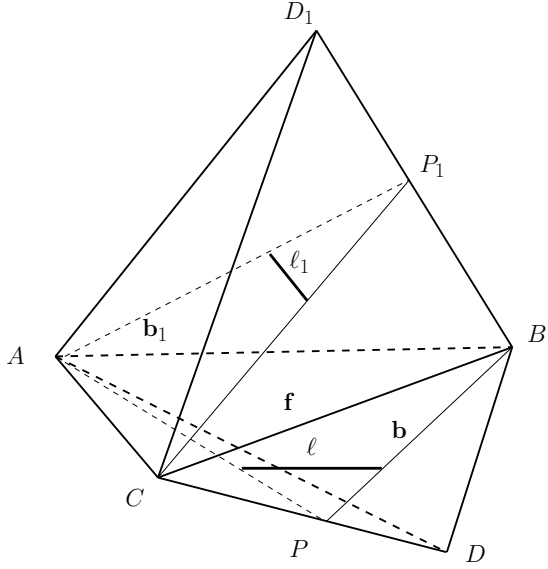

Edges of $G_\eps$ joining pairs of Steiner points on bisectors are naturally partitioned into groups corresponding to ordered triples $(\b,\bb,\f)$, where $\b$ and $\bb$ are neighboring bisectors with respect to the face $\f$ (see Section \ref{discretepath-section} for the definitions). The edges of the initial tetrahedralization $\D$ are assumed to belong to the bisectors incident to them. So, the group of edges corresponding to an ordered triple $(\b,\bb,\f)$ consists of all edges from a node on $\b$ to a node on $\bb$ that cross $\f$. Recall that the nodes (Steiner points) on any bisector $\b$ were placed on a set of segments parallel to the edge of $\D$ incident to $\b$. In our discussion below, we refer to these segments, including the edge of $\D$, as {\em Steiner segments}. We further partition the group of edges associated with the triple $(\b,\bb,\f)$ into subgroups corresponding to pairs
of Steiner segments $(\ell,\ell_1)$ on $\b$ and $\bb$, respectively, see Figure \ref{fig:F65} (a).
In this way, the edges of $G_\eps$ are partitioned into groups corresponding to ordered triples $(\ell,\ell_1,\f)$, where $\ell$ and $\ell_1$ are Steiner segments parallel to $\f$ on two neighboring bisectors sharing $\f$. The group corresponding to $(\ell,\ell_1,\f)$ is denoted by $E(\ell,\ell_1,\f)$ and consists of all oriented edges from a node
on $\ell$ to a node on $\ell_1$ that cross $\f$.

A fixed bisector $\b$ has either three or six neighboring bisectors ($\b$ itself and two or five others, respectively) with respect to each of the two faces forming the dihedral angle bisected by $\b$. Hence, the total number of ordered triples $(\b,\bb,\f)$ does not exceed $36n$. By Lemma \ref{number-of-points}, the number of Steiner segments on any
bisector is $O(\frac{1}{\sqrt{\eps}}\log\frac{1}{\eps})$ and thus the number of subgroups of a group corresponding to a triple $(\b,\bb,\f)$ is $O(\frac{1}{\eps}\log^2\frac{1}{\eps})$. In total, the number of groups $E(\ell,\ell_1,\f)$ is $O(\frac{n}{\eps}\log^2\frac{1}{\eps})$.

A node $u$ lying on a Steiner segment $\ell$ will have a set of representatives for each group $E(\ell,\ell_1,\f)$ corresponding to a triple, where $\ell$ is the first element and where its propagation set is non-empty. We denote this set by $\boldrho(u,\ell_1, \f)$. The set of representatives $\boldrho(u,\ell_1, \f)$ will correspond to the structure of the propagation set $I(u;\ell,\ell_1, \f)$, as we  will detail in the next subsection.

Consider an iteration of our  algorithm. Let $v$ be the node extracted  from priority queue $R$, containing all representatives. Let $\T(v)$ be the set of triples $(\ell,\ell_1,\f)$ such that $v$ lies on $\ell$. First,
we need to move $v$ from $S^a$ to $S$, since the distance from $v$ to the source $s$ has been found. Nodes that
are targets of the edges originating from $v$ need to be added to $S^a$. Then, we need to compute representatives of $v$ for each group of edges, where edges originating at $v$ participate and where its propagation set is non-empty. Finally, we need to compute new representatives for all nodes in the set of predecessors of $v$, which we denote by  $R^{-1}(v)$. The outline of our algorithm is as follows.
\medskip
\begin{boxit}
{\bf ALGORITHM:} {\sf SSSP($G_\eps$, $s$) }

{\bf While} $S\not=V_\eps$ {\bf do}

 1.  $v$ $\longleftarrow$ Extract\_min($R$);

 2. Insert $v$ in  $S$ and update $S^a$;

 3. {\bf For} each triple $(\ell,\ell_1,\f)\in {\cal T}(v)$  {\bf do}

      \quad  3.1 Update the data structures related to the Propagation Diagram $\I(\ell, \ell_1,\f)$;

      \quad 3.2 Find new representatives for nodes whose propagation set has changed in 3.1;

      \quad 3.3 Update sets of representatives $\boldrho(u;\ell',\ell ,\f')$, for all $u\in R^{-1}(v) $;

     \quad 3.4 Update R with respect to 3.2 and 3.3.
\end{boxit}
\medskip
\noindent In the remainder of this section, we address the implementation of this algorithm and analyze its complexity.
First, we observe that the number of iterations is $|V_\eps|$.  The total number of representatives cannot exceed
the number of oriented edges in $G_\eps$, which is less than $|V_\eps|^2$ and so, the size of the priority queue $R$ is bounded by $|V_\eps|^2$ (later we show that it is actually $O(\frac{|V_\eps|}{\sqrt{\eps}}\log^2\frac{1}{\eps})$). Therefore, a single priority queue operation takes
$O(\log |V_\eps|)$ time and the total time for Step 1 is  $O(|V_\eps|\log |V_\eps|)$. The total time for Step 2 is $O(|V_\eps|\log\frac{1}{\eps})$.

In Section \ref{step_3.1}, we describe the structure and maintenance of the data structures related to the propagation diagrams $\I(\ell,\ell_1,\f)$. Computation and updates of the sets of representatives are described in Section
\ref{step_3.x}. We conclude our discussion in Section \ref{main_res} by summarizing the time complexity of the algorithm and by establishing our main result.

\subsubsection{Implementation of Step 3.1}\label{step_3.1}
We consider a fixed triple $(\l,\l_1, f)$, where $\ell$ and $\ell_1$ are Steiner segments on neighboring bisectors $\b$ and $\bb$ sharing $\f$. The propagation diagram   $\I(\l,\l_1,f)$, was defined as the set consisting of the propagation sets of the active nodes on $\ell$. Instead of explicitly computing the propagation diagram,  we construct and maintain a number of data structures that allow efficient computation and updates of representatives.

Consider an iteration of our algorithm. Denote the currently active nodes on $\ell$ by $u_1,\dots,u_k$, and assume that they are listed by their order of entering $S$. We denote this set by $S(\ell)$ and assume that it is stored and maintained as a doubly linked list ordered according to the position of the nodes on $\ell$. In Step 3.1, we update the data structures related to the propagation diagram $\I(\l,\l_1, f)$.
According to our definition, the propagation set $I(u)=I(u;\ell,\ell_1, f)$ of a node $u\in\ell$ consists of all edges $(u,v_1)$ in $E(\ell,\ell_1, f)$ such that $\delta(u)+c(u,v_1) < \delta(u_i)+c(u_i,v_1)$, for $i=1,\dots,k$.
Clearly, the set $I(u)$ can be viewed and described as a subset of the set of nodes $v_1$ on $\ell_1$ that satisfy the following three conditions:\\[.5ex]
\indent {\bf C1.} The nodes $u$ and $v_1$ are adjacent in $G_\eps$ by an edge that crosses $\f$;\\[.5ex]
\indent {\bf C2.} $\delta(u)+c(u,v_1) < \delta(u_i)+c(u_i,v_1)$, for $i=1,\dots,k$;\\[.5ex]
\indent {\bf C3.} The node $v_1$ is in $S^a$.\\[.5ex]
We construct and maintain separate data structures for the nodes on $\ell_1$ satisfying each of these three conditions: The data structure related to {\bf C1} is called {\em Adjacency Diagram} and
is denoted by $\A(\ell,\ell_1, f)$. It consists of sets $A(u,\ell_1)$, for all nodes $u$ on $\ell$, where the set $A(u,\ell_1)$ consists of the nodes on $\ell_1$ that satisfy {\bf C1}. This data structure is static. The data structure related to {\bf C2} is, in fact, a dynamic additive Voronoi diagram on $\ell_1$ for the active nodes on $\ell$ with respect to the weighted distance function $c(u,x)$ defined and studied in Section \ref{model}, see
(\ref{c(v,x)}). Finally, the nodes  on $\ell_1$ that are in $S^a$ are stored in a dynamic doubly-linked list
and organized in a binary search tree with respect to their position on $\ell_1$. We denote this data structure by $S^a(\ell_1)$. The lists $S(\ell)$ and $S^a(\ell_1)$ are readily maintained throughout the algorithm in logarithmic time per operation. Next, we describe in detail the construction and maintenance of these data structures.

\noindent {\bf Adjacency Diagram:} The Adjacency Diagram $\A(\ell,\ell_1, f)$ consists of sets
$A(u,\ell_1)$, for all nodes $u$ on $\ell$.  We assume that the nodes on $\ell_1$ are stored in an ordered
list $V(\ell_1)$ according to their position on that segment. For any fixed node $u\in\ell$, the adjacency
set $A(u,\ell_1)$ will be computed and stored as a sublist of the list $V(\ell_1)$.
We denote this sublist by $\bar{A}(u,\ell_1)$.

We reduce the size of $\bar{A}(u,\ell_1)$ by replacing each portion of consecutive nodes in them by a pair of pointers to the first and to the last node in that portion. (Isolated nodes are treated as portions of length one.)
Hence, each sublist $\bar{A}(u,\ell_1)$ is an ordered list of pairs of pointers identifying portions of consecutive nodes in the underlying list  $V(\ell_1)$. The size of the sublists implemented in this way is proportional to the number of the consecutive portions they contain. Next, we discuss the structure of the lists $\bar{A}(u,\ell_1)$ and show that their size is bounded by a small constant.

According to our definitions (Section \ref{discretepath-section}), an edge $(u,u_1)$ is present in $A(u,\l_1)$ if the local shortest path $\hat{\pi}(u,u_1;f)$ does not touch the boundary of $f$, where the path $\hat{\pi}(u,u_1;f)$ was defined in (\ref{local-path}).
We refer to intervals on $\ell_1$ with both of their end-points being Steiner points as {\em Steiner intervals}. Furthermore, we say that a Steiner interval is covered by the set $A(u,\ell_1)$ if all Steiner points, including its end-points, are in $A(u,\ell_1)$. Clearly, each maximal interval covered by $A(u,\ell_1)$ corresponds to and
defines a portion of consecutive nodes on $\ell_1$ that are adjacent to $u$. Moreover, by our definition, the list $\bar{A}(u,\ell_1)$ consists of the pairs of pointers to the end-points of the maximal intervals covered by $A(u,\ell_1)$. In the next lemma, we show that there are at most seven maximal Steiner intervals covered by $A(u,\ell_1)$.

\begin{lemma} \label{adj-int}
The number of the maximal intervals covered by $A(u,\l_1)$ is at most seven. The corresponding ordered list $\bar{A}(u,\ell_1)$ can be computed in $O(\log K(\ell_1))$ time, where $K(\ell_1)$ denotes the number of Steiner points on $\ell_1$.
\end{lemma}
{\bf Proof:} Presented in Appendix 3. $\Box$\\

We assume that the nodes that are end-points of the maximal Steiner intervals covered by the sets $A(u,\ell_1)$, for all nodes $u\in\ell_1$, are pre-computed in a preprocessing step and stored in the lists $\bar{A}(u,\ell_1)$ as discussed above. Lemma \ref{adj-int} implies that this preprocessing related to the group $(\ell,\ell_1,f)$ takes $O(K(\ell)\log K(\ell_1))$ time, where $K(\ell)$ and $K(\ell_1)$ denote the number of the nodes on $\ell$ and $\ell_1$, respectively. Next, we discuss the Voronoi diagram data structure related to condition {\bf C2}.

\noindent {\bf Dynamic Additive Voronoi Diagram:}
We assumed that the currently active nodes, $u_1,\dots,u_k$ on $\ell$, are listed by order of their insertion into $S$. So, for the distances of these nodes to the source, we have $\delta(u_1)\leq\dots\leq\delta(u_k)$.
We view the distance $\delta(u_i)$  as an additive weight assigned to the node $u_i$, and consider the additive Voronoi diagram of  $u_1,\dots ,u_k$ on $\ell_1$ with respect to the weighted distance function, introduced and studied in Section \ref{model} and defined by (\ref{c(v,x)}). From the definition (see (\ref{c(v,x)})), the weighted distance $c(u,x)$ for a node $u$ on $\ell$ and a point $x\in \ell_1$ is given by
$$c(u,x)=c(u,x;\f)=\min_{a,a_1\in F}\{w|ua|+w_f|aa_1|+w_1|a_1x|\},$$
where $F$ is the plane containing the face $\f$; $w$, $w_1$ are the weights of the cells containing $\ell$ and $\ell_1$, respectively, and $w_f$ is the weight associated to the face $f$. An important observation for our discussion here is that if $x$ is a node on $\ell_1$ adjacent to $u$, then the cost of the edge $(u,x)$ is $c(u,x)$.

We denote the end-points of the segment $\ell_1$ by  $A_1$ and $B_1$ and assume
that it is oriented so that $A_1<B_1$. For $i=1,\dots ,k$,  the Voronoi cell ${\cal V}(u_i)$
is defined as the set of points  on $\ell_1$
$${\cal V}(u_i)=\{x\in(A_1,B_1)\ :\ \delta(u_i) + c(u_i,x)\leq
\delta(u_j)+c(u_j,x)\, {\rm for}\ j\not=i\},$$ where ties are resolved in favor of the node that has entered $S$ earlier. Clearly, the Voronoi diagram ${\cal V} (u_1,\dots,u_k)$ is a partitioning of $(A_1,B_1)$ into a set of intervals, where each interval belongs to exactly one of the Voronoi cells. Hence, ${\cal V} (u_1,\dots,u_k)$ is completely described by a set of points $A_1=x_0<x_1<\dots <x_m<x_{m+1}=B_1$ and an assignment between the
intervals $(x_j,x_{j+1})$, for $j=0,\dots, m$, and the cells of the diagram.

We assume that ${\cal V}(u_1, \dots ,u_k)$ is known and stored. We further assume that a node $v$ on $\ell$ has been extracted by the extract-min operation in Step 1 of our algorithm.  In Step 3.1, we need to add the new site $v$
and to compute the Voronoi diagram ${\cal V}(u_1, \dots , u_k,v)$. Next we show how this can be achieved in
$O(\log^2 \frac{1}{\eps})$ time. First, the following lemma shows that the Voronoi cell of $v$ has a simple structure.
\begin{lemma}\label{cell}
Let $u_1,\dots, u_k$  be the active nodes on $\ell$ and let $v$ be the last node inserted in $S$. Then the Voronoi cell ${\cal V}(v)$, in the Voronoi diagram ${\cal V}(u_1, \dots , u_k, v)$, is either empty or consists of a single interval on $\ell_1$.
\end{lemma}
{\bf Proof:} By our assumptions $\delta(u_i)\leq \delta(v)$, for $i=1,\dots ,k$. The Voronoi cell  ${\cal V}(v)$ can be represented as an intersection ${\cal V}(v)=\cap_{i=1}^k{\cal V}_i(v)$, where the sets ${\cal V}_i(v)$ are defined by ${\cal V}_i(v)=\{x\in\ell_1\ :\ \delta(v)-\delta(u_i) +c(v,x)<c(u_i,x)\}$. By Theorem \ref{the_thm}, each of  ${\cal V}_i(v)$ is either empty or is an interval on $\ell_1$, and thus the same is true for their intersection.
\hfill $\Box$\\[1ex]
Using the above lemma, we easily obtain a bound on the size of the Voronoi diagrams.
\begin{corollary} \label{|X|}
The number of the intervals comprising the diagram ${\cal V}(u_1,\dots ,u_k)$ does not exceed $2k-1$.
\end{corollary}
Next, we present and analyze an efficient procedure which, given the Voronoi diagram ${\cal V}(u_1,\dots ,u_k)$ and a new node $v$ inserted in $S$, determines the Voronoi diagram ${\cal V}(u_1,\dots ,u_k, v)$. This includes computation of the Voronoi cell ${\cal V}(v)$, update of the set of points $x_1,\dots ,x_m$ describing ${\cal V}(u_1,\dots ,u_k)$ to another set describing ${\cal V}(u_1,\dots ,u_k,v)$ and update of the assignment information between intervals and Voronoi cells.

According to Lemma \ref{cell}, the Voronoi cell ${\cal V}(v)$ is an interval, which we denote by $(x^-,x^+)$.
Let $M$ be any of the points $x_1,\dots ,x_m$ characterizing the diagram ${\cal V}(u_1,\dots ,u_k)$. The following claim shows that the relative position of $M$ with respect to the interval $(x^-,x^+)$ can be determined in constant time.
\begin{claim}\label{bsearch}
The relative position of $M$ with respect to the interval $(x^-,x^+)$ can be determined in $O(1)$ time.
\end{claim}
{\bf Proof:} By the definition of  point $M$, it follows that there are two nodes  $u_{i_1}$ and $u_{i_2}$ such that $\delta(u_{i_1})+c(u_{i_1},M)=\delta(u_{i_2})+c(u_{i_2},M)$. We denote the latter value by $d(M)$ and note that $d(M)\leq \delta(u_i)+c(u_i,M)$, for $i=1,\dots ,k$. Then, we compute the value $d(v,M)=\delta(v)+c(v,M)$ and compare it with $d(M)$.

If $d(v,M)<d(M)$, then we have $M\in(x^-,x^+)$ and thus $x^-<M<x^+$. In the case where $d(v,M)\geq d(M)$, we compute the Voronoi cell $\triangle(v)$ of $v$ in the three cites diagram ${\cal V}(u_{i_1},u_{i_2},v)$. By Lemma \ref{cell}, the cell $\triangle(v)$ is an interval on $\ell_1$. Since $M$ must be outside $\triangle(v)$ and
$(x^-,x^+)\subset \triangle(v)$, it follows that the relative position between $M$ and $(x^-,x^+)$ is the same as
the relative position between $M$ and $\triangle(v)$.

The claimed time bound follows from the described procedure, which besides the constant number of simple
computations, involves a constant number of evaluations of the function $c(\cdot,\cdot)$ and eventually solving of
the equations $c(u_i,x)-c(v,x)=\delta(v)-\delta(u_i)$, for $i=i_1,i_2$.\hfill $\Box$

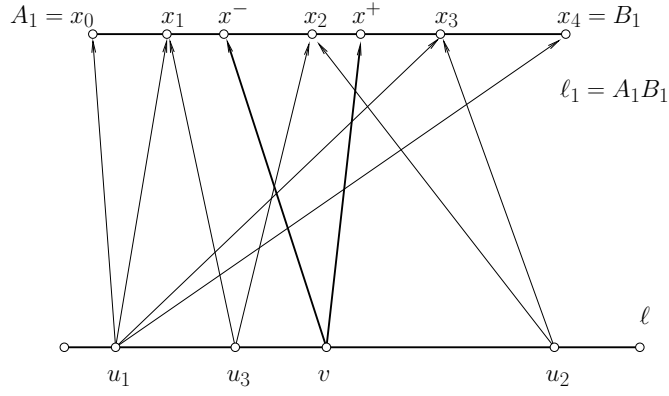
\begin{figure}
\begin{center}
\resizebox{8.5cm}{!}{
\input{F67.pstex_t}}
\end{center}
\caption{The figure illustrates updates of the diagram ${\cal V}$. The Voronoi diagram
${\cal V}(u_1,u_2,u_3)$ for the nodes $u_1$, $u_2$, and $u_3$
is characterized by the sequence $\{x_0<x_1<x_2<x_3<x_4\}$ and the assignment
${\cal V}(u_1)=(x_0,x_1)\cup (x_3,x_4)$, ${\cal V}(u_2)=(x_2,x_3)$, ${\cal V}(u_3)=(x_1,x_2)$.
After computation of the Voronoi cell ${\cal V}(v)=(x^-,x^+)$ the Voronoi diagram
${\cal V}(u_1,u_2,u_3,v)$ is characterized by the sequence $\{x_0<x_1<x^-<x^+<x_3<x_4\}$ and
the assignment ${\cal V}(u_1)=(x_0,x_1)\cup (x_3,x_4)$, ${\cal V}(u_2)=(x^+,x_3)$,
 ${\cal V}(u_3)=(x_1,x^-)$, ${\cal V}(v)=(x^-,x^+)$.}
\label{fig:F67}
\end{figure}

We derive the following binary search procedure, which computes the Voronoi cell  ${\cal V}(v)$.

\medskip
\begin{boxit}

{\bf ALGORITHM:} {\sf Voronoi cell}  ${\cal V}(v)$

{\em Input:} The sequence $X=\{A_1=x_0<x_1< \dots <x_m<x_{m+1}=B_1\}$.

{\em Output:} Points $x^-$ and $x^+$ such that ${\cal V}(v)=(x^-,x^+)$.

A. Compute the point $x^-$ first by the following:

\quad 1. {\bf While } $|X|>2$ {\bf do} Steps 1.1 -- 1.3 below

\qquad 1.1. Find the median $M$ of the sequence $X$.

\qquad 1.2. Determine the relative position  between $M$ and $x^-$.

\qquad 1.3 {\bf If} $x^-<M$  {\bf then} set $X=\{x_0<\dots <M\}$
 {\bf else} set $X=\{M<\dots <x_{m+1}\}$.

\quad 2. {\bf If}  $|X|= 2$  compute $x^-$ directly.

B. Compute the point $x^+$ in the same way.

\end{boxit}
\medskip
Once the cell ${\cal V}(v)=(x^-,x^+)$ has been computed, the update of diagram ${\cal V}(u_1,\dots ,u_k)$ to diagram ${\cal V}(u_1,\dots ,u_k,v)$ can be done in a natural way. The sorted sequence of points $X(u_1,\dots,u_k,v)$ characterizing the diagram ${\cal V}(u_1,\dots ,u_k,v)$ is obtained from the sequence
$X(u_1,\dots,u_k)=\{x_0<\dots <x_{m+1}\}$ by inserting the points $x^-$ and $x^+$ at their positions and by deleting  points (if any) $x^-<x_{j^-+1}<\dots <x_{j^+-1}<x^+$ lying inside the interval $(x^-,x^+)$,
where  $x_{j^-}$ and $x_{j^+}$ are the left and the right neighbors of the points $x^-$ and $x^+$, respectively.
%
We need to delete each of the intervals $(x_j,x_{j+1})$, for $j=j^-, \dots ,j^+-1$, from the cell that contains it and to add intervals $(x_{j^-},x^-)$ and $(x^+,x_{j^+})$ to the cells that have contained intervals $(x_{j^-},x_{j^-+1})$ and $(x_{j^+-1},x_{j^+})$, respectively. Indeed, if the cell of some of the nodes $u_1,\dots, u_k$ becomes empty
then this node is removed from the set of active nodes in the group $E(\ell,\ell_1, f)$.

To implement all of these updates efficiently, we maintain the sequence $X$ of points characterizing the Voronoi diagram of the currently active nodes on $\ell$ in an order-statistics tree, allowing us to report order statistics as well as insertions and deletions  in $O(\log |X|)$ time.  Based on this data structure, computation of the
interval $(x^-,x^+)$ takes $O(\log^2 |X|)$ time, since it takes $O(\log |X|)$ iterations, and each iteration
takes $O(\log |X|)$ time. The  update of the Voronoi diagram requires two insertions and $j^+-j^-+1$ deletions in $X$, where insertions take $O(\log|X|)$ time and deletions are done in amortized $O(1)$ time.

Let us estimate the time for the maintenance of the Voronoi diagram of the active nodes in the group $E(\ell,\ell_1,\f)$. We denoted the total number of the nodes on $\ell$ by $K(\ell)$. Each of the nodes on $\ell$ becomes active  once during the execution. Thus, each node on $\ell$ becomes  subject of the procedure {\sf Voronoi cell} exactly once. According to Corollary \ref{|X|}, the sizes of the sequences $X$ characterizing Voronoi diagrams in the group $E(\ell,\ell_1,\f)$ are bounded by $2K(\ell)+1$. Therefore, the total time spent by the procedure
{\sf Voronoi cell} in the group $E(\ell,\ell_1,\f)$ is $O(K(\ell)\log^2K(\ell))$. In total, there are $O(K(\ell))$ insertions in the sequence $X$, and the total number of deletions, clearly, is at most the number of insertions.
Hence, the total time spent for insertions and deletions is $O(K(\ell)\log K(\ell))$. Thus, the time
spent for the maintenance of the Voronoi diagram in a fixed group $E(\ell, \ell_1,\f)$ is
$O(K(\ell)\log^2 K(\ell))$. Next, we discuss the computation and maintenance of a data structure that combines the adjacency diagram and the Voronoi diagram.

\noindent {\bf Propagation Diagram:}
As discussed above, the propagation set $I(u)$ of an active node $u$ on $\ell$ is  described completely
by the set of nodes on $\ell_1$ satisfying conditions {\bf C1}, {\bf C2}, and {\bf C3}. We denote the set of nodes
on $\ell_1$ satisfying  {\bf C1} and {\bf C2} with respect to $u$ by $I'(u)$. Slightly abusing
our terminology, we refer to this set again as propagation set of $u$. Similarly, we refer to the set
consisting of the sets $I'(u)$ for all currently active nodes as propagation diagram and denote it
by $\I'(u_1,\dots ,u_k)$, where, as above, $u_1,\dots ,u_k$ are the currently active nodes on $\ell_1$.

The difference between the originally defined propagation set $I(u)$ and the set $I'(u)$ is that the
elements of $I(u)$ are the edges joining $u$ to the nodes on $\ell_1$ satisfying  {\bf C1}, {\bf C2},
and {\bf C3}, whereas the elements of $I'(u)$ are the nodes on $\ell_1$ that satisfy {\bf C1} and {\bf C2},
but not necessarily {\bf C3}. Indeed, the set $I'(u)$ is closely related to  $I(u)$ and
when combined with the list $S^a(\ell_1)$ describes it completely. Based on this observation, we compute
and maintain the propagation diagram $\I'(u_1,\dots ,u_k)$ instead of the originally defined diagram.

We describe the sets $I'(u_i)$ by specifying the maximal Steiner intervals they cover. We implement these sets as ordered lists of pairs of pointers to the end-points of these intervals in the underlying list $V(\ell_1)$. The propagation sets of different active nodes do not intersect, and hence, the end-points of the maximal Steiner intervals of the propagation sets $I'(u_1), \dots, I'(u_k)$ form a sequence, $\I'=\{A_1\leq y_1 \leq z_1\leq \dots \leq y_{m_1}\leq z_{m_1} \leq B_1\}$, where $\ell_1=(A_1,B_1)$. The points $y_j$ and $z_j$, for $j=1,\dots,m_1$, are Steiner points (nodes) on $\ell_1$. Any of the Steiner intervals $(y_j,z_j)$ is a maximal Steiner interval covered by one of the sets $I'(u_1),\dots,I'(u_k)$, whereas the Steiner points inside the intervals $(z_j,y_{j+1})$ do not belong to any of the sets $I'(u_i)$. Clearly, the sequence $\I'$ plus the assignment of the intervals $(y_j,z_j)$ to the sets $\I'(u_i)$ covering them determine the diagram $\I'(u_1,\dots,u_k)$. We implement sequence $\I'$ as an ordered list of pointers to the underlying list $V(\ell_1)$. In addition, we associate with it a binary search tree based on the position of the Steiner points on the segment $\ell_1$. The diagram $\I'(u_1,\dots,u_k)$ is maintained in Step 3.1 and details are as follows.

Let, as above, $v$ be the node extracted by the extract-min operation in Step 1 in the current iteration of the algorithm. We assume that the diagram $\I'(u_1,\dots,u_k)$ is known -- i.e., we know the sequence $\I'$ as well as the assignment of the intervals $(y_j,z_j)$ to the propagation sets $I'(u_i)$. Next, we describe the update of $\I'$ and the assignment information specifying  $\I'(u_1,\dots,u_k,v)$.
By  definition,  $I'(v)$ consists of the nodes on $\ell_1$ that lie in the Voronoi cell ${\cal V}(v)$ and belong to the adjacency set $A(v,\ell_1)$. By Lemma \ref{cell}, ${\cal V}(v)$ is either empty or a single interval, which we have denoted by $(x^-,x^+)$. We denote by $(v^-,v^+)$, the largest Steiner interval inside the interval $(x^-,x^+)$. The interval $(v^-,v^+)$ is easily found using binary search in $O(\log K(\ell_1))$ time, where as above $K(\ell_1)$ denotes the number of  Steiner points on $\ell_1$.
On the other hand (see Lemma \ref{adj-int}),  the adjacency set $A(v,\ell_1)$ consists of the nodes lying inside constant number (at most seven) of Steiner intervals, which were computed and stored as the list $\bar{A}(v,\ell_1)$. Hence, the maximal Steiner intervals specifying the propagation set $I'(v)$ can be obtained
as the intersection of  intervals in $\bar{A}(v,\ell_1)$ with $(v^-,v^+)$. This is  done in
constant time by identifying the position of the points $v^-$ and $v^+$ with respect to the elements of the list
$\bar{A}(v,\ell_1)$. Clearly, the so-computed maximal Steiner intervals covered by $I'(v)$ are at most seven.
We update the sequence $\I'$ by inserting each of the maximal Steiner intervals covered by $I'(v)$ in the
same way as we inserted the interval $(x^-,x^+)$ into the sequence $X$ describing the Voronoi diagram.
More precisely, let $(y,z)$ be any of the maximal Steiner intervals covered by $\I'(v)$. We insert the points $y$ and $z$ at their positions in the ordered sequence $\I'$, and then we delete the points of $\I'$ between $y$ and $z$. If the interval containing $y$ is $(y_j,z_j)$, we set new $z_j$ to be the  Steiner point preceding $y$ on $\ell_1$. Similarly, if the interval containing $z$ is $(y_j,z_j)$, then we set  $y_j$ to be the Steiner point following $z$ on $\ell_1$.

At each iteration of the algorithm, the endpoints of at most seven intervals are inserted into the sequence $\I'$.
Hence, the size of the sequence $\I'$ is bounded by $14K(\ell)$ and insertions in $\I'$ are implemented in $O(\log K(\ell))$ time. Deletions are implemented in $O(1)$ time. The total number of insertions is $O(K(\ell))$ and the total number of deletions is at most the number of insertions. Therefore, the total time spent for the maintenance of  $\I'$ and the propagation diagram is $O(K(\ell)(\log K(\ell)+K(\ell_1)))$.

Finally, we summarize our discussion on the implementation of Step 3.1. The computations and times related to a fixed
triple $(\ell,\ell_1,f)$ are as follows. First, in a preprocessing step  the lists $\bar{A}(u,\ell_1)$, for all nodes $u$ on $\ell$, are computed in $O(K(\ell)\log K(\ell_1))$ time (Lemma \ref{adj-int}).
Times spent for the maintenance of the lists $S(\ell)$ and $S^a(\ell_1)$ are $O(K(\ell)\log K(\ell)$ and
$O(K(\ell_1)\log K(\ell_1)$, respectively. The time spent for maintenance of the Voronoi diagram for the
active nodes on $\ell$ requires $O(K(\ell)\log^2K(\ell))$ time. The time for the maintenance of the Propagation Diagram is $O(K(\ell)(\log K(\ell) +\log K(\ell_1)))$.
Therefore, the total time for the implementation of Step 3.1 is
\begin{eqnarray*}
&&\sum_{(\ell,\ell_1, f)}\left(O(K(\ell)(\log^2K(\ell)+\log K(\ell_1))+O(K(\ell_1)\log K(\ell_1))\right)\\
&&\leq
O(\frac{1}{\sqrt{\eps}}\log\frac{1}{\eps})\left(\sum_{\ell}K(\ell)(\log^2K(\ell)+\log K(\ell_1))+
\sum_{\ell_1}K(\ell_1)\log K(\ell_1)\right) \\
&&\leq
O(\frac{1}{\sqrt{\eps}}\log^3\frac{1}{\eps})
\left(\sum_{\ell}K(\ell)+\sum_{\ell_1}K(\ell_1)\right)=
O(\frac{|V_\eps|}{\sqrt{\eps}}\log^3\frac{1}{\eps}),
\end{eqnarray*}
where we have used Lemma \ref{number-of-points} to estimate that the number of triples
$(\ell,\ell_1,\f)$ with a fixed first or second element is  $O(\frac{1}{\sqrt{\eps}}\log\frac{1}{\eps})$,
and that $\log K(\ell)$ and $\log K(\ell_1)$ are $O(\log\frac{1}{\eps})$.
\begin{lemma} \label{Step3.1}
The total time spent by the algorithm implementing Step 3.1 is $O(\frac{|V_\eps|}{\sqrt{\eps}}\log^3\frac{1}{\eps})$.
\end{lemma}

\subsubsection{Computation and updates of set of representatives}\label{step_3.x}
Next, we concentrate on the computation of representatives in Steps 3.2, 3.3 and 3.4.
The set of representatives $\boldrho(v;\ell,\ell_1,e)$  of an active  node $v$ on $\ell$ in
a group $E(\ell,\ell_1,\f)$ contains one representative for each interval $(y_j,z_j)$ in the propagation set $I(v)$. Recall that $I(v)$ consists of a set of intervals $(y_j,z_j)$ stored in the sequence $\I'$, characterizing the propagation diagram of the currently active nodes on $\ell$. The representative in  $\boldrho(v;\ell,\ell_1,e)$, corresponding to $(y_j,z_j)\in I(v)$, is the target of the minimum cost edge from $v$ to a node in $S^a\cap (y_j,z_j)$. By Lemma \ref{wdf-convex}, the function $c(v,x)$ is convex and thus in any interval it has a single minimum. Let $x^*(v)$ be the point on $\ell_1$, where $c(v,x)$ achieves its minimum. To efficiently compute the representatives, we compute in a preprocessing step the points $x^*(v)$, for all nodes on $\ell$.
From the definition of the function $c(v,x)$ and Snell's law, it follows that $x^*(v)$ is the point on $\ell_1$ that is closest to $v$. So, each of $x^*(v)$ can be computed in constant time, which leads to  $O(K(\ell))$  preprocessing time  for the group $E(\ell,\ell_1,\f)$, where $K(\ell)$ is the number of  nodes on $\ell$. Thus, the total time for preprocessing in all groups is $O(\frac{|V_\eps|}{\sqrt{\eps}}\log \frac{1}{\eps})$.

We have associated two data structures to the set of nodes in $S^a$ that lie on a fixed Steiner segment $\ell_1$. First, we maintain them in a doubly-linked list  and second, we maintain them in a binary-search tree,
with respect to their position on $\ell_1$. We show that finding a representative $\rho(v)\in\boldrho(v;\ell,\ell_1,e)$ takes $O(\log\frac{1}{\eps})$ time. There are three situations, where we need to compute or update $\rho(v)$:
\begin{enumerate}
\item New representatives $\rho(v)$ are computed  when $v$ becomes
active and  its propagation set is non-empty. We need to compute one new representative for
each maximal Steiner interval $(y,z)$ in the propagation set $I(v)$. Recall that there are
at most seven such intervals and they were computed and stored in the sequence $\I'$.

To compute $\rho(v)$ in the interval $(y,z)$, we determine
the leftmost and rightmost nodes from $S^a$ inside the interval $(y,z)$.
This is done  by finding the position of the points $y$ and
$z$ in the sequence of nodes currently in $S^a$. Let the
leftmost and the rightmost  nodes  from $S^a$ in $(y,z)$ be
$y^a$ and $z^a$, respectively. Then, we determine
the position of the  point $x^*(v)$ with respect to $y^a$ and $z^a$.

If it is to the left of $y^a$, then $\rho(v)=y^a$. If it is to the right of $z^a$, then $\rho(v)=z^a$.
If  $x^*(v)$ is inside $(y^a,z^a)$, we determine the two nodes in $S^a$ immediately to the left and to the right of $x^*(v)$, and $\rho(v)$ is one of these two nodes. Using the binary-search tree on $S^a$, the nodes $y^a$ and $z^a$ and eventually the nodes neighboring $x^*(v)$ are determined in
$O(\log\frac{1}{\eps})$ time.
\item When some representative $\rho(v)$ is removed from $S^a$,
a new representative for $v$ is one of the neighbors of $\rho(v)$
in the doubly-linked list $S^a$ that lie in the same interval $(y,z)$ as
$\rho(v)$. This is done in $O(1)$ time.
\item When some interval of the propagation set $I(v)$ shrinks and the
current representative $\rho(v)$ is no longer inside this interval, then $\rho(v)$
is updated as follows. Let, as above, $y^a$ and $z^a$ be the
leftmost and the rightmost nodes from $S^a$, respectively, in the updated
interval. Then, if $\rho(v)$ lies to the left of $y^a$, we set
$\rho(v)=y^a$. If $\rho(v)$ is to the right of $z^a$, we set
$\rho(v)=z^a$. As above, determination of the nodes $y^a$ and $z^a$ is done
in $O(\log\frac{1}{\eps})$ time.
\end{enumerate}
To complete our analysis, we need to estimate the total number of representatives which are computed by our algorithm. Each pair (representative, predecessor) relates  to the edge joining them. Since such a pair can be
computed at most once by the algorithm, the total number of representatives related to nodes that are vertices of $\D$ is bounded by the total number of edges incident to these nodes, which is $O(|V_\eps|)$. It remains to estimate the  number of representatives which are related to nodes that are Steiner points.
Consider an iteration for a node $v$ that is a Steiner point. There are $O(\frac{1}{\sqrt{\eps}}\log\frac{1}{\eps})$ triples in ${\cal T}(v)$, and at most nine new representatives are computed in Step 3.2. For each predecessor in $R^{-1}(v)$ that is a Steiner point, a single representative is computed. The number of predecessors
$|R^{-1}(v)|$ is $O(\frac{1}{\sqrt{\eps}}\log\frac{1}{\eps})$. Hence, in a single iteration, $O(\frac{1}{\sqrt{\eps}}\log\frac{1}{\eps})$ representatives related to Steiner points are computed.
Since the number of iterations is $O(|V_\eps|)$ and the computation of a single representatives takes $O(\log\frac{1}{\eps})$ time, we obtain that the total time for the execution of Steps 3.2 and 3.3 is
$O(\frac{|V_\eps|}{\sqrt{\eps}}\log^2\frac{1}{\eps})$.

Finally, the number of  priority queue operations executed in Step
3.4  is bounded by the number of computed representatives. Thus, the total time for Step 3.4 is
$O(\frac{|V_\eps|}{\sqrt{\eps}}\log\frac{n}{\eps}\log \frac{1}{\eps})$.
\subsubsection{Complexity of the algorithm and the main result}\label{main_res}
Here, we summarize our discussion from the previous three subsections and state our main result. Step 1 of our algorithm takes $O(|V_\eps|\log\frac{n}{\eps})$ time. Step 2 requires $O(|V_\eps|)$ time. By Lemma \ref{Step3.1},
Step 3.1 takes in total $O(\frac{|V_\eps|}{\sqrt{\eps}}\log^3\frac{1}{\eps})$ time. The total time for
implementation of Steps 3.2 and 3.3 is $O(\frac{|V_\eps|}{\sqrt{\eps}}\log^2\frac{1}{\eps})$ and the total
time for Step 3.4 is $O(\frac{|V_\eps|}{\sqrt{\eps}}\log\frac{n}{\eps}\log \frac{1}{\eps})$.
By Lemma \ref{number-of-points}, we have that $|V_\eps|=O(\frac{n}{\eps^2}\log\frac{1}{\eps})$.
We have thus established the following:
\begin{theorem}\label{AlgCompl}
The SSSP problem in the approximation graph $G_\eps$ can be solved in\\
$O(\frac{n}{\eps^{2.5}}\log \frac{n}{\eps}\log^3\frac{1}{\eps})$ time.
\end{theorem}

Consider the polyhedral domain $\D$. Starting from a vertex $v_0$ of $\D$, our algorithm solves the SSSP problem in the corresponding graph $G_\eps$ and constructs a shortest paths tree rooted at $v_0$. According to Theorem \ref{approx}, the computed distances from $v_0$ to all other vertices of $\D$ (and to all Steiner points)
are within a factor of $1+\eps$ of the cost of the corresponding shortest paths. Using the definition of the
edges of $G_\eps$, an approximate shortest path can be output by simply replacing the edges in the discrete path with the corresponding local shortest paths used to define their costs. This can be done  in time proportional to the number of segments in this path, because computation of the local shortest paths takes $O(1)$ time. The approximate shortest paths tree rooted at $v_0$ and containing all Steiner points and vertices of $\D$ can be
output in $O(|V_\eps|)$ time.  Thus, the algorithm we described solves the WSP3D problem and the following theorem states the result.
\begin{theorem} \label{Main}
Let $\D$ be a weighted polyhedral domain consisting of $n$ tetrahedra and $\eps\in(0,1)$. The weighted shortest path problem in three dimensions (WSP3D), requiring the computation of approximate shortest paths from a source vertex  to all other vertices of $\D$, can be solved in $O(\frac{n}{\eps^{2.5}}\log\frac{n}{\eps}\log^3\frac{1}{\eps})$ time.
\end{theorem}

\section{Conclusions}\label{conclusions}
This paper generalizes the weighted region problem, originally studied in 1991 by Mitchell and Papadimitriou \cite{Mit91} for the planar setting, to 3-d weighted domains. We present the first polynomial time approximation scheme for the WSP3D problem. 
The complexity of our algorithm is independent of the weights, but depends upon the geometric features of the given tetrahedra as stated in Lemma \ref{number-of-points}. 

There are some fairly standard techniques which can be employed here to remove the dependence on geometry (cf., \cite{ASY09}), provided that there is an estimate known on the maximum number of segments (i.e., the combinatorial complexity) in weighted  shortest paths in three dimensions. It can be shown that the combinatorial complexity of weighted shortest paths in planar case is $\Theta(n^2)$ \cite{Mit91}. We conjecture that the same bound holds in three dimensions, but the proof techniques in \cite{Mit91} do not seem to apply here, since they use planarity.
If the combinatorial complexity of these paths in three dimensions is a polynomial in $n$, then we can remove the dependence on the geometry by increasing the run time by a polynomial factor in $n$. We do not recommend this approach, since the increase in the running time will be significant. Already, in the planar case (and in terrains), in an experimental study \cite{LMS01}, it was shown that a constant number of Steiner points suffice to produce high-quality paths.  We believe that the same holds here and this merits further investigation.

This paper also investigated additive Voronoi diagrams in heterogeneous media. We studied a fairly simple scenario and already the analysis of that was very technical and cumbersome. It is desirable to find simpler and more elegant ways to understand the combinatorics of these diagrams. Nevertheless, we believe that the discretization scheme and the algorithms presented here can be used successfully for efficient computation of approximate Voronoi diagrams in heterogeneous media.

Our algorithm does not require any complex data structures or primitives and as such should be implementable and even practical. Its structure allows Steiner points to be generated ``on the fly'' as the shortest path wavefront propagates though the weighted domain. This feature allows the design of more compact and adaptive implementation schemes that can be of high practical value.

One of the classical problems that motivated this study is the unweighted version of this problem, namely the ESP3D problem. There, we need to find a shortest path between a source and a target point, lying completely in the free space, avoiding three-dimensional polyhedral obstacles. We can use our techniques to solve this problem, though this will require triangulating (i.e., tetrahedralization) the free space. As outlined above, the complexity of our algorithm depends upon the geometry of these tetrahedra; so it is natural to ask whether the free space can be partitioned into nice tetrahedra? Unfortunately, there is no simple answer to this question which has been an important topic of study in computational and combinatorial geometry for several decades. Nevertheless, our algorithm provides a much simpler and so far the fastest method for solving the ESP3D problem, provided the free space is already partitioned into non-degenerate tetrahedra.

Combining the techniques of answering weighted shortest path queries on polyhedral surfaces \cite{ADGMNS10} and the existence of nice separators for well-shaped meshes \cite{MTTV93}, we believe that our construction presented in this paper can be used for answering (approximate) weighted shortest path queries in 3-d.

\bibliography{previous-biblio}
\bibliographystyle{plain}


\newpage
\appendix
%
\section{Appendix}
\subsection{Proof of Proposition \ref{prop:p2}}  \label{appendix1}
\noindent{\bf Proposition \ref{prop:p2}:} {\em The second mixed derivative of the function
$x=x(y,\alpha)$ is negative, i.e. $x_{y\alpha}<0$.}\\[1ex]
{\bf Proof:} First, we consider the case
where $w^-=w^+$. In this case, the function $x(y,\alpha)$ can be
represented and differentiated explicitly. Recall, that the path
$\bar{\pi}(v,\x)$ in this case either consists of a single segment or
is a three segment path as shown in Figure \ref{snells1} (b). So, in
the case where the path consists of a single segment, we have
$x(y,\alpha)=y \cot\alpha$. In the case where the path consists of
three segments, $x(y,\alpha)=y \cos\alpha/\sqrt{\k2-\c2}$, where
$\kappa=w/w^-$. The mixed derivatives $x_{y\alpha}$ of these two
functions are $-1/\sin^2\alpha$ and
$-\k2\sin\alpha/(\k2-\c2)^\frac{3}{2}$, respectively, and both are
readily negative.

Next, we  consider the case where $w^-\not=w^+$. We introduce some
additional notation as necessary for our presentation below
(Figure \ref{fig:wdf}). We denote the coordinates of the bending point
$a$ of the path $\bar{\pi}(v,\x)$ by $a=(x^-,y^+)$. Furthermore,
we set $x^+=x-x^-$ and $y^-=y-y^+$. Clearly, $x^-$, $x^+$,
$y^-$, and $y^+$ can be viewed as functions of the independent
variables $y$ and $\alpha$. We have $ x =
x^-(y^-(y,\alpha),\alpha) + x^+(y^+(y,\alpha),\alpha)$ and thereby
\begin{equation}\label{110}
x_y=x^-_{y^-}y^-_y + x^+_{y^+}y^+_y.
\end{equation}
We differentiate(\ref{110}) with respect to $\alpha$ and obtain
$x_{y\alpha} = A + A_1 + A_2 $, where
$$
A = x^-_{y^-}y^-_{y\alpha} + x^+_{y^+}y^+_{y\alpha},\ A_1=
x^-_{y^-y^-}y^-_{\alpha}y^-_y +x^+_{y^+y^+}y^+_{\alpha}y^+_y,\
{\rm and} \  A_2 = x^-_{y^-\alpha}y^-_y  + x^+_{y^+\alpha}y^+_y.
$$
We complete the proof by showing that the terms $A$, $A_1$, and $A_2$,
are negative.

We begin with the term $A = x^-_{y^-}y^-_{y\alpha} + x^+_{y^+}y^+_{y\alpha}$.
From the identity $y = y^- + y^+$, it follows that $y^-_{y\alpha} +
y^+_{y\alpha} = 0$ and hence, $A=(x^-_{y^-} -
x^+_{y^+})y^-_{y\alpha}$. From our notation, Snell's law, and
the relation $ \cos\alpha=\cos\theta\sin\varphi$, we derive the
following:
\begin{eqnarray} \label{120}
&&x^+ = \frac{z^+ \cos \alpha}{\sqrt{\k2 - \s2}}, \qquad y^+ =
\frac{z^+\sqrt{\s2 - \c2}}{\sqrt{\k2 - \s2}},\nonumber\\
&&\hspace*{-2cm}{\rm and}\\
&& x^- = \frac{z^-\cos \alpha}{\sqrt{1 - \s2}}, \qquad y^- =
\frac{z^-\sqrt{\s2 - \c2}}{\sqrt{1 - \s2}}.\nonumber
\end{eqnarray}
First, we compute $x^-_{y^-}$ and $x^+_{y^+}$. We denote $\s2$ by
$\sigma$, differentiate $x^+$ and $y^+$ with respect to $\sigma$
and obtain $x^+_{y^+}$ as the ratio $x^+_\sigma /y^+_\sigma$. We
differentiate expressions (\ref{120}) with respect to $\sigma$ and
obtain
\begin{equation} \label{123}\{x\in {\cal H}: c(v,x)+C < c(v',x)\}
x^+_\sigma = \frac{z^+\cos\alpha}{2(\k2 -
\sigma)^{\frac{3}{2}}},\qquad
y^+_\sigma = \frac{z^+(\k2 - \c2)}{2\sqrt{\sigma
-\c2}(\k2-\sigma)^{\frac{3}{2}}}
\end{equation}
and hence
\begin{equation}
x^+_{y^+} = x^+_\sigma /y^+_\sigma = \frac{\cos\alpha\sqrt{\s2
-\c2}}{\k2 - \c2}. \label{125}
\end{equation}
Similarly,
\begin{eqnarray}\label{128}
x^-_\sigma = \frac{z^-\cos\alpha}{2(1 -
\sigma)^{\frac{3}{2}}},\qquad
y^-_\sigma= \frac{z^-(1 -
\c2)}{2\sqrt{\sigma-\c2}(1-\sigma)^{\frac{3}{2}}},\qquad {\rm and}\nonumber\\
 x^-_{y^-} = x^-_\sigma/y^-_\sigma = \frac{\cos\alpha\sqrt{\s2 -\c2}}{ 1 -
\c2}.
\end{eqnarray}
So, for the difference $x^-_{y^-} - x^+_{y^+}$ we have
\begin{eqnarray}\label{129}
x^-_{y^-} - x^+_{y^+}= \cos\alpha\sqrt{\s2-\c2} (\frac{1}{1-\c2} -
\frac{1}{\k2 - \c2}) \nonumber\\
= (\k2 -1) \frac{\cos\alpha
\sqrt{\s2-\c2}}{(1-\c2)(\k2-\c2)} .
\end{eqnarray}
The latter shows that
\begin{equation} \label{130}
{\rm sign}(x^-_{y^-} - x^+_{y^+}) = {\rm sign} (\k2-1).
\end{equation}
To prove the negativity of  $A=(x^-_{y^-} -
x^+_{y^+})y^-_{y\alpha}$, we show that ${\rm
sign}(y^-_{y\alpha})={\rm sign} (1- \k2)$. We have
%
%
\begin{eqnarray*}
y^-_y=y^-_\sigma /y_\sigma =y^-_\sigma /(y^-_\sigma + y^+_\sigma)=
\frac{1}{1+y^+_\sigma/y^-_\sigma}\quad {\rm and\ thus}\quad
y^-_{y\alpha}=-\frac{(y^+_\sigma /
y^-_\sigma)_\alpha}{(1+y^+_\sigma /y^-_\sigma)^2}.
\end{eqnarray*}
Hence, it is sufficient to show that
\begin{equation} \label{140}
{\rm sign}((y^+_\sigma /y^-_\sigma)_\alpha)={\rm sign}(\k2 - 1) .
\end{equation}
So, we continue by establishing the sign of the derivative
$(y^+_\sigma /y^-_\sigma)_\alpha$. We use (\ref{123}) and
(\ref{128}) and compute the ratio $y^+_\sigma /y^-_\sigma$ as
follows
\begin{eqnarray}
y^+_\sigma /y^-_\sigma = \frac{z^+(\k2-\c2)(1-\sigma)^{3/2}}{z^-(1-\c2)(\k2-\sigma)^{3/2}}
= (z^+/z^-)BC,\quad {\rm where}\nonumber \\
B=\frac{\k2-\c2}{1-\c2} \qquad {\rm and} \qquad
C=\left(\frac{1-\sigma}{\k2-\sigma}\right)^{3/2}=\left(\frac{1-\s2}{\k2-\s2}\right)^{3/2}.
\label{150}
\end{eqnarray}
Then, we compute the derivatives $B_\alpha$ and $C_\alpha$ using
the expressions (\ref{150})
\begin{eqnarray*}
B_\alpha=\frac{\sin 2\alpha(1-\c2) - \sin
2\alpha(\k2-\c2)}{(1-\c2)^2} = \frac{(1-\k2)\sin
2\alpha}{(1-\c2)^2} \qquad {\rm and}\\
C_\alpha = \frac{3}{2}\left(\frac{1-\s2}{\k2-\s2}\right)^{1/2}
\frac{(-\sin 2\varphi)(\k2-\s2)-(-\sin
2\varphi)(1-\s2)}{(\k2-\s2)^2}\varphi_\alpha  =\\
\frac{3}{2}\left(\frac{1-\s2}{\k2-\s2}\right)^{1/2} \frac{\sin
2\varphi(1-\k2)}{(\k2-\c2)^2}\varphi_\alpha
\end{eqnarray*}
and obtain
\begin{eqnarray}\label{160}
\lefteqn{(z^-/z^+)(y^+_\sigma /y^-_\sigma)_\alpha = B_\alpha C + BC_\alpha} \\
&&=\left(\frac{1-\s2}{\k2-\s2}\right)^{\frac{3}{2}} \frac{(1-\k2)\sin
2\alpha}{(1-\c2)^2} +
\frac{3}{2}\left(\frac{1-\s2}{\k2-\s2}\right)^{\frac{1}{2}}\frac{(\k2-\c2)\sin2\varphi(1-\k2)\varphi_\alpha}
{(1-\c2)(\k2-\s2)^2} \nonumber \\
&&=(1-\k2)\frac{\sqrt{1-\s2}}{(\k2-\s2)^{3/2}(1-\c2)} D, \nonumber \\
&& {\rm where} \quad D = \frac{\sin2\alpha(1-\s2)}{1-\c2} +
\frac{3\sin2\varphi(\k2-\c2)}{2(\k2-\s2)}\varphi_\alpha .\nonumber
\end{eqnarray}
Omitting the positive multiplicative terms in (\ref{160}), we
derive that ${\rm sign} (y^+_\sigma /y^-_\sigma)_\alpha = {\rm
sign}((1-\k2)D)$ and continue with the evaluation of ${\rm sign}(D)$.
We compute the derivative $\varphi_\alpha$ using the identity
$y=y^-+y^+$, which implies $0=y^-_\alpha + y^+_\alpha$. We
differentiate expressions from (\ref{120}) with respect to
$\alpha$ and obtain
\begin{eqnarray}\label{165}
y^-_\alpha = (z^-/2)\frac{(1-\s2)\sin2\alpha +
\sin2\varphi(1-\c2)\varphi_\alpha}
{(\s2-\c2)^{1/2}(1-\s2)^{3/2}},\nonumber \\ \\
y^+_\alpha = (z^+/2)\frac{(\k2-\s2)\sin2\alpha +
\sin2\varphi(\k2-\c2)\varphi_\alpha}
{(\s2-\c2)^{1/2}(\k2-\s2)^{3/2}},\nonumber
\end{eqnarray}
From these two, we obtain
\begin{equation} \label{170}
\varphi_\alpha = - \frac{I\sin2\alpha(1-\s2)(\k2-\s2)}
{J\sin2\varphi}, \quad{\rm where}
\end{equation}
\begin{eqnarray}\label{175}
&& I=z^-(\k2-\s2)^{1/2} + z^+(1-\s2)^{1/2}\qquad {\rm and} \\
&& J=z^-(1-\c2)(\k2-\s2)^{3/2} + z^+(\k2-\c2)(1-\s2)^{3/2}. \nonumber
\end{eqnarray}
Next, we substitute $\varphi_\alpha$ from (\ref{170}) in the
expression $D$ given in (\ref{160}) and obtain
\begin{eqnarray*}
D = \sin\alpha\cos\alpha(1-\s2)\frac{2J- 3I(\k2-\c2)(1-\c2)}{J(1-\c2)}.
\end{eqnarray*}
The term $\sin\alpha\cos\alpha(1-\s2)$ and the denominator in this
expression are positive and by expanding the numerator we have
\begin{eqnarray}\label{180}
&&{\rm sign}(D) = \sign[2J- 3I(\k2-\c2)(1-\c2)]\nonumber\\
&&={\rm sign}\left[ 2z^-(1-\c2)(\k2-\s2)^{3/2} +
2z^+(\k2-\c2)(1-\s2)^{3/2}\right. \nonumber \\
&&\left. - 3z^-(1-\c2)(\k2-\c2)(\k2-\s2)^{1/2}
-3z^+(1-\c2)(\k2-\c2)(1-\s2)^{1/2} \right]\nonumber \\
&&={\rm sign}\left[z^-(1-\c2)(\k2-\s2)^{1/2} D^- +
z^+(\k2-\c2)(1-\s2)^{1/2} D^+ \right] ,\nonumber\\
&&{\rm where} \quad D^- = 3\c2-2\s2-\k2 \quad {\rm and} \quad D^+ = 3\c2-2\s2-1 .
\end{eqnarray}
Now, we observe that the terms multiplied by $D^+$ and by $D^-$ are positive and
show that $D^-$ and $D^+$ are negative. We use
$\cos\alpha=\cos\theta\sin\varphi$ and
$\cos\alpha_\kappa=\cos\theta\sin\varphi_\kappa$, where $\sin\varphi=\kappa\sin\varphi_\kappa$ and obtain
\begin{eqnarray*}
D^+ = 3\c2-2\s2-1 = 2\c2-2\s2-\sin^2\alpha \\ = 2\cos^2\theta\s2 -
2\s2 - \sin^2\alpha = -2\s2\sin^2\theta - \sin^2\alpha < 0\\
{\rm and} \quad
D^- = 3\c2-2\s2-\k2 = \k2(3\cos^2\alpha_\kappa -
2\sin^2\varphi_\kappa - 1) = \\
\k2( -2\sin^2\varphi_\kappa\sin^2\theta - \sin^2\alpha_\kappa) < 0
.
\end{eqnarray*}
From  (\ref{180}) and $D^+, D^- <O$, we get ${\rm sign}(D)=-1$.
From (\ref{160}) it follows that ${\rm sign} (y^+_u/y^-_u)_\alpha
= -{\rm sign}(1-\k2)$ and thus ${\rm sign}(y^-_{y\alpha})={\rm
sign}(1-\k2)$. The latter implies that $A<0$.

\medskip

Next, we consider the term $A_1$. From the identity $y^-_\alpha +
y^+_\alpha=y_\alpha=0$, we get $A_1=y^-_\alpha( x^-_{y^-y^-}y^-_y -
x^+_{y^+y^+}y^+_y)$. To evaluate the sign of $y^-_\alpha$,
we substitute $\varphi_\alpha$ from (\ref{170}) in the expression
(\ref{165}) and by omitting positive multiplicative term, we obtain
\begin{eqnarray}\label{190}
&&\sign(y^-_\alpha)= \sign[(J - (1-\c2)(\k2-\s2)I)\cos\alpha]\\
&&=\sign\{z^+(1-\s2)^{1/2}[(\k2-\c2)(1-\s2)-(1-\c2)(\k2-\s2)]\cos\alpha\} \nonumber\\
&&=\sign[(1-\k2)\cos\alpha].\nonumber
\end{eqnarray}
Next, we evaluate the sign of the difference $x^-_{y^-y^-}y^-_y -
x^+_{y^+y^+}y^+_y$. We compute
$x^+_{y^+y^+}$ as follows
\begin{eqnarray*}
x^+_{y^+y^+}=(x^+_{y^+})_\sigma /y^+_\sigma =\frac{\cos\alpha}{2(\k2-\c2)\sqrt{\sigma -\c2}}\
/\ \frac{z^+(\k2 - \c2)}{2\sqrt{\sigma -\c2}(\k2-\sigma )^{\frac{3}{2}}},
\end{eqnarray*}
where we have differentiated (\ref{125}) with respect to $\sigma =\s2$ and
used (\ref{123}). We compute $x^-_{y^-y^-}$ in the same way and
obtain
\begin{equation} \label{200}
x^+_{y^+y^+} = \frac{\cos\alpha
(\k2-\s2)^{\frac{3}{2}}}{z^+(\k2-\c2)^2}\quad {\rm and} \quad
 x^-_{y^-y^-}=\frac{\cos\alpha
(1-\s2)^{\frac{3}{2}}}{z^-(1-\c2)^2}.
\end{equation}
Furthermore, we have $y^+_y=y^+_\sigma /y_\sigma =y^+_\sigma /(y^+_\sigma +y^-_\sigma )$ and
$y^-_\sigma =y^+_\sigma /(y^+_\sigma +y^-_\sigma )$. So, we compute $y^+_y$ and $y^-_y$
using (\ref{123}) as follows
\begin{eqnarray}\label{210}
y^+_y=\frac{z^+(\k2-\c2)(1-\s2)^{\frac{3}{2}}}{J}\quad {\rm
and}\quad
y^-_y=\frac{z^-(1-\c2)(\k2-\s2)^{\frac{3}{2}}}{J},
\end{eqnarray}
where $J$ was defined in (\ref{175}).
Using (\ref{200}) and (\ref{210}), we determine
$$
\sign(x^-_{y^-y^-}y^-_y -x^+_{y^+y^+}y^+_y)=
\sign[(1/(1-\c2)-1/(\k2-\c2))\cos\alpha]=\sign[(\k2-1)\cos\alpha].
$$
The latter and (\ref{190}) imply $A_1<0$.

\medskip

Finally, we show that $A_2= x^-_{y^-\alpha}y^-_y + x^+_{y^+\alpha}y^+_y$
is negative too. We first compute the derivative
$x^+_{y^+\alpha}$ by differentiating the expression (\ref{125}) with respect to $\alpha$.
We have
\begin{equation}\label{220}
 x^+_{y^+\alpha}=\frac{P_\kappa\sin\alpha +
\cos\alpha\sin\varphi\cos\varphi(\k2-\c2)\varphi_\alpha}
{(\s2-\c2)^{\frac{1}{2}}(\k2-\c2)^2},
\end{equation}
where $P_\kappa =2\c2(\k2-\s2) - \s2(\k2-\c2)$.  We
substitute $\varphi_\alpha$ using the expression (\ref{170}) and multiply by $y^+_y$
using the expression (\ref{210}).  After simplification, we obtain
\begin{equation}\label{230}
x^+_{y^+\alpha}y^+_y = z^+\sin\alpha(1-\s2)^{3/2}\frac
{JP_\kappa - \c2(\k2-\c2)(1-\s2)(\k2-\s2)I}
{J^2(\s2-\c2)^{\frac{1}{2}}(\k2-\c2)}.
\end{equation}
Analogously, we obtain the following
\begin{equation}\label{240}
x^-_{y^-\alpha}y^-_y = z^-\sin\alpha(\k2-\s2)^{3/2}\frac
{JP_1 - \c2(1-\c2)(1-\s2)(\k2-\s2)I}
{J^2(\s2-\c2)^{\frac{1}{2}}(1-\c2)},
\end{equation}
where $P_1 =2\c2(1-\s2) - \s2(1-\c2)$.
We sum (\ref{230}) and (\ref{240}), simplify and omit the positive multiplicative
terms obtaining
\begin{eqnarray}\label{250}
&&\sign(x^-_{y^-\alpha}y^-_y +x^+_{y^+\alpha}y^+_y)=\\
&&\sign[z^-(\k2-\s2)^{3/2}(\k2-\c2) Q_1 + z^+(1-\s2)^{3/2}(1-\c2) Q_\kappa], \nonumber \\
&&{\rm where}\nonumber \\
&& Q_1=JP_1-\c2(1-\c2)(1-\s2)(\k2-\s2)I\quad {\rm and} \nonumber\\
&& Q_\kappa= JP_\kappa - \c2(\k2-\c2)(1-\s2)(\k2-\s2)I\ . \nonumber
\end{eqnarray}
We denote the expression in the square brackets by $R$. Finally, evaluate
$R$ and show that it is negative. First, we evaluate and simplify $Q_\kappa$ and $Q_1$ .
We substitute the expressions $I$ and $J$ from (\ref{175})
in $Q_\kappa$  and group the terms containing $z^-$ and $z^+$. Then, we substitute
the expression for $P_\kappa$ from (\ref{220}) and by simplification we get
\begin{eqnarray*}
&&Q_\kappa=z^-(\k2-\s2)^{3/2}[(1-\c2)P_k - \c2(\k2-\c2)(1-\s2)] + \\
&&\mbox{}\hspace*{3cm} z^+(1-\s2)^{3/2}[(\k2-\c2)P_k - \c2(\k2-\c2)(\k2-\s2)]\\
&&\quad =z^-(\k2-\s2)^{3/2}(\c2-\s2)(\k2+\c2-2\k2\c2) + \\
&&\mbox{}\hspace*{3cm}z^+(1-\s2)^{3/2}(\c2-\s2)\k2(\k2-\c2).
\end{eqnarray*}
In the same way, we obtain the following representation for $Q_1$
\begin{eqnarray*}
&&Q_1=z^-(\k2-\s2)^{3/2}(\c2-\s2)(1-\c2) +\\
&&\mbox{}\hspace*{3cm}z^+(1-\s2)^{3/2}(\c2-\s2)(\k2 + \k2\c2 - 2\c2).
\end{eqnarray*}
Substitution of $Q_\kappa$ and $Q_1$ in (\ref{250}) produces an expression for $R$ of the form
\begin{eqnarray*}
&&R=(z^-)^2 R_1+ (z^+)^2R_2+z^-z^+R_3,\quad  {\rm where}\\
&&R_1=(\k2-\s2)^3(\k2-\c2)(1-\c2)(\c2-\s2)\\
&&R_2=(1-\s2)^3(1-\c2)\k2(\k2-\c2)(\c2-\s2)\\
&&R_3=(1-\s2)^{3/2}(\k2-\s2)^{3/2}(\c2-\s2)\ \times\\
&&\mbox{}\hspace*{1cm}[(\k2-\c2)(\k2+\k2\c2-2\c2)+(1-\c2)(\k2+\c2-2\k2\c2)]\\
&&\qquad=(1-\s2)^{3/2}(\k2-\s2)^{3/2}(\c2-\s2)\ \times\\
&&\mbox{}\hspace*{5cm}[(\k2-\c2)^2+\k2(1-\c2)^2+(1-\k2)^2\c2]
\end{eqnarray*}
From these expressions, it is clear that terms $R_1$, $R_2$, and
$R_3$ are negative, since $\c2-\s2<0$ and all other terms are positive.
Thus, $R$ and consequently $A_2$ are negative. The proposition is proved.\hfill $\Box$

\subsection{Upper bound on the constant $C_{ABP}(t)$}\label{appendix2}
Next, we show that
$$
C_{ABP}(t)\leq \frac{11|AB|}{r(e)\sin^2(\gamma/2)}
\log_2 \frac{4|AB|^2 h}{r(e)r(A)r(B)}.
$$
Recall that $\lambda=(1+\sqrt{\eps/8}\sin(\gamma/2))^{-1}$ and
$0<\eps\leq 1$. We use the following two inequalities, which are
easily derived from the
properties of logarithms:\\
For any $X\geq 1$
\begin{eqnarray}\label{A30}
\log_{\lambda^{-1}} X < \frac{3.44 \ln X}{\sqrt{\eps} \sin(\gamma/2)}, \qquad
 \ln\frac{X}{\eps}\leq \ln\frac{2}{\eps}\log_2 X
\end{eqnarray}
By our definition of the constant $C_{ABP}(t)$ and
(\ref{e80}) it follows that
\begin{eqnarray}\label{A40}
C_{ABP}(t)\leq
\frac{2\eps|AB|}{r(e)\lambda(1-\lambda)\log \frac{2}{\eps}}
\log_{\lambda^{-1}}\frac{|AB|}{\varepsilon \lambda \sqrt{r(A)r(B)}}
&+& \frac{\eps|AB|}{r(e)(1-\lambda)^2 \log \frac{2}{\eps}}
\\
+ \frac{2\eps^2}{\log \frac{2}{\eps}}\log_{\lambda^{-1}}\frac{h}{\eps r(e)}
&+&
\frac{4\eps^2}{\log \frac{2}{\eps}}\log_{\lambda^{-1}}\frac{|AB|}
{\varepsilon \lambda \sqrt{r(A)r(B)}}.\nonumber
\end{eqnarray}
We estimate the terms on the right-hand side of this
inequality using inequalities (\ref{A30}) above.
For the first one, we have
\begin{eqnarray}\label{A50}
\frac{2\eps|AB|}{ r(e)\lambda(1-\lambda)\log \frac{2}{\eps}}
\log_{\lambda^{-1}}\frac{|AB|}{\varepsilon \lambda
\sqrt{r(A)r(B)}}\leq
\frac{(2\sqrt{2}+1)^2\sqrt{\eps}|AB|}
{\sqrt{2}r(e)\sin(\gamma/2)\log \frac{2}{\eps}}
\log_{\lambda^{-1}}\frac{2|AB|}{\varepsilon
\sqrt{r(A)r(B)}} \nonumber \\
\leq \frac{3.44 (2\sqrt{2}+1)^2 |AB|}
{\sqrt{2}r(e)\sin^2(\gamma/2)\log\frac{2}{\eps}}
\ln\frac{2|AB|}{\varepsilon \sqrt{r(A)r(B)}}
<25\frac{|AB|}{ r(e)\sin^2(\gamma/2)} \log_2\frac{2|AB|}{\sqrt{r(A)r(B)}}.
\end{eqnarray}
For the second one, we have
\begin{eqnarray}\label{A60}
\frac{\eps |AB|}{ r(e) (1-\lambda)^2\log \frac{2}{\eps}}\leq
\frac{(2\sqrt{2}+1)^2 |AB|}{ r(e) \sin^2(\gamma/2) \log \frac{2}{\eps}}<
15\frac{|AB|}{ r(e) \sin^2(\gamma/2)\log \frac{2}{\eps}}.
\end{eqnarray}
The sum of the third and fourth terms is estimated by
\begin{eqnarray}\label{A70}
\frac{2\eps^2}{\log \frac{2}{\eps}}\left(
\log_{\lambda^{-1}}\frac{h}{\eps r(e)} +
2\log_{\lambda^{-1}}\frac{|AB|}{\varepsilon
\lambda \sqrt{r(A)r(B)}} \right) &\leq&
\frac{14\eps^{1.5}}{\sin(\gamma/2)\log \frac{2}{\eps}}
\left(\ln\sqrt{\frac{h}{\eps r(e)}} +
\ln\frac{2|AB|}{\varepsilon \sqrt{r(A)r(B)}}\right)\nonumber \\
\leq \frac{14 \eps^{1.5}\ln 2}{\sin(\gamma/2)}
\log_2\frac{2|AB|\sqrt{h}}{\sqrt{r(e)r(A)r(B)}} &<&
5\frac{\eps^{1.5}}{\sin(\gamma/2)}\log_2\frac{4|AB|^2h}{r(e)r(A)r(B)}.
\end{eqnarray}
We substitute (\ref{A50}), (\ref{A60}), and (\ref{A70}) in
(\ref{A40}), use $2r(e)\leq |AB|$, $r(e)\leq h$ and obtain
\begin{eqnarray*}
C_{ABP}(t)<23\frac{|AB|}{r(e)\sin^2(\gamma/2)}\log
\frac{4|AB|^2h}{r(e)r(A)r(B)}.
\end{eqnarray*}

\subsection{Proof of Lemma \ref{adj-int}}  \label{appendix3}
{\bf Lemma \ref{adj-int}}
{\em The number of the maximal intervals covered by $A(u,\l_1)$ is at most seven.
The corresponding list $\bar{A}(u,\ell_1)$ is computed in $O(\log K(\ell_1))$,
where $K(\ell_1)$ denotes the number of Steiner points on $\ell_1$.}\\[1ex]
{\bf Proof:} First consider the case when the segments $\l$
and $\l_1$ lie in different tetrahedra. We denote the weights of the
tetrahedra containing $\l$ and $\l_1$ by $w^-$ and $w^+$,
respectively. Let $F$ be the plane defined by the face $f$ and let
$F^-$ and $F^+$ be the two half-spaces defined by $F$, where we
assume that $\l$ is in $F^-$ and $\l_1$ is in $F^+$. Furthermore, we
assign weights $w^-$ and $w^+$ to $F^-$ and $F^+$, respectively, and
we consider shortest weighted path $\bar{\pi}(u,y)$ between $u$, that
is on $\l$, and an arbitrary point $y$ on $\l_1$.

As discussed in Section \ref{model}, in this case, the path $\bar{\pi}(u,y)$
has the form $\bar{\pi}(u,y)=\{u,a(y),y\}$, where the point $a(y)$ lies in
$F$ and is uniquely defined by  Snell's law (Figure \ref{snells1} (a)).
By our definition, there is an edge joining $u$ to a Steiner point $u_1\in\l_1$
if and only if the point $a(u_1)$ lies in the interior of the triangle $f$.
The interior can be represented as intersection of three half-planes
defined by the lines containing the sides of $f$. So, we first obtain
an upper bound on the number of maximal intervals covered by $A(u,\l_1)$ in
the case where $f$ is a half-plane defined by an arbitrary line $L$ in $F$.

There is one-to-one correspondence between end-points of the maximal
intervals and the points $y$ on $\l_1$ for which $a(y)$ lies on $L$.
Hence, the number of maximal intervals covered by $A(u,\l_1)$ can be
estimated by counting the number of points $y$ for which $a(y)$ lies
on $L$.
When the point $y$ traverses the segment $\l_1$, the bending point
$a(y)$ defines a curve in the plane $F$, which we denote by $\a(y)$.
We consider Cartesian coordinate system $O_{\mu,\nu}$ in $F$, such
that segment $\l_1$ projects onto the segment $(\mu_1,\mu_2)$ of
the $\mu$-axis,  and $u$ projects onto $\nu$-axis, say, at the point
$u'=(0,\nu_0)$. We denote the $\mu$-coordinate of the projection of $y$
by $\mu(y)$ or simply by $\mu$ when no ambiguity arises. Then, as we
discussed in Section \ref{model}, the curve $\a(y)$ has a representation
$\a(\mu(y))=(\tau\mu(y),(1-\tau)\nu_0)$, where $\tau$ is the unique solution
of the equation (\ref{tau}).

Let $L(\mu,\nu)$ be the linear function, such that $L(\mu,\nu)=0$ represents
the line $L$ in $O_{\mu,\nu}$. Then, a point $a(y)$ belongs to $L$ if
$L(\tau\mu(y),(1-\tau)\nu_0)=0$. Thus, we obtain the following system
of algebraic equations for $\tau$ and $\mu(y)$
\begin{equation}\label{explicit}
\left\{
\begin{array}{l}
\frac{w^-\tau}{\sqrt{\tau^2(\mu^2(y)+\nu_0^2)+(z^-)^2}} =
\frac{w^+(1-\tau)}{\sqrt{(1-\tau)^2(\mu^2(y)+\nu_0^2)+(z^+)^2}}\\
L(\tau\mu(y),(1-\tau)\nu_0)=0,
\end{array}
\right.
\end{equation}
where $z^+$ is the Euclidean distance from $\l_1$ to $F$ and $z^-$ is the
Euclidean distance from $u$ to $F$. Excluding $\tau$ from this system
leads to a degree four algebraic equation for $\mu(y)$. Therefore, there
may be no more than four intersections between $\a(y)$ and $L$ and, hence,
the number of the maximal intervals covered by $A(u,\l_1)$ in the case where
$f$ is a half-plane is at most three.

In the case where $f$ is a triangle, we denote by $f_1$, $f_2$, and $f_3$
the three half-planes defining $f$ and by $I_1$, $I_2$, and $I_3$ the corresponding
sets of maximal intervals. Then, any maximal interval $\triangle$ defined by $f$
is obtained as an intersection $\triangle_1\cap\triangle_2\cap\triangle_3$, where
$\triangle_i\in I_i$, $i=1,2,3$. Each of the sets $I_i$ contains at most three
intervals and it is easily seen that the number of intervals that are intersections
of the type $\triangle_1\cap\triangle_2\cap\triangle_3$ does not exceed seven.

The list $\bar{A}(u,\l_1)$ is computed by finding the position of the solutions of the systems
(\ref{explicit}) with respect to the elements of $V(\ell_1)$. This can be done by performing binary search and
hence the computation of the list $\bar{A}(u,\l_1)$ in this case takes $O(\log K(\l_1))$ time.

\begin{figure}
\begin{center}
\resizebox{15cm}{!}{
\input{f_app3_2.pstex_t}}
\caption{The figure illustrates the curves $\a(y)$ and $\a_1(y)$
in the case (a) when $\nu^-+\nu^+>\nu_0$ and in the case (b)
when  $\nu^-+\nu^+\leq \nu_0$. A line $L$ and its intersections with the
curves $\a(y)$ and $\a_1(y)$ are shown. The number of maximal intervals
covered by $E(v,\l_1)$ in the case when $f$ is the lower half-plane
defined by $L$, is 2 for both instances (a) and (b).}
\label{f_aap3_1}
\end{center}
\end{figure}
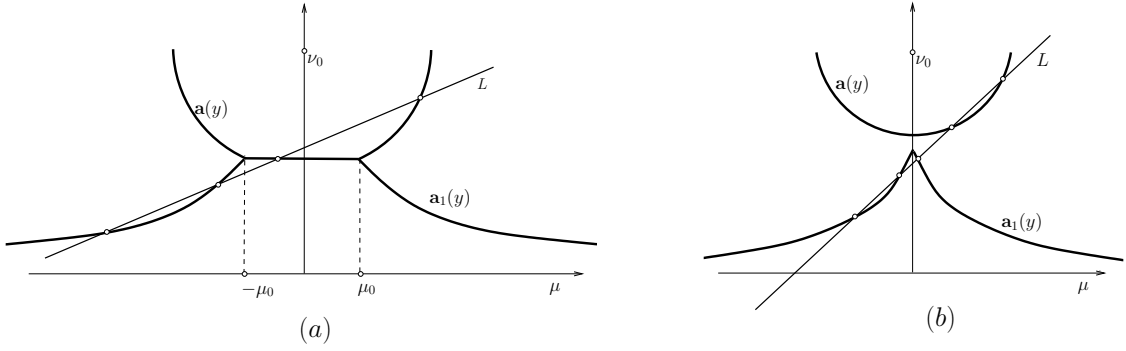

Next, we consider the case where $u$ and $\l_1$ lie in the same tetrahedron,
say the one that is in $F^-$. If the weight $w^-$ is smaller than $w^+$, then
the path $\bar{\pi}(u,y)$ has the form $\{u,a(y),y\}$.  The point $a(y)$
is the point in $F$ that lies on the segment $(u,y')$, where
$y'$ is the point symmetric to $y$ with respect to $F$. So, the curve $\a(y)$,
in this case, is a segment. Hence, each of the sets $I_i$, for $i=1,2,3$, in this case,
is either empty or consists of a single interval. Consequently, there can be at
most one interval obtained as intersection of intervals in these sets. Thus,
in this case, there can be no more than one maximal interval covered by $A(u,\l_1)$.

Finally, we consider the case where $w^-$ is greater than $w^+$. In this case,
the shortest path $\bar{\pi}(u,y)$ has the form $\{u,a(y),a_1(y),y\}$, where the
segment $(a(y),a_1(y))$ lies in $F$. We discussed the structure of this path
in Section \ref{model} and illustrated it in (Figure \ref{snells1} (b)). The curves
$\a(y)$ and $\a_1(y)$ have explicit representations as we detail below.
We set $\nu^-=z^-\tan\varphi^*$ and $\nu^+=z^+\tan\varphi^*$, where the critical angle
$\varphi^*$ is defined by $\sin\varphi^*=w^+/w^-$ and $z^-$, $z^+$ are the distances from
$v$ and $\l_1$ to $F$, respectively.

In this notation the curves $a(y)$ and $a_1(y)$ have the following representations
in $O_{\mu,\nu}$

\begin{equation} \label{explicit-a}
\a(y)=\left\{
\begin{array}{lcl}
\nu = \frac{z^+\nu_0}{z^-+z^+} & \quad {\rm for} \quad & |\mu| < \mu_0\\
\nu = \nu_0 - \sqrt{(\nu^-)^2 - \mu^2} & \quad {\rm for} \quad &
\mu_0 \geq |\mu| \leq \nu^-,
\end{array}
\right.
\end{equation}
\begin{equation} \label{explicit-a1}
\a_1(y)=\left\{
\begin{array}{lcl}
\nu = \frac{z^+\nu_0}{z^-+z^+} & \quad {\rm for} \quad & |\mu| < \mu_0\\
(\nu_0-\nu)\sqrt{(\nu^+)^2\ - \nu^2}=\mu\nu & \quad {\rm for} \quad &
|\mu| \geq \mu_0,
\end{array}
\right.,
\end{equation}
where $\mu_0=\frac{z^-}{z^-+z^+}\sqrt{(\nu^-+\nu^+)^2 - \nu_0^2}$ if
$\nu^-+\nu^->\nu_0$ and $\mu_0=0$ if $\nu^-+\nu^-\leq \nu_0$.

In the case where $\nu^-+\nu^->\nu_0$ (Figure \ref{f_aap3_1} (b)),
the curve $\a(y)$ is a half-circle centered at the point $(0,\nu_0)$
with radius $\nu^-$. The curve $\a_1(y)$ is symmetric with respect
to the $\nu$-axis. The part of $\a_1(y)$ to the right of $O_\nu$
is monotonically decreasing. It is convex and approaches the
$\mu$-axis at infinity. In the case where $\nu^-+\nu^-\leq \nu_0$
(Figure \ref{f_aap3_1} (a)), the curves $\a(y)$ and $\a_1(y)$ have
a common part -- a horizontal segment that projects at $(-\mu_0,\mu_0)$
on the $\mu$-axis. For $|\mu|>\mu_0$, the curves are the same as in
the case $\nu^-+\nu^->\nu_0$.

Again, we consider, first, the case when $f$ is a half-plane defined by a line
$L$ in $F$. There is an edge between $u$ and a point $y$ on $\l_1$ if
and only if the segment $(a(y),a_1(y))$ lies entirely inside $f$,
i.e. in one of the half-planes defined by $L$. By a simple case analysis,
we determine that in the case where $f$ is a half-plane, the number of
maximal intervals covered by $A(u,\l_1)$ is at most 2.

Hence, in the general case when $f$ is a triangle, each of the sets $I_1$, $I_2$,
and $I_3$, as defined above, contains at most 2 intervals. The number of
intervals that can be intersections of the type $\cap_{i=1}^3\triangle_i$
with $\triangle_i\in I_i$ is at most 4. The latter proves that the number of maximal
intervals covered by  $A(u,\l_1)$ in the case where $\l$ and $\l_1$ are in the same
tetrahedron, whose weight $w^-$ is bigger than the weight $w^+$ of the neighboring
tetrahedron, is at most 4.

Computation of the list $\bar{A}(u,\l_1)$ in this case is done again by binary search and takes
$O(K(\l_1))$ time. The lemma is proved.\hfill $\Box$

\end{document}

%% file: snells.pstex_t
\begin{picture}(0,0)%
\includegraphics{snells.pstex}%
\end{picture}%
\setlength{\unitlength}{4144sp}%
\begingroup\makeatletter\ifx\SetFigFont\undefined%
\gdef\SetFigFont#1#2#3#4#5{%
  \reset@font\fontsize{#1}{#2pt}%
  \fontfamily{#3}\fontseries{#4}\fontshape{#5}%
  \selectfont}%
\fi\endgroup%
\begin{picture}(7452,4362)(1152,-4563)
\put(4951,-2221){\makebox(0,0)[lb]{\smash{{\SetFigFont{14}{16.8}{\familydefault}{\mddefault}{\updefault}{\color[rgb]{0,0,0}$\theta^+$}%
}}}}
\put(6526,-826){\makebox(0,0)[lb]{\smash{{\SetFigFont{14}{16.8}{\familydefault}{\mddefault}{\updefault}{\color[rgb]{0,0,0}$s^+$}%
}}}}
\put(3106,-2491){\makebox(0,0)[lb]{\smash{{\SetFigFont{14}{16.8}{\familydefault}{\mddefault}{\updefault}{\color[rgb]{0,0,0}$\theta^-$}%
}}}}
\put(4141,-2446){\makebox(0,0)[lb]{\smash{{\SetFigFont{14}{16.8}{\familydefault}{\mddefault}{\updefault}{\color[rgb]{0,0,0}$a$}%
}}}}
\put(4231,-1771){\makebox(0,0)[lb]{\smash{{\SetFigFont{14}{16.8}{\familydefault}{\mddefault}{\updefault}{\color[rgb]{0,0,0}$\varphi^+$}%
}}}}
\put(3376,-3661){\makebox(0,0)[lb]{\smash{{\SetFigFont{14}{16.8}{\familydefault}{\mddefault}{\updefault}{\color[rgb]{0,0,0}$s^-$}%
}}}}
\put(3781,-3076){\makebox(0,0)[lb]{\smash{{\SetFigFont{14}{16.8}{\familydefault}{\mddefault}{\updefault}{\color[rgb]{0,0,0}$\varphi^-$}%
}}}}
\put(1756,-2671){\makebox(0,0)[lb]{\smash{{\SetFigFont{14}{16.8}{\familydefault}{\mddefault}{\updefault}{\color[rgb]{0,0,0}$f$}%
}}}}
\end{picture}%

%% file: f_wdf.pstex_t
\begin{picture}(0,0)%
\includegraphics{f_wdf.pstex}%
\end{picture}%
\setlength{\unitlength}{4144sp}%
\begingroup\makeatletter\ifx\SetFigFont\undefined%
\gdef\SetFigFont#1#2#3#4#5{%
  \reset@font\fontsize{#1}{#2pt}%
  \fontfamily{#3}\fontseries{#4}\fontshape{#5}%
  \selectfont}%
\fi\endgroup%
\begin{picture}(10420,6581)(1889,-7144)
\put(4816,-3606){\makebox(0,0)[lb]{\smash{{\SetFigFont{20}{24.0}{\rmdefault}{\mddefault}{\updefault}{\color[rgb]{0,0,0}$O$}%
}}}}
\put(7739,-6399){\makebox(0,0)[lb]{\smash{{\SetFigFont{20}{24.0}{\rmdefault}{\mddefault}{\updefault}{\color[rgb]{0,0,0}$F$}%
}}}}
\put(4689,-4836){\makebox(0,0)[lb]{\smash{{\SetFigFont{14}{16.8}{\rmdefault}{\mddefault}{\updefault}{\color[rgb]{0,0,0}$a$}%
}}}}
\put(5221,-2616){\makebox(0,0)[lb]{\smash{{\SetFigFont{14}{16.8}{\rmdefault}{\mddefault}{\updefault}{\color[rgb]{0,0,0}$z^+$}%
}}}}
\put(3751,-4571){\makebox(0,0)[lb]{\smash{{\SetFigFont{14}{16.8}{\rmdefault}{\mddefault}{\updefault}{\color[rgb]{0,0,0}$y^+$}%
}}}}
\put(3376,-5136){\makebox(0,0)[lb]{\smash{{\SetFigFont{14}{16.8}{\rmdefault}{\mddefault}{\updefault}{\color[rgb]{0,0,0}$y=y^++y^-$}%
}}}}
\put(7651,-3786){\makebox(0,0)[lb]{\smash{{\SetFigFont{14}{16.8}{\rmdefault}{\mddefault}{\updefault}{\color[rgb]{0,0,0}$\theta$}%
}}}}
\put(8439,-3849){\makebox(0,0)[lb]{\smash{{\SetFigFont{14}{16.8}{\rmdefault}{\mddefault}{\updefault}{\color[rgb]{0,0,0}$x=x^++x^-$}%
}}}}
\put(5301,-5434){\makebox(0,0)[lb]{\smash{{\SetFigFont{14}{16.8}{\rmdefault}{\mddefault}{\updefault}{\color[rgb]{0,0,0}$z^-$}%
}}}}
\put(6064,-3546){\makebox(0,0)[lb]{\smash{{\SetFigFont{14}{16.8}{\rmdefault}{\mddefault}{\updefault}{\color[rgb]{0,0,0}$x^-$}%
}}}}
\put(3376,-2611){\makebox(0,0)[lb]{\smash{{\SetFigFont{14}{16.8}{\rmdefault}{\mddefault}{\updefault}{\color[rgb]{0,0,0}$\ell$}%
}}}}
\put(9073,-2579){\makebox(0,0)[lb]{\smash{{\SetFigFont{14}{16.8}{\rmdefault}{\mddefault}{\updefault}{\color[rgb]{0,0,0}$\alpha_\kappa$}%
}}}}
\put(8106,-2617){\makebox(0,0)[lb]{\smash{{\SetFigFont{14}{16.8}{\rmdefault}{\mddefault}{\updefault}{\color[rgb]{0,0,0}${\mathbf x}$}%
}}}}
\put(3654,-6749){\makebox(0,0)[lb]{\smash{{\SetFigFont{14}{16.8}{\rmdefault}{\mddefault}{\updefault}{\color[rgb]{0,0,0}$\alpha$}%
}}}}
\put(4743,-4254){\makebox(0,0)[lb]{\smash{{\SetFigFont{14}{16.8}{\rmdefault}{\mddefault}{\updefault}{\color[rgb]{0,0,0}$\varphi_\kappa$}%
}}}}
\put(4391,-5328){\makebox(0,0)[lb]{\smash{{\SetFigFont{14}{16.8}{\rmdefault}{\mddefault}{\updefault}{\color[rgb]{0,0,0}$\varphi$}%
}}}}
\put(5308,-841){\makebox(0,0)[lb]{\smash{{\SetFigFont{14}{16.8}{\rmdefault}{\mddefault}{\updefault}{\color[rgb]{0,0,0}$z$}%
}}}}
\put(3016,-7071){\makebox(0,0)[lb]{\smash{{\SetFigFont{14}{16.8}{\rmdefault}{\mddefault}{\updefault}{\color[rgb]{0,0,0}$v$}%
}}}}
\end{picture}%

%% file: snells1.pstex_t
\begin{picture}(0,0)%
\includegraphics{snells1.pstex}%
\end{picture}%
\setlength{\unitlength}{4144sp}%
\begingroup\makeatletter\ifx\SetFigFont\undefined%
\gdef\SetFigFont#1#2#3#4#5{%
  \reset@font\fontsize{#1}{#2pt}%
  \fontfamily{#3}\fontseries{#4}\fontshape{#5}%
  \selectfont}%
\fi\endgroup%
\begin{picture}(13601,5039)(1236,-4896)
\put(13445,-52){\makebox(0,0)[lb]{\smash{{\SetFigFont{14}{16.8}{\rmdefault}{\mddefault}{\updefault}{\color[rgb]{0,0,0}$\mathbf{x}$}%
}}}}
\put(10162,-68){\makebox(0,0)[lb]{\smash{{\SetFigFont{14}{16.8}{\rmdefault}{\mddefault}{\updefault}{\color[rgb]{0,0,0}$\ell$}%
}}}}
\put(9766,-2319){\makebox(0,0)[lb]{\smash{{\SetFigFont{14}{16.8}{\rmdefault}{\mddefault}{\updefault}{\color[rgb]{0,0,0}$a$}%
}}}}
\put(12553,-1263){\makebox(0,0)[lb]{\smash{{\SetFigFont{14}{16.8}{\rmdefault}{\mddefault}{\updefault}{\color[rgb]{0,0,0}$a_1$}%
}}}}
\put(9410,-2840){\makebox(0,0)[lb]{\smash{{\SetFigFont{14}{16.8}{\rmdefault}{\mddefault}{\updefault}{\color[rgb]{0,0,0}$\varphi^*$}%
}}}}
\put(8245,-4670){\makebox(0,0)[lb]{\smash{{\SetFigFont{14}{16.8}{\rmdefault}{\mddefault}{\updefault}{\color[rgb]{0,0,0}$v$}%
}}}}
\put(12871,-1801){\makebox(0,0)[lb]{\smash{{\SetFigFont{14}{16.8}{\rmdefault}{\mddefault}{\updefault}{\color[rgb]{0,0,0}$\varphi^*$}%
}}}}
\put(4125,-2329){\makebox(0,0)[lb]{\smash{{\SetFigFont{14}{16.8}{\rmdefault}{\mddefault}{\updefault}{\color[rgb]{0,0,0}$a$}%
}}}}
\put(3437,-3417){\makebox(0,0)[lb]{\smash{{\SetFigFont{14}{16.8}{\rmdefault}{\mddefault}{\updefault}{\color[rgb]{0,0,0}$s^-$}%
}}}}
\put(6375,-817){\makebox(0,0)[lb]{\smash{{\SetFigFont{14}{16.8}{\rmdefault}{\mddefault}{\updefault}{\color[rgb]{0,0,0}$s^+$}%
}}}}
\put(4162,-1704){\makebox(0,0)[lb]{\smash{{\SetFigFont{14}{16.8}{\rmdefault}{\mddefault}{\updefault}{\color[rgb]{0,0,0}$\varphi^+$}%
}}}}
\put(3789,-3002){\makebox(0,0)[lb]{\smash{{\SetFigFont{14}{16.8}{\rmdefault}{\mddefault}{\updefault}{\color[rgb]{0,0,0}$\varphi^-$}%
}}}}
\put(4957,-2124){\makebox(0,0)[lb]{\smash{{\SetFigFont{14}{16.8}{\rmdefault}{\mddefault}{\updefault}{\color[rgb]{0,0,0}$\theta^+$}%
}}}}
\put(3076,-2371){\makebox(0,0)[lb]{\smash{{\SetFigFont{14}{16.8}{\rmdefault}{\mddefault}{\updefault}{\color[rgb]{0,0,0}$\theta^-$}%
}}}}
\put(4608,-118){\makebox(0,0)[lb]{\smash{{\SetFigFont{14}{16.8}{\rmdefault}{\mddefault}{\updefault}{\color[rgb]{0,0,0}$\ell$}%
}}}}
\put(7696,-106){\makebox(0,0)[lb]{\smash{{\SetFigFont{14}{16.8}{\rmdefault}{\mddefault}{\updefault}{\color[rgb]{0,0,0}$\mathbf{x}$}%
}}}}
\put(2596,-4693){\makebox(0,0)[lb]{\smash{{\SetFigFont{14}{16.8}{\rmdefault}{\mddefault}{\updefault}{\color[rgb]{0,0,0}$v$}%
}}}}
\put(1858,-2606){\makebox(0,0)[lb]{\smash{{\SetFigFont{14}{16.8}{\rmdefault}{\mddefault}{\updefault}{\color[rgb]{0,0,0}$F$}%
}}}}
\put(7285,-2620){\makebox(0,0)[lb]{\smash{{\SetFigFont{14}{16.8}{\rmdefault}{\mddefault}{\updefault}{\color[rgb]{0,0,0}$F$}%
}}}}
\end{picture}%

%% file: alpha.pstex_t
\begin{picture}(0,0)%
\includegraphics{alpha.pstex}%
\end{picture}%
\setlength{\unitlength}{4144sp}%
\begingroup\makeatletter\ifx\SetFigFont\undefined%
\gdef\SetFigFont#1#2#3#4#5{%
  \reset@font\fontsize{#1}{#2pt}%
  \fontfamily{#3}\fontseries{#4}\fontshape{#5}%
  \selectfont}%
\fi\endgroup%
\begin{picture}(8477,4575)(952,-5173)
\put(9289,-1105){\makebox(0,0)[lb]{\smash{{\SetFigFont{14}{16.8}{\familydefault}{\mddefault}{\updefault}{\color[rgb]{0,0,0}$\overrightarrow{e}_1$}%
}}}}
\put(2878,-4292){\makebox(0,0)[lb]{\smash{{\SetFigFont{14}{16.8}{\familydefault}{\mddefault}{\updefault}{\color[rgb]{0,0,0}$\alpha({x_0})$}%
}}}}
\put(3869,-4642){\makebox(0,0)[lb]{\smash{{\SetFigFont{14}{16.8}{\familydefault}{\mddefault}{\updefault}{\color[rgb]{0,0,0}$\alpha'({x_0})$}%
}}}}
\put(1246,-3623){\makebox(0,0)[lb]{\smash{{\SetFigFont{17}{20.4}{\familydefault}{\mddefault}{\updefault}{\color[rgb]{0,0,0}$F$}%
}}}}
\put(3089,-4724){\makebox(0,0)[lb]{\smash{{\SetFigFont{14}{16.8}{\familydefault}{\mddefault}{\updefault}{\color[rgb]{0,0,0}$\overrightarrow{e}_1$}%
}}}}
\put(4151,-5086){\makebox(0,0)[lb]{\smash{{\SetFigFont{14}{16.8}{\familydefault}{\mddefault}{\updefault}{\color[rgb]{0,0,0}$\overrightarrow{e}_1$}%
}}}}
\put(5561,-805){\makebox(0,0)[lb]{\smash{{\SetFigFont{14}{16.8}{\familydefault}{\mddefault}{\updefault}{\color[rgb]{0,0,0}$\alpha'_\kappa(x_0)$}%
}}}}
\put(5213,-1565){\makebox(0,0)[lb]{\smash{{\SetFigFont{14}{16.8}{\familydefault}{\mddefault}{\updefault}{\color[rgb]{0,0,0}$\alpha_\kappa(x_0)$}%
}}}}
\put(4461,-3178){\makebox(0,0)[lb]{\smash{{\SetFigFont{14}{16.8}{\familydefault}{\mddefault}{\updefault}{\color[rgb]{0,0,0}$a'(x_0)$}%
}}}}
\put(2812,-1247){\makebox(0,0)[lb]{\smash{{\SetFigFont{14}{16.8}{\familydefault}{\mddefault}{\updefault}{\color[rgb]{0,0,0}$\ell$}%
}}}}
\put(3360,-2662){\makebox(0,0)[lb]{\smash{{\SetFigFont{14}{16.8}{\familydefault}{\mddefault}{\updefault}{\color[rgb]{0,0,0}$a(x_0)$}%
}}}}
\put(2557,-4661){\makebox(0,0)[lb]{\smash{{\SetFigFont{14}{16.8}{\familydefault}{\mddefault}{\updefault}{\color[rgb]{0,0,0}$v$}%
}}}}
\put(3613,-4962){\makebox(0,0)[lb]{\smash{{\SetFigFont{14}{16.8}{\familydefault}{\mddefault}{\updefault}{\color[rgb]{0,0,0}$v'$}%
}}}}
\put(6592,-1236){\makebox(0,0)[lb]{\smash{{\SetFigFont{14}{16.8}{\familydefault}{\mddefault}{\updefault}{\color[rgb]{0,0,0}$\mathbf x_0$}%
}}}}
\end{picture}%

%% file: diagonals.pstex_t
\begin{picture}(0,0)%
\includegraphics{diagonals.pstex}%
\end{picture}%
\setlength{\unitlength}{4144sp}%
\begingroup\makeatletter\ifx\SetFigFont\undefined%
\gdef\SetFigFont#1#2#3#4#5{%
  \reset@font\fontsize{#1}{#2pt}%
  \fontfamily{#3}\fontseries{#4}\fontshape{#5}%
  \selectfont}%
\fi\endgroup%
\begin{picture}(8284,3396)(952,-3745)
\put(1246,-3623){\makebox(0,0)[lb]{\smash{{\SetFigFont{17}{20.4}{\rmdefault}{\mddefault}{\updefault}{\color[rgb]{0,0,0}$F$}%
}}}}
\put(5561,-805){\makebox(0,0)[lb]{\smash{{\SetFigFont{14}{16.8}{\rmdefault}{\mddefault}{\updefault}{\color[rgb]{0,0,0}$\alpha'_\kappa(x_1^\delta)$}%
}}}}
\put(7336,-1636){\makebox(0,0)[lb]{\smash{{\SetFigFont{14}{16.8}{\rmdefault}{\mddefault}{\updefault}{\color[rgb]{0,0,0}$\alpha_\kappa(x_2^\delta)$}%
}}}}
\put(7111,-556){\makebox(0,0)[lb]{\smash{{\SetFigFont{14}{16.8}{\rmdefault}{\mddefault}{\updefault}{\color[rgb]{0,0,0}$\alpha'_\kappa(x_2^\delta)$}%
}}}}
\put(4906,-1771){\makebox(0,0)[lb]{\smash{{\SetFigFont{14}{16.8}{\rmdefault}{\mddefault}{\updefault}{\color[rgb]{0,0,0}$\alpha_\kappa(x_1^\delta)$}%
}}}}
\put(2812,-1247){\makebox(0,0)[lb]{\smash{{\SetFigFont{14}{16.8}{\rmdefault}{\mddefault}{\updefault}{\color[rgb]{0,0,0}$\ell$}%
}}}}
\put(5176,-2851){\makebox(0,0)[lb]{\smash{{\SetFigFont{14}{16.8}{\rmdefault}{\mddefault}{\updefault}{\color[rgb]{0,0,0}$a'(x_2^\delta)$}%
}}}}
\put(3196,-2491){\makebox(0,0)[lb]{\smash{{\SetFigFont{14}{16.8}{\rmdefault}{\mddefault}{\updefault}{\color[rgb]{0,0,0}$a(x_1^\delta)$}%
}}}}
\put(6796,-1231){\makebox(0,0)[lb]{\smash{{\SetFigFont{14}{16.8}{\rmdefault}{\mddefault}{\updefault}{\color[rgb]{0,0,0}$\mathbf x_0$}%
}}}}
\put(6391,-1231){\makebox(0,0)[lb]{\smash{{\SetFigFont{14}{16.8}{\rmdefault}{\mddefault}{\updefault}{\color[rgb]{0,0,0}$\mathbf x_1^\delta$}%
}}}}
\put(7381,-1231){\makebox(0,0)[lb]{\smash{{\SetFigFont{14}{16.8}{\rmdefault}{\mddefault}{\updefault}{\color[rgb]{0,0,0}$\mathbf x_2^\delta$}%
}}}}
\put(3691,-2986){\makebox(0,0)[lb]{\smash{{\SetFigFont{14}{16.8}{\rmdefault}{\mddefault}{\updefault}{\color[rgb]{0,0,0}$a(x_2^\delta)$}%
}}}}
\put(4276,-3256){\makebox(0,0)[lb]{\smash{{\SetFigFont{14}{16.8}{\rmdefault}{\mddefault}{\updefault}{\color[rgb]{0,0,0}$a'(x_1^\delta)$}%
}}}}
\end{picture}%

%% file: f20.pstex_t
\begin{picture}(0,0)%
\includegraphics{f20.pstex}%
\end{picture}%
\setlength{\unitlength}{3947sp}%
\begingroup\makeatletter\ifx\SetFigFont\undefined%
\gdef\SetFigFont#1#2#3#4#5{%
  \reset@font\fontsize{#1}{#2pt}%
  \fontfamily{#3}\fontseries{#4}\fontshape{#5}%
  \selectfont}%
\fi\endgroup%
\begin{picture}(10744,6489)(1686,-6783)
\put(1701,-3899){\makebox(0,0)[lb]{\smash{{\SetFigFont{20}{24.0}{\rmdefault}{\mddefault}{\updefault}{\color[rgb]{0,0,0}$A$}%
}}}}
\put(3489,-6674){\makebox(0,0)[lb]{\smash{{\SetFigFont{20}{24.0}{\rmdefault}{\mddefault}{\updefault}{\color[rgb]{0,0,0}$B$}%
}}}}
\put(5126,-911){\makebox(0,0)[lb]{\smash{{\SetFigFont{20}{24.0}{\rmdefault}{\mddefault}{\updefault}{\color[rgb]{0,0,0}$D$}%
}}}}
\put(6314,-3886){\makebox(0,0)[lb]{\smash{{\SetFigFont{20}{24.0}{\rmdefault}{\mddefault}{\updefault}{\color[rgb]{0,0,0}$C$}%
}}}}
\put(11214,-5849){\makebox(0,0)[lb]{\smash{{\SetFigFont{20}{24.0}{\rmdefault}{\mddefault}{\updefault}{\color[rgb]{0,0,0}$H$}%
}}}}
\put(11928,-3182){\makebox(0,0)[lb]{\smash{{\SetFigFont{14}{16.8}{\rmdefault}{\mddefault}{\updefault}{\color[rgb]{0,0,0}$B_2$}%
}}}}
\put(12005,-3609){\makebox(0,0)[lb]{\smash{{\SetFigFont{14}{16.8}{\rmdefault}{\mddefault}{\updefault}{\color[rgb]{0,0,0}$B_3$}%
}}}}
\put(11322,-561){\makebox(0,0)[lb]{\smash{{\SetFigFont{20}{24.0}{\rmdefault}{\mddefault}{\updefault}{\color[rgb]{0,0,0}$P$}%
}}}}
\put(7207,-5479){\makebox(0,0)[lb]{\smash{{\SetFigFont{20}{24.0}{\rmdefault}{\mddefault}{\updefault}{\color[rgb]{0,0,0}$A$}%
}}}}
\put(12415,-5470){\makebox(0,0)[lb]{\smash{{\SetFigFont{20}{24.0}{\rmdefault}{\mddefault}{\updefault}{\color[rgb]{0,0,0}$B$}%
}}}}
\put(11800,-2499){\makebox(0,0)[lb]{\smash{{\SetFigFont{14}{16.8}{\rmdefault}{\mddefault}{\updefault}{\color[rgb]{0,0,0}$B_1$}%
}}}}
\put(12101,-6215){\makebox(0,0)[lb]{\smash{{\SetFigFont{20}{24.0}{\rmdefault}{\mddefault}{\updefault}{\color[rgb]{0,0,0}$D_\varepsilon(B)$}%
}}}}
\put(7247,-6297){\makebox(0,0)[lb]{\smash{{\SetFigFont{20}{24.0}{\rmdefault}{\mddefault}{\updefault}{\color[rgb]{0,0,0}$D_\varepsilon(A)$}%
}}}}
\put(9750,-6160){\makebox(0,0)[lb]{\smash{{\SetFigFont{20}{24.0}{\rmdefault}{\mddefault}{\updefault}{\color[rgb]{0,0,0}$D_\varepsilon(AB)$}%
}}}}
\put(12103,-4195){\makebox(0,0)[lb]{\smash{{\SetFigFont{14}{16.8}{\rmdefault}{\mddefault}{\updefault}{\color[rgb]{0,0,0}$B_5$}%
}}}}
\put(12058,-3922){\makebox(0,0)[lb]{\smash{{\SetFigFont{14}{16.8}{\rmdefault}{\mddefault}{\updefault}{\color[rgb]{0,0,0}$B_4$}%
}}}}
\put(6069,-2684){\makebox(0,0)[lb]{\smash{{\SetFigFont{20}{24.0}{\rmdefault}{\mddefault}{\updefault}{\color[rgb]{0,0,0}$P$}%
}}}}
\put(9691,-2508){\makebox(0,0)[lb]{\smash{{\SetFigFont{14}{16.8}{\rmdefault}{\mddefault}{\updefault}{\color[rgb]{0,0,0}$A_1$}%
}}}}
\put(9179,-3182){\makebox(0,0)[lb]{\smash{{\SetFigFont{14}{16.8}{\rmdefault}{\mddefault}{\updefault}{\color[rgb]{0,0,0}$A_2$}%
}}}}
\put(8855,-3626){\makebox(0,0)[lb]{\smash{{\SetFigFont{14}{16.8}{\rmdefault}{\mddefault}{\updefault}{\color[rgb]{0,0,0}$A_3$}%
}}}}
\put(8453,-4164){\makebox(0,0)[lb]{\smash{{\SetFigFont{14}{16.8}{\rmdefault}{\mddefault}{\updefault}{\color[rgb]{0,0,0}$A_5$}%
}}}}
\put(8616,-3934){\makebox(0,0)[lb]{\smash{{\SetFigFont{14}{16.8}{\rmdefault}{\mddefault}{\updefault}{\color[rgb]{0,0,0}$A_4$}%
}}}}
\end{picture}%

%% file: f30.pstex_t
\begin{picture}(0,0)%
\includegraphics{f30.pstex}%
\end{picture}%
\setlength{\unitlength}{3947sp}%
\begingroup\makeatletter\ifx\SetFigFont\undefined%
\gdef\SetFigFont#1#2#3#4#5{%
  \reset@font\fontsize{#1}{#2pt}%
  \fontfamily{#3}\fontseries{#4}\fontshape{#5}%
  \selectfont}%
\fi\endgroup%
\begin{picture}(4630,6051)(1711,-6708)
\put(5314,-924){\makebox(0,0)[lb]{\smash{{\SetFigFont{20}{24.0}{\rmdefault}{\mddefault}{\updefault}{\color[rgb]{0,0,0}$D$}%
}}}}
\put(1726,-3936){\makebox(0,0)[lb]{\smash{{\SetFigFont{20}{24.0}{\rmdefault}{\mddefault}{\updefault}{\color[rgb]{0,0,0}$A$}%
}}}}
\put(5989,-2749){\makebox(0,0)[lb]{\smash{{\SetFigFont{20}{24.0}{\rmdefault}{\mddefault}{\updefault}{\color[rgb]{0,0,0}$P$}%
}}}}
\put(3614,-6599){\makebox(0,0)[lb]{\smash{{\SetFigFont{20}{24.0}{\rmdefault}{\mddefault}{\updefault}{\color[rgb]{0,0,0}$B$}%
}}}}
\put(6326,-3911){\makebox(0,0)[lb]{\smash{{\SetFigFont{20}{24.0}{\rmdefault}{\mddefault}{\updefault}{\color[rgb]{0,0,0}$C$}%
}}}}
\put(5354,-4369){\makebox(0,0)[lb]{\smash{{\SetFigFont{14}{16.8}{\rmdefault}{\mddefault}{\updefault}{\color[rgb]{0,0,0}$x_1$}%
}}}}
\put(3809,-2969){\makebox(0,0)[lb]{\smash{{\SetFigFont{14}{16.8}{\rmdefault}{\mddefault}{\updefault}{\color[rgb]{0,0,0}$x_2$}%
}}}}
\put(4936,-3686){\makebox(0,0)[lb]{\smash{{\SetFigFont{14}{16.8}{\rmdefault}{\mddefault}{\updefault}{\color[rgb]{0,0,0}$q$}%
}}}}
\put(4588,-4111){\makebox(0,0)[lb]{\smash{{\SetFigFont{14}{16.8}{\rmdefault}{\mddefault}{\updefault}{\color[rgb]{0,0,0}$x_0$}%
}}}}
\put(2613,-4098){\makebox(0,0)[lb]{\smash{{\SetFigFont{14}{16.8}{\rmdefault}{\mddefault}{\updefault}{\color[rgb]{0,0,0}$\gamma$}%
}}}}
\end{picture}%

%% file: Step1.pstex_t
\begin{picture}(0,0)%
\includegraphics{Step1.pstex}%
\end{picture}%
\setlength{\unitlength}{3947sp}%
\begingroup\makeatletter\ifx\SetFigFont\undefined%
\gdef\SetFigFont#1#2#3#4#5{%
  \reset@font\fontsize{#1}{#2pt}%
  \fontfamily{#3}\fontseries{#4}\fontshape{#5}%
  \selectfont}%
\fi\endgroup%
\begin{picture}(6770,2418)(1527,-1703)
\put(1542,  9){\makebox(0,0)[lb]{\smash{{\SetFigFont{12}{14.4}{\familydefault}{\mddefault}{\updefault}{\color[rgb]{0,0,0}$x_1$}%
}}}}
\put(2472,-84){\makebox(0,0)[lb]{\smash{{\SetFigFont{12}{14.4}{\familydefault}{\mddefault}{\updefault}{\color[rgb]{0,0,0}$p$}%
}}}}
\put(2482,-941){\makebox(0,0)[lb]{\smash{{\SetFigFont{12}{14.4}{\familydefault}{\mddefault}{\updefault}{\color[rgb]{0,0,0}$b$}%
}}}}
\put(2240,-404){\makebox(0,0)[lb]{\smash{{\SetFigFont{12}{14.4}{\familydefault}{\mddefault}{\updefault}{\color[rgb]{0,0,0}$s$}%
}}}}
\put(3237,-419){\makebox(0,0)[lb]{\smash{{\SetFigFont{12}{14.4}{\familydefault}{\mddefault}{\updefault}{\color[rgb]{0,0,0}$x_2$}%
}}}}
\put(2364,-1630){\makebox(0,0)[lb]{\smash{{\SetFigFont{12}{14.4}{\familydefault}{\mddefault}{\updefault}(a)}}}}
\put(6414,-1630){\makebox(0,0)[lb]{\smash{{\SetFigFont{12}{14.4}{\familydefault}{\mddefault}{\updefault}(b)}}}}
\put(7545,-1129){\makebox(0,0)[lb]{\smash{{\SetFigFont{12}{14.4}{\familydefault}{\mddefault}{\updefault}$\tp_2$}}}}
\put(4945,-1129){\makebox(0,0)[lb]{\smash{{\SetFigFont{12}{14.4}{\familydefault}{\mddefault}{\updefault}$\tp$}}}}
\put(6251,-1129){\makebox(0,0)[lb]{\smash{{\SetFigFont{12}{14.4}{\familydefault}{\mddefault}{\updefault}$\tp_1$}}}}
\put(4541,-308){\makebox(0,0)[lb]{\smash{{\SetFigFont{12}{14.4}{\familydefault}{\mddefault}{\updefault}$\tp_1$}}}}
\put(4925,-467){\makebox(0,0)[lb]{\smash{{\SetFigFont{12}{14.4}{\familydefault}{\mddefault}{\updefault}$\tp$}}}}
\put(6228,-559){\makebox(0,0)[lb]{\smash{{\SetFigFont{12}{14.4}{\familydefault}{\mddefault}{\updefault}$x_3$}}}}
\put(7272,-182){\makebox(0,0)[lb]{\smash{{\SetFigFont{12}{14.4}{\familydefault}{\mddefault}{\updefault}$x_4$}}}}
\put(8116,-22){\makebox(0,0)[lb]{\smash{{\SetFigFont{12}{14.4}{\familydefault}{\mddefault}{\updefault}$x_5$}}}}
\put(7666,385){\makebox(0,0)[lb]{\smash{{\SetFigFont{12}{14.4}{\familydefault}{\mddefault}{\updefault}$p_4$}}}}
\put(6610, 14){\makebox(0,0)[lb]{\smash{{\SetFigFont{12}{14.4}{\familydefault}{\mddefault}{\updefault}$p_3$}}}}
\put(5913,144){\makebox(0,0)[lb]{\smash{{\SetFigFont{12}{14.4}{\familydefault}{\mddefault}{\updefault}$p_2$}}}}
\put(4870,  8){\makebox(0,0)[lb]{\smash{{\SetFigFont{12}{14.4}{\familydefault}{\mddefault}{\updefault}$p_1$}}}}
\put(4450,-801){\makebox(0,0)[lb]{\smash{{\SetFigFont{12}{14.4}{\familydefault}{\mddefault}{\updefault}$x_1$}}}}
\put(5406,213){\makebox(0,0)[lb]{\smash{{\SetFigFont{12}{14.4}{\familydefault}{\mddefault}{\updefault}$x_2$}}}}
\put(7055,163){\makebox(0,0)[lb]{\smash{{\SetFigFont{12}{14.4}{\familydefault}{\mddefault}{\updefault}$\tp_2$}}}}
\end{picture}%

%% file: Step4.pstex_t
\begin{picture}(0,0)%
\includegraphics{Step4.pstex}%
\end{picture}%
\setlength{\unitlength}{3947sp}%
\begingroup\makeatletter\ifx\SetFigFont\undefined%
\gdef\SetFigFont#1#2#3#4#5{%
  \reset@font\fontsize{#1}{#2pt}%
  \fontfamily{#3}\fontseries{#4}\fontshape{#5}%
  \selectfont}%
\fi\endgroup%
\begin{picture}(5391,1155)(1093,-583)
\put(4713, 44){\makebox(0,0)[lb]{\smash{{\SetFigFont{12}{14.4}{\familydefault}{\mddefault}{\updefault}$a'_{i,j}$}}}}
\put(1793,324){\makebox(0,0)[lb]{\smash{{\SetFigFont{12}{14.4}{\familydefault}{\mddefault}{\updefault}$b_{i,j}$}}}}
\put(1691,-270){\makebox(0,0)[lb]{\smash{{\SetFigFont{12}{14.4}{\familydefault}{\mddefault}{\updefault}$b'_{i,j}$}}}}
\put(2011,-243){\makebox(0,0)[lb]{\smash{{\SetFigFont{12}{14.4}{\familydefault}{\mddefault}{\updefault}$q_{i,j}$}}}}
\put(4782,-505){\makebox(0,0)[lb]{\smash{{\SetFigFont{12}{14.4}{\familydefault}{\mddefault}{\updefault}$a_{i,j}$}}}}
\put(4352, 44){\makebox(0,0)[lb]{\smash{{\SetFigFont{12}{14.4}{\familydefault}{\mddefault}{\updefault}$p_{i,j}$}}}}
\put(3751,239){\makebox(0,0)[lb]{\smash{{\SetFigFont{12}{14.4}{\familydefault}{\mddefault}{\updefault}$\tp_3$}}}}
\put(5626, 14){\makebox(0,0)[lb]{\smash{{\SetFigFont{12}{14.4}{\familydefault}{\mddefault}{\updefault}$e_j$}}}}
\put(3376,-286){\makebox(0,0)[lb]{\smash{{\SetFigFont{12}{14.4}{\familydefault}{\mddefault}{\updefault}$\tp_4$}}}}
\end{picture}%

%% file: F6.pstex_t
\begin{picture}(0,0)%
\includegraphics{F6.pstex}%
\end{picture}%
\setlength{\unitlength}{1579sp}%
\begingroup\makeatletter\ifx\SetFigFont\undefined%
\gdef\SetFigFont#1#2#3#4#5{%
  \reset@font\fontsize{#1}{#2pt}%
  \fontfamily{#3}\fontseries{#4}\fontshape{#5}%
  \selectfont}%
\fi\endgroup%
\begin{picture}(7929,8211)(2079,-8171)
\put(9001,-1411){\makebox(0,0)[lb]{\smash{{\SetFigFont{9}{10.8}{\familydefault}{\mddefault}{\updefault}{\color[rgb]{0,0,0}$G$}%
}}}}
\put(7126,-3736){\makebox(0,0)[lb]{\smash{{\SetFigFont{9}{10.8}{\familydefault}{\mddefault}{\updefault}{\color[rgb]{0,0,0}$e(S)$}%
}}}}
\put(6751,-6136){\makebox(0,0)[lb]{\smash{{\SetFigFont{9}{10.8}{\familydefault}{\mddefault}{\updefault}{\color[rgb]{0,0,0}$S$}%
}}}}
\put(6451,-4636){\makebox(0,0)[lb]{\smash{{\SetFigFont{9}{10.8}{\familydefault}{\mddefault}{\updefault}{\color[rgb]{0,0,0}$u$}%
}}}}
\put(5326,-5161){\makebox(0,0)[lb]{\smash{{\SetFigFont{9}{10.8}{\familydefault}{\mddefault}{\updefault}{\color[rgb]{0,0,0}$s$}%
}}}}
\put(4501,-2911){\makebox(0,0)[lb]{\smash{{\SetFigFont{9}{10.8}{\familydefault}{\mddefault}{\updefault}{\color[rgb]{0,0,0}$E(S)$}%
}}}}
\put(5251,-4461){\makebox(0,0)[lb]{\smash{{\SetFigFont{9}{10.8}{\familydefault}{\mddefault}{\updefault}{\color[rgb]{0,0,0}$\delta(u)$}%
}}}}
\put(7364,-3074){\makebox(0,0)[lb]{\smash{{\SetFigFont{9}{10.8}{\familydefault}{\mddefault}{\updefault}{\color[rgb]{0,0,0}$v$}%
}}}}
\end{picture}%

%% file: F65.pstex_t
\begin{picture}(0,0)%
\includegraphics{F65.pstex}%
\end{picture}%
\setlength{\unitlength}{4144sp}%
\begingroup\makeatletter\ifx\SetFigFont\undefined%
\gdef\SetFigFont#1#2#3#4#5{%
  \reset@font\fontsize{#1}{#2pt}%
  \fontfamily{#3}\fontseries{#4}\fontshape{#5}%
  \selectfont}%
\fi\endgroup%
\begin{picture}(8310,9156)(3766,-8482)
\put(3781,-5191){\makebox(0,0)[lb]{\smash{{\SetFigFont{29}{34.8}{\familydefault}{\mddefault}{\updefault}{\color[rgb]{0,0,0}$A$}%
}}}}
\put(12061,-4831){\makebox(0,0)[lb]{\smash{{\SetFigFont{29}{34.8}{\familydefault}{\mddefault}{\updefault}{\color[rgb]{0,0,0}$B$}%
}}}}
\put(11071,-8341){\makebox(0,0)[lb]{\smash{{\SetFigFont{29}{34.8}{\familydefault}{\mddefault}{\updefault}{\color[rgb]{0,0,0}$D$}%
}}}}
\put(10351,-2131){\makebox(0,0)[lb]{\smash{{\SetFigFont{29}{34.8}{\familydefault}{\mddefault}{\updefault}{\color[rgb]{0,0,0}$P_1$}%
}}}}
\put(8281,-3661){\makebox(0,0)[lb]{\smash{{\SetFigFont{29}{34.8}{\familydefault}{\mddefault}{\updefault}{\color[rgb]{0,0,0}$\ell_1$}%
}}}}
\put(8551,-6631){\makebox(0,0)[lb]{\smash{{\SetFigFont{29}{34.8}{\familydefault}{\mddefault}{\updefault}{\color[rgb]{0,0,0}$\ell$}%
}}}}
\put(8191,-5911){\makebox(0,0)[lb]{\smash{{\SetFigFont{29}{34.8}{\familydefault}{\mddefault}{\updefault}{\color[rgb]{0,0,0}$\f$}%
}}}}
\put(9901,-6361){\makebox(0,0)[lb]{\smash{{\SetFigFont{29}{34.8}{\familydefault}{\mddefault}{\updefault}{\color[rgb]{0,0,0}$\b$}%
}}}}
\put(8191,299){\makebox(0,0)[lb]{\smash{{\SetFigFont{29}{34.8}{\familydefault}{\mddefault}{\updefault}{\color[rgb]{0,0,0}$D_1$}%
}}}}
\put(5941,-4741){\makebox(0,0)[lb]{\smash{{\SetFigFont{29}{34.8}{\familydefault}{\mddefault}{\updefault}{\color[rgb]{0,0,0}$\b_1$}%
}}}}
\put(5671,-7441){\makebox(0,0)[lb]{\smash{{\SetFigFont{29}{34.8}{\familydefault}{\mddefault}{\updefault}{\color[rgb]{0,0,0}$C$}%
}}}}
\put(8281,-8251){\makebox(0,0)[lb]{\smash{{\SetFigFont{29}{34.8}{\familydefault}{\mddefault}{\updefault}{\color[rgb]{0,0,0}$P$}%
}}}}
\end{picture}%

%% file: F67.pstex_t
\begin{picture}(0,0)%
\includegraphics{F67.pstex}%
\end{picture}%
\setlength{\unitlength}{4144sp}%
\begingroup\makeatletter\ifx\SetFigFont\undefined%
\gdef\SetFigFont#1#2#3#4#5{%
  \reset@font\fontsize{#1}{#2pt}%
  \fontfamily{#3}\fontseries{#4}\fontshape{#5}%
  \selectfont}%
\fi\endgroup%
\begin{picture}(10035,6195)(-14,-6061)
\put(9946,-5011){\makebox(0,0)[lb]{\smash{{\SetFigFont{29}{34.8}{\familydefault}{\mddefault}{\updefault}{\color[rgb]{0,0,0}$\ell$}%
}}}}
\put(8686,-1411){\makebox(0,0)[lb]{\smash{{\SetFigFont{29}{34.8}{\familydefault}{\mddefault}{\updefault}{\color[rgb]{0,0,0}$\ell_1=A_1B_1$}%
}}}}
\put(8641,-241){\makebox(0,0)[lb]{\smash{{\SetFigFont{29}{34.8}{\familydefault}{\mddefault}{\updefault}{\color[rgb]{0,0,0}$x_4=B_1$}%
}}}}
\put(  1,-241){\makebox(0,0)[lb]{\smash{{\SetFigFont{29}{34.8}{\familydefault}{\mddefault}{\updefault}{\color[rgb]{0,0,0}$A_1=x_0$}%
}}}}
\put(2386,-241){\makebox(0,0)[lb]{\smash{{\SetFigFont{29}{34.8}{\familydefault}{\mddefault}{\updefault}{\color[rgb]{0,0,0}$x_1$}%
}}}}
\put(3286,-241){\makebox(0,0)[lb]{\smash{{\SetFigFont{29}{34.8}{\familydefault}{\mddefault}{\updefault}{\color[rgb]{0,0,0}$x^-$}%
}}}}
\put(4636,-241){\makebox(0,0)[lb]{\smash{{\SetFigFont{29}{34.8}{\familydefault}{\mddefault}{\updefault}{\color[rgb]{0,0,0}$x_2$}%
}}}}
\put(5401,-241){\makebox(0,0)[lb]{\smash{{\SetFigFont{29}{34.8}{\familydefault}{\mddefault}{\updefault}{\color[rgb]{0,0,0}$x^+$}%
}}}}
\put(6706,-241){\makebox(0,0)[lb]{\smash{{\SetFigFont{29}{34.8}{\familydefault}{\mddefault}{\updefault}{\color[rgb]{0,0,0}$x_3$}%
}}}}
\put(8461,-5911){\makebox(0,0)[lb]{\smash{{\SetFigFont{29}{34.8}{\familydefault}{\mddefault}{\updefault}{\color[rgb]{0,0,0}$u_2$}%
}}}}
\put(4861,-5911){\makebox(0,0)[lb]{\smash{{\SetFigFont{29}{34.8}{\familydefault}{\mddefault}{\updefault}{\color[rgb]{0,0,0}$v$}%
}}}}
\put(1531,-5911){\makebox(0,0)[lb]{\smash{{\SetFigFont{29}{34.8}{\familydefault}{\mddefault}{\updefault}{\color[rgb]{0,0,0}$u_1$}%
}}}}
\put(3421,-5911){\makebox(0,0)[lb]{\smash{{\SetFigFont{29}{34.8}{\familydefault}{\mddefault}{\updefault}{\color[rgb]{0,0,0}$u_3$}%
}}}}
\end{picture}%

%% file: f_app3_2.pstex_t
\begin{picture}(0,0)%
\includegraphics{f_app3_2.pstex}%
\end{picture}%
\setlength{\unitlength}{4144sp}%
\begingroup\makeatletter\ifx\SetFigFont\undefined%
\gdef\SetFigFont#1#2#3#4#5{%
  \reset@font\fontsize{#1}{#2pt}%
  \fontfamily{#3}\fontseries{#4}\fontshape{#5}%
  \selectfont}%
\fi\endgroup%
\begin{picture}(13650,4194)(227,-5840)
\put(3926,-2404){\makebox(0,0)[lb]{\smash{{\SetFigFont{14}{16.8}{\rmdefault}{\mddefault}{\updefault}{\color[rgb]{0,0,0}$\nu_0$}%
}}}}
\put(5992,-2712){\makebox(0,0)[lb]{\smash{{\SetFigFont{12}{14.4}{\rmdefault}{\mddefault}{\updefault}{\color[rgb]{0,0,0}$L$}%
}}}}
\put(3128,-5205){\makebox(0,0)[lb]{\smash{{\SetFigFont{14}{16.8}{\rmdefault}{\mddefault}{\updefault}{\color[rgb]{0,0,0}$-\mu_0$}%
}}}}
\put(13289,-5166){\makebox(0,0)[lb]{\smash{{\SetFigFont{14}{16.8}{\rmdefault}{\mddefault}{\updefault}{\color[rgb]{0,0,0}$\mu$}%
}}}}
\put(11317,-2418){\makebox(0,0)[lb]{\smash{{\SetFigFont{14}{16.8}{\rmdefault}{\mddefault}{\updefault}{\color[rgb]{0,0,0}$\nu_0$}%
}}}}
\put(12361,-4399){\makebox(0,0)[lb]{\smash{{\SetFigFont{14}{16.8}{\rmdefault}{\mddefault}{\updefault}{\color[rgb]{0,0,0}${\mathbf a_1}(y)$}%
}}}}
\put(10346,-2722){\makebox(0,0)[lb]{\smash{{\SetFigFont{14}{16.8}{\rmdefault}{\mddefault}{\updefault}{\color[rgb]{0,0,0}${\mathbf a}(y)$}%
}}}}
\put(12790,-2434){\makebox(0,0)[lb]{\smash{{\SetFigFont{14}{16.8}{\rmdefault}{\mddefault}{\updefault}{\color[rgb]{0,0,0}$L$}%
}}}}
\put(11439,-5603){\makebox(0,0)[lb]{\smash{{\SetFigFont{20}{24.0}{\rmdefault}{\mddefault}{\updefault}{\color[rgb]{0,0,0}$(b)$}%
}}}}
\put(3826,-5731){\makebox(0,0)[lb]{\smash{{\SetFigFont{20}{24.0}{\rmdefault}{\mddefault}{\updefault}{\color[rgb]{0,0,0}$(a)$}%
}}}}
\put(5413,-4100){\makebox(0,0)[lb]{\smash{{\SetFigFont{14}{16.8}{\rmdefault}{\mddefault}{\updefault}{\color[rgb]{0,0,0}${\mathbf a_1}(y)$}%
}}}}
\put(2555,-3016){\makebox(0,0)[lb]{\smash{{\SetFigFont{14}{16.8}{\rmdefault}{\mddefault}{\updefault}{\color[rgb]{0,0,0}${\mathbf a}(y)$}%
}}}}
\put(4539,-5182){\makebox(0,0)[lb]{\smash{{\SetFigFont{14}{16.8}{\rmdefault}{\mddefault}{\updefault}{\color[rgb]{0,0,0}$\mu_0$}%
}}}}
\put(6870,-5171){\makebox(0,0)[lb]{\smash{{\SetFigFont{14}{16.8}{\rmdefault}{\mddefault}{\updefault}{\color[rgb]{0,0,0}$\mu$}%
}}}}
\end{picture}%